# HOW IS GENE-REGULATORY EVOLUTION AFFECTED BY CELL-TO-CELL VARIABILITY?

**Leonardo I. Estrella Dzib**
Minerva University
14 Mint Plaza,
San Francisco, CA, 94103, USA

**James Holehouse**
jamesholehouse1@gmail.com
The Santa Fe Institute
1399 Hyde Park Road,
Santa Fe, NM, 87501, USA

## ABSTRACT

The evolutionary origins of structural features in reconstructed gene-regulatory networks (GRNs) remain poorly understood, especially given the random aspects of gene expression. Here, we extend a classical model of GRN evolution to allow a single network to express a distribution of phenotypes through noisy developmental dynamics. Inspired by Hopfield networks, we introduce an alignment score that quantifies the cohesion of gene-gene interactions in the network to support a target stable phenotype. Overall, evolved populations optimized their fitness and reduced the length of their developmental paths. Increased noise levels promoted alignment, enriched coherent feedforward and positive feedback loops relative to non-evolved and noiseless controls, and buffered against mutational perturbations. Alignment provides intuitive interpretations because an increased number of appropriately signed gene-gene interactions is more redundant and thus more robust against developmental noise and mutations. Together, these results demonstrate that cell-to-cell variability exerts strong selective pressure, driving the evolution of aligned, robust, and motif-enriched GRN architectures.

## 1 Introduction

Gene-regulatory networks (GRNs) govern how cells process signals and execute gene expression programs, and across organisms they display striking, highly non-random structural features [1–3]. Their study is essential for understanding cellular identity [4], disease mechanisms [5], and even for identifying regulatory targets in cancer therapies [6]. A major challenge in GRN research is the immense combinatorial space of possible interactions among thousands of genes, which has motivated the development of computational methods that exploit empirical structural regularities to infer network topology from data [7–12]. These empirically reconstructed GRNs consistently exhibit properties, such as sparsity [13], high clustering [13], heavy-tailed degree distributions [13], enrichment of specific network motifs like feedforward loops (FFLs) [14], and hierarchical organization [15], that cannot be reproduced by simple random network models. Understanding the evolutionary origins of these topological features and their effect on their function is therefore a central question in systems biology.

Nevertheless, there is little work on the impact of environmental variability on the evolution of GRN structures. The inherent stochasticity in GRN dynamics caused by multiple sources of noise in phenotype expression has long been recognized [17–19]. Their effects might allow populations to exhibit multiple stable phenotypes simultaneously [20] and cause heavy tails in RNA copy-number distributions [21]. Notably, numerous studies have proposed that network structures buffer against environmental noise by means of increased plasticity [22], negative auto-regulation [23], redundancy [24], and canalization [25, 26]. Whether these features appear in response to environmental variability and how they emerge are open questions. Whereas previous research in the field commonly models environmental change as either permanent [27, 28] or cyclical shifts in the optimal phenotype [29, 30], we instead conceptualize environmental variability as the set of mechanisms that cause the expressed phenotype to spread around a fixed average stochastically. For example, Weinreich et al. [31] provide numerous sources of variability in the development of phenotype, ranging from intracellular processes to environmental sensitivity. Through this change in perspective, we ask whether evolutionary processes are reliant not only on genetic variation but also on the interplay between phenotypic expression and cell-to-cell variability in regulatory interactions.

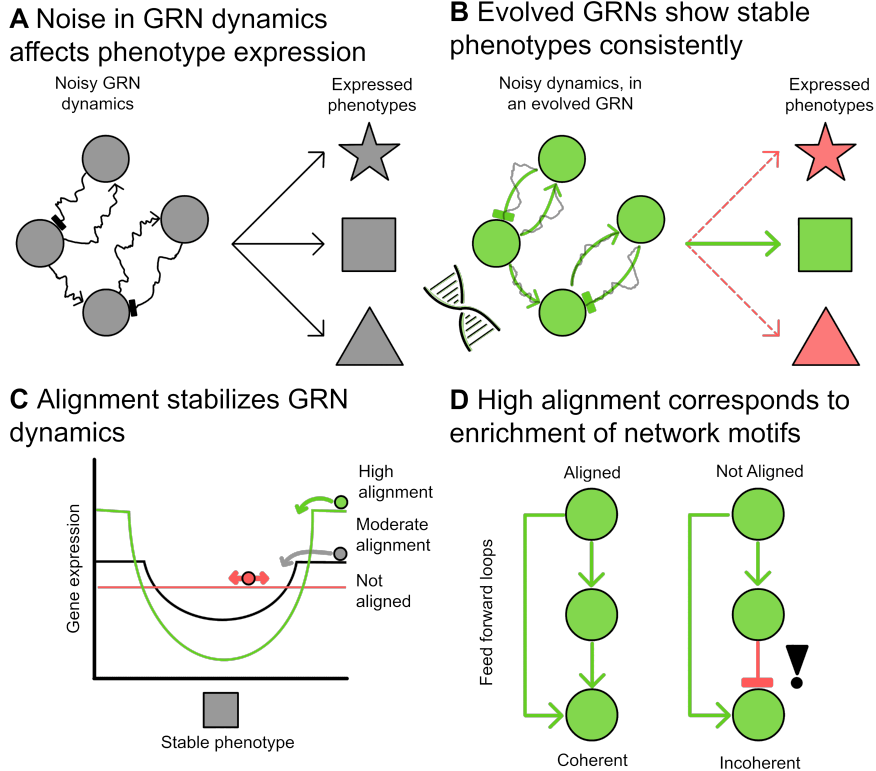

Figure 1: **Summary of the findings of our study.** (A) Stochasticity in the regulatory connections comprising the genotype can lead to a diverse set of phenotypes. (B) GRNs evolved in stochastic environments develop mechanisms to express optimal phenotypes even in highly variable environments. (C) Successful GRNs become highly aligned with the optimal phenotype, meaning that very few regulatory connections lead to deviations from achieving the optimal phenotype. Alignment can be seen as the GRNs developing highly robust solutions that mimic the Hopfield learning rule for storing memories [16] in an energy landscape. (D) Alignment is found to promote specific types of network motifs in the GRN, in particular coherent FFLs and positive FBLs.

Traditionally, the study of gene regulatory networks has emphasized the relationship between structure and function, focusing on how specific architectural features give rise to stable or robust expression dynamics [9, 23, 25, 37–39]. In contrast, evolutionary algorithms have largely focused on how traits such as mutational robustness or evolvability emerge under selective pressure when phenotype is completely determined by genotype, often abstracting away mechanistic network structure [29, 32, 33, 40] (but see [41] for a study on topology affecting these properties). Guo and Amir exemplify the first perspective [37] by using an ODE-based transcription–translation model in *E. coli* to show that protein-level stability is primarily driven by self-regulation, precise network configurations, and transcription-factor interactions, with degradation rates playing a comparatively minor role. Kumawat et al. [29] exemplifies the second by co-evolving linear sequences of computer instructions to study adaptation under environmental change, demonstrating that rapid adaptation depends on the rate of environmental shifts and the access to alternative phenotypes within a few mutations. In this work, we highlight the role of environmental variability (not to be confused with environmental change) in developmental dynamics as a significant driver of regulatory architectures. In other words, we evolve a population while holding the optimal phenotype (the desired behavior of an organism) constant and manipulating the phenotypic variability of its members. We summarize our findings in Figure 1.

In Figure 2, we provide an overview of the model developed in our study. In the main text below, we focus on an evolutionary model of GRNs introduced by Wagner in refs. [32, 33]. In Wagner's model, within each evolutionary time step, the evolution equation determining the mapping between the *genotype* (the gene-gene dynamics for a fixed number of $N$ genes) and the *phenotype* (the limiting expression vector $p_i(t \to \infty)$) is shown in Fig. 2(a). The main contribution of our paper is a modification of Wagner's model to account for cell-to-cell variability within these evolutionary dynamics. At each generation, the connections in the genotype are assumed to be perturbed by a stochastic multiplicative pre-factor (see Fig. 2(b)–(c)). This effectively models cell-to-cell variability—instigated by a noisy environment—in the gene-gene relationships that constitute the genotype. These multiplicative pre-factors account for processes inside the cell that are not explicitly included in the model, such as differential expression of other genes not included in our model, distinct metabolic states between cells, differing cellular morphologies, and variability in intracellular pH. In sum, our study unifies prior literature exploring the evolution of *in silico* GRNs [29, 32, 33, 42] with studies on the stochasticity inherent to gene expression [18, 43–46], and uncovers the genetic mechanisms that may be at play in allowing genomes to tolerate large levels of environmental noise, as may be hypothesized in extreme environments such as salt lakes where salinity and water levels are variable throughout the year [47].

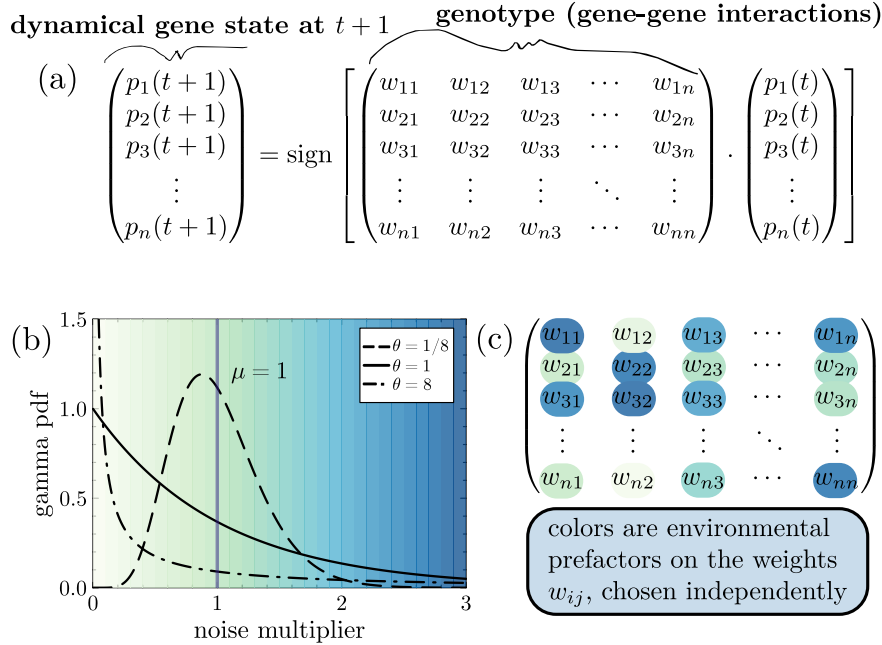

Figure 2: **Overview of the GRN evolution model used in this study.** (c) The model studied in the main text builds on refs. [32, 33] in which the iterative equation shown constitutes the developmental dynamics that determine the eventual phenotype expressed by a given genotype. Between evolutionary time steps, elements of the genotype may mutate. (b)–(c) In our adaptation of the model from refs. [32, 33], we assume that elements of the genotype are affected by a noise that modifies elements of the genotype by a multiplicative pre-factor. Here, the different color shades indicate the strength of the noise multiplier. In line with other literature, we assume gamma-distributed environmental noise which has two parameters [34–36]. Throughout our study, we assume the distribution's mean is $\mu = 1$, leaving us with one free parameter, $\theta$, to tune the distribution's variance from no-noise scenarios to pathological levels of noise.

Our paper is structured as follows. In Section 2.1, we extend Wagner's model from ref. [33] by introducing stochasticity into phenotype expression with a parameter that allows us to manipulate the level of variation, rendering the genotype-to-phenotype map and fitness probabilistic. In Section 2.2, we describe the evolutionary algorithm implementing directional selection toward a fixed optimal phenotype. We then introduce, in Section 2.3, an alignment measure that quantifies how regulatory interactions jointly support stable phenotype expression, and relate it to the enrichment of positive feedback loops (FBLs) and coherent feedforward loops (FFLs) in Section 2.4. Section 2.5 defines two complementary quantities to assess the mutational robustness of evolved GRNs by comparing phenotype distribution means after a perturbation. The results in Section 3 show that, despite high variability in phenotype expression, the model robustly reproduces the key findings of Wagner [33], while shaping the space of viable network configurations, driving higher alignment, motif enrichment, and enhanced robustness to mutational perturbations. Finally, in Section 4, we discuss the limitations and implications of the model and outline directions for future research.

## 2 Materials and Methods

### 2.1 Gene Network Model

To explore the effects of stochastic gene expression on evolutionary algorithms, we extend the model presented in [33], in which GRN dynamics are simplified via matrix multiplication and gene expression is represented as a Boolean vector. The evolutionary processes are based on standard population genetics. We define an organism as having an associated initial gene state $\boldsymbol{P}_0$ as a boolean vector with $N$ entries valued either $\pm 1$, a genotype $\boldsymbol{W}$ as a real-valued matrix encoding the GRN with $N^2$ entries, and a target phenotype $\boldsymbol{P}_{\text{opt}}$ as a boolean vector with $N$ entries valued either $\pm 1$. We now outline the process for evaluating an organism's fitness in the population through stochastic expression.

In a generation, each organism has a deterministic genotype $\boldsymbol{W}$, which encodes its GRN. To incorporate stochasticity, we multiply each entry of the genotype by an independently sampled random variable $X_{ij}$, which we set to follow a gamma distribution $\Gamma(\alpha = 1/\theta, \theta)$ with fixed average $E(X_{ij}) = 1$ and a parametrized variance $\text{Var}(X_{ij}) = \theta$. This choice stems from the understanding that biological noise does not affect the average expression of a phenotype, but its variance [18, 31]. Furthermore, this choice reflects the fact that noise in regulatory networks is usually caused by fluctuations in reactant concentrations at low copy numbers, resulting in varying reaction rates or bursting [17, 48]; since the Gamma distribution has positive support, the stochastic part does not change the direction of interactions nor creates spurious connections between genes.

To incorporate the effects of the deterministic and stochastic parts, we create the combined regulatory matrix $\boldsymbol{W}'$ by setting $W'_{ij} = W_{ij} X_{ij}$. At the beginning of a generation, we sample each weight $X_{ij}$, making the stochasticity non-heritable. In contrast, the genotype $\boldsymbol{W}$ might be passed on to the next generation and is subject to mutation; this is the target of selection. However, it is both the deterministic and stochastic parts that determine the regulatory dynamics through $\boldsymbol{W}'$.

To determine the phenotype expression of an organism in a single generation, we sequentially update the initial gene state $\boldsymbol{P}_0$ to $\boldsymbol{P}_t$ until a stable state is reached. We use the combined matrix $\boldsymbol{W}'$ to multiply the previous state,

$$\boldsymbol{P}_{t+1} = \text{sign}\left(\boldsymbol{W}' \boldsymbol{P}_t\right). \qquad (1)$$

In other words, to update the state of the $i$-th gene at time $t$, written as $P_{i,t}$, we add up all the contributions of all the other genes by multiplying the $i$-th row of $\boldsymbol{W}'$ with the current expression $\boldsymbol{P}_t$, and then take the sign of the resultant sum. To associate a phenotype with an organism, we update its expression vector up to 100 steps, which were more than enough to find stable states in our simulations. If we observe that $\boldsymbol{P}_{t+1} = \boldsymbol{P}_t$, we declare the phenotype as stable and store the expressed phenotype $\boldsymbol{P} = \boldsymbol{P}_t$; we call the number of steps it took to find a stable state the *path length*. Otherwise, we declare the organism's phenotype as unstable. These dynamics have been thoroughly explored using recurrent neural networks with asynchronous updates as a mechanism for storing and retrieving information in matrices; they are Hopfield networks [16] (we replicate our results with asynchronous update dynamics in Supplementary Material S2.1).

### 2.2 Population Genetics

Through these simulations, we generate a population to be `pop_size` individuals with a shared random initial expression $\boldsymbol{P}_0$ and a shared target optimal phenotype $\boldsymbol{P}_{\text{opt}}$ (but $\boldsymbol{P}_0 \neq \boldsymbol{P}_{\text{opt}}$), where each element in the vectors has an equal probability of being either $\pm 1$. To generate each individual, we sample the elements of the genotype $\boldsymbol{W}$ from independent normal distributions, $W_{ij} \sim \mathcal{N}(0, 1)$. Since we do this for every $i, j \leq N$, we create a fully connected network, which restricts all evolved populations to have fully connected networks with density $c = 1$ (we check the robustness of our results in different initial densities in Supplementary Material S2.2; qualitative differences are discussed). We include an organism in the initial population only if it shows a stable phenotype (not necessarily $\boldsymbol{P}_{\text{opt}}$) from deterministic dynamics, which is obtained by fixing the random components to $X_{ij} = 1$, resulting in the combined matrix to be identical to the deterministic one $\boldsymbol{W}' = \boldsymbol{W}$ in Equation 1 (we check the robustness of our results in different initial populations in Supplementary Material S2.3).

We implement free sexual recombination with mutations in non-overlapping generations. In each generation, we will obtain a new population from the previous one by iteratively creating and selecting candidates until the population size reaches `pop_size`. We build candidates in two steps.

First, we implement *sexual reproduction* through free recombination by randomly selecting two parents from the current population. From one parent, we will randomly choose each of its rows with probability $1/2$ and insert them into its offspring, while simultaneously picking the other parent's rows not chosen by the first. Note that if both parents are fully connected, the offspring will be too because any choice of rows will be full. This keeps matrices with a constant density $c = 1$ throughout the evolutionary process in our simulations. The offspring is then subject to mutation. We choose each entry in the matrix with probability $p_m$ and replace it with a new sample of the normal distribution $\mathcal{N}(0, 1)$. We chose $p_m$ such that, on average, every matrix has $N^2 p_m = 1$ mutation per generation. The candidate now has a deterministic $\boldsymbol{W}$, the result of recombination and mutation.

Second, we develop the candidate's phenotype and pass it on to the next generation with a probability equal to its fitness. The candidate's phenotype comes from a single realization of the stochastic dynamics outlined in Section 2.1. After obtaining the phenotype, we evaluate fitness using

$$\text{Fitness} = \begin{cases} \exp(-s\, d(\boldsymbol{P}, \boldsymbol{P}_{\text{opt}})) & \text{if stable} \\ \exp(-s) & \text{if unstable} \end{cases} \quad (2)$$

where we set $s = 10$ as the selection strength and $d$ is the Hamming distance between two vectors. Larger values of $s$ imply stronger selective pressure. Note that the maximum distance evaluates to 1, so unstable phenotypes have the lowest fitness. We incorporate a candidate into the next generation with a probability equal to its fitness. The evolutionary algorithm runs simulations with $G_{\max}$ generations. To see the parameters, see Appendix A, and for pseudocode of the process incorporating Sections 2.1 and 2.2, see Appendix B.

In summary, the evolutionary algorithm works by repeatedly coupling noisy gene expression with selection at the population level. A population of gene regulatory networks is initialized with a shared initial dynamical gene state and a common target phenotype. In each generation, individuals experience stochastic gene expression: random fluctuations are applied to their regulatory interactions, and gene activity is updated step by step until the system either settles into a stable pattern or fails to stabilize. This final pattern is taken as the individual's phenotype and compared with the target. Stable, closer matches receive higher fitness, and unstable outcomes are strongly penalized. New individuals are then created by combining regulatory interactions from two parents and introducing occasional mutations. Offspring are accepted into the next generation with a probability equal to their fitness. Over many generations, this process favors gene networks that reliably produce stable, target-like phenotypes despite the noisy dynamics.

### 2.3 Alignment Score

Hopfield networks implement essentially equivalent dynamics for storing and retrieving information in matrices using recurrent neural networks [16]. An important metric in investigating their dynamics around a specific attractor is the *energy* of a state. The lower the energy, the larger the basin of attraction around it. We develop a similar metric in the context of our model since we consider the target phenotypes as attractors or "memories to recall." First, we notice that the dynamics of Equation 1 are unchanged after multiplying any row of the matrix $\boldsymbol{W}'$ by a positive scalar (because $\text{sign}(kx) = \text{sign}(x)$, for $k > 0$), so we standardize each row of the deterministic $\boldsymbol{W}$ by dividing by the sum of it s absolute values, such that

$$W_{ij}^{\text{norm}} = \frac{W_{ij}}{\sum_{j=1}^{N} |W_{ij}|}. \quad (3)$$

Then, the alignment score of an organism with a GRN $\boldsymbol{W}$ is

$$A(\boldsymbol{W}) = \frac{1}{N} \sum_{i=1}^{N} \sum_{j=1}^{N} P_{\text{opt},i} P_{\text{opt},j} W_{ij}^{\text{norm}}. \quad (4)$$

We interpret the alignment score $A(\boldsymbol{W})$ as a measure of how well a gene regulatory network is structurally tuned to produce a target phenotype $\boldsymbol{P}_{\text{opt}}$. Each term $P_i P_j$ encodes the ideal sign of the regulatory interaction from gene $i$ to gene $j$: genes that should be oppositely expressed require inhibitory interactions, while co-expressed genes require activating ones. We exhibit this idea in Figure 3. An interaction contributes positively to alignment when $\text{sign}(W_{ij}) = \text{sign}(P_i P_j)$, indicating that the network's regulatory logic supports the target phenotype. Row normalization in Equation 3 ensures $|A(\boldsymbol{W})| \leq 1$. Notably, the Hopfield learning rule for storing a target pattern [16] constructs matrices that achieve the maximal alignment $A(\boldsymbol{W}) = 1$, corresponding to the target phenotype being an energetic minimum.

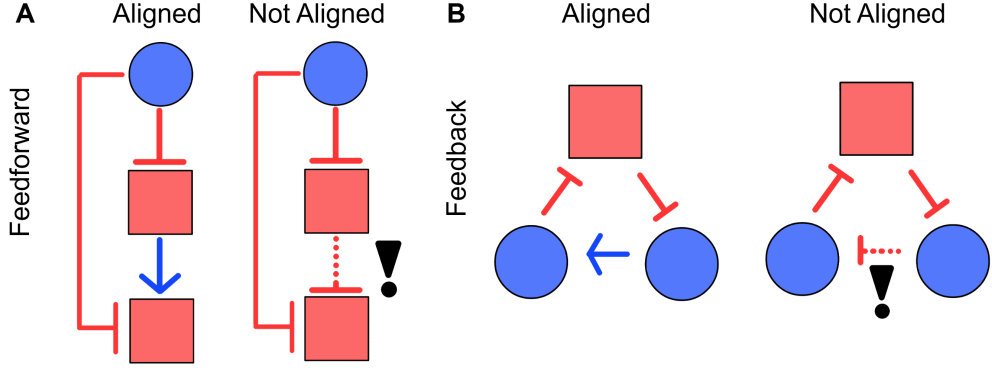

Figure 3: **Examples of motifs in some aligned and non-aligned network motifs**. The blue circles represent a positive and the red squares a negative entry in the target phenotype $\boldsymbol{P}_{\text{opt}}$. Positive edges (promoting interactions) are represented with a blue arrow, and negative ones (inhibitory interactions) with a squared red tip. By changing the direction of one edge (marked with a dashed line and an exclamation mark), we transformed a (A) coherent FFL into an incoherent one and a (B) positive FBL into a negative one. These changes decrease the overall alignment score $A(\boldsymbol{W})$ because similarly signed pairs of nodes in the target phenotype should have a positive interaction, whereas differently signed pairs should have a negative one.

### 2.4 Network Motifs

Some network motifs have been associated with robustness in their dynamics against noise of various sources [24, 49, 50]. In particular, [23] demonstrated that FFLs are associated with functional responses, such as accelerated dynamics or time delays, through their coherence. A coherent FFL is a 3-node network in which an upstream gene exerts the same influence on a downstream target via an intermediary gene; an incoherent one is one in which it does not. FBLs are characterized by one or more cycles within the network. FBLs are positive if the feedback enhances the output signal, leading to amplification, whereas they are negative if they act to stabilize the system by reducing fluctuations.

Network motifs are closely related to our measure of alignment. In fact, the maximal alignment $A(\boldsymbol{W}) = 1$ corresponds to an architecture that only allows coherent FFLs and positive FBLs (see Appendix C for a small proof). This is because maximal alignment indicates that any pathway between two nodes in the network will have the same overall sign as the direct connection between those nodes. To visualize this, we built two examples in Figure 3. In it, we see that an aligned motif would have inhibitory relationships between two differently signed optimal states, leading to a coherent feedforward or an (overall) positive FBL. When we change the sign of interaction, they become incoherent or negative, respectively. We evaluate the concentration of these motifs in a network based on [51]. For all network motifs of a given type (i.e., all FFLs of size 3), we define the *concentration* of a particular network motif as its number of occurrences divided by the total number of networks of a similar type (e.g., number of coherent FFLs of type I divided by the total number of FFLs).

### 2.5 Mutational Robustness

One of the main results of Wagner[1] [33] is that mutational robustness increases over evolutionary time, where robustness is defined as a genotype's ability to preserve its phenotype under mutations. However, in our stochastic setting, a single genotype $\boldsymbol{W}$ may express multiple phenotypes due to the noisy developmental dynamics described in Section 2.1. As a result, mutational robustness must be assessed at the level of *phenotypic distributions* rather than at the level of individual phenotypic outcomes.

The challenge lies in comparing *the* phenotype distribution from $\boldsymbol{W}$ to *many* phenotype distributions generated by mutational perturbations $\tilde{\boldsymbol{W}}$. If a genotype $\boldsymbol{W}$ is mutationally robust, each instance of the latter should be indistinguishable from the former.

For a given genotype $\boldsymbol{W}$, we can find the probabilities of each unique Boolean vector $\boldsymbol{v}_i \in \{\pm 1\}^N$: each genotype can have many stable phenotypes. We associate each with a probability $p_i$. Because the dynamics may fail to reach a stable state, we define $p_u(\boldsymbol{W}) = 1 - \sum_i p_i(\boldsymbol{W})$ as the probability of showing an unstable phenotype. We evaluate all phenotype distributions at a fixed noise variance, $\theta_{\text{fixed}} = 1$, to provide a baseline for GRNs evolved across noise levels. To empirically estimate the probabilities $p_i(\boldsymbol{W})$ and $p_u(\boldsymbol{W})$, we generate $k = 30$ independent realizations of the stochastic matrix $\boldsymbol{W}'$ and simulate their developmental dynamics. To compare stable and unstable expressed phenotypes, we consider each phenotype type separately.

---

[1] We use the word "mutational" instead of "epistatic" since gene-gene interactions determine phenotype in this model. Epistasis might be understood from a developmental or mutational perspective. Following the original publication, we focus on mutational.

First, we sample the stable phenotypes of a genotype $\boldsymbol{W}$ by drawing $k$ samples from the phenotype expression distribution.[2] Then, we approximate the *average stable phenotype*

$$\bar{\boldsymbol{v}}(\boldsymbol{W}) = \frac{1}{(1-p_u)k} \sum_{i=1}^{k} \boldsymbol{P}^i, \quad (5)$$

where $\bar{\boldsymbol{v}}(\boldsymbol{W}) \in \mathbf{R}^N$ is a real-valued vector of $N$ dimensions, and $\boldsymbol{P}^i$ is the $i$-th phenotype expression sample. We set $\boldsymbol{P}^i = \boldsymbol{0}$ if the phenotype is unstable since it does not contribute to the average, and we normalize by the factor $1/(1 - p_u)$. When we sample the stable expressed phenotypes, if all Boolean vectors $\boldsymbol{v}_i \in \{\pm 1\}^N$ are equally likely, the average should be the zero vector $\bar{\boldsymbol{v}} = \sum_i \boldsymbol{v}_i = \boldsymbol{0}$. In contrast, evolutionary dynamics are expected to bias expression toward the target phenotype, resulting in $\bar{\boldsymbol{v}} \approx \boldsymbol{P}_{\text{opt}}$. A mutationally-robust GRN $\boldsymbol{W}$ should yield a phenotype distribution mostly unchanged after a mutational perturbation. We therefore expect both $\bar{\boldsymbol{v}}(\boldsymbol{W}) \approx \bar{\boldsymbol{v}}(\tilde{\boldsymbol{W}})$ and $p_u(\boldsymbol{W}) \approx p_u(\tilde{\boldsymbol{W}})$.

We generate the corresponding phenotype distribution samples of $K = 5$ perturbations for each matrix $\boldsymbol{W}$ in the population. For the perturbed genotypes $\tilde{\boldsymbol{W}}$, we sample each of their average stable behavior, and average them across mutations to obtain $\langle \bar{\boldsymbol{v}}(\tilde{\boldsymbol{W}}) \rangle$. We define the *stable expression shift* as the $\ell_1$ distance from the original average stable behavior and the average across perturbations

$$\Delta \bar{\boldsymbol{v}}(\boldsymbol{W}) = \|\bar{\boldsymbol{v}}(\boldsymbol{W}) - \langle \bar{\boldsymbol{v}}(\tilde{\boldsymbol{W}}) \rangle\|_1 \in [0, 2N],$$

which quantifies the sensitivity of the average stable phenotype to genetic perturbations. Small values indicate that mutations tend to preserve the network's stable mean phenotype. In contrast, larger values imply that stable phenotypes change substantially after mutations. However, if differences in the average tend to cancel each other out, we might miss the extent to which a mutation produces an alternative stable phenotype. Therefore, we also want to investigate the spread of the average stable behavior upon mutation. We take the sample *stable phenotype variance* across mutations as

$$\sigma^2(\boldsymbol{W}) = \frac{1}{K-1} \sum_{i=1}^{N} \sum_{j=1}^{K} \left( \bar{\boldsymbol{v}}(\tilde{\boldsymbol{W}})_i^j - \langle \bar{\boldsymbol{v}}(\tilde{\boldsymbol{W}}) \rangle_i \right)^2,$$

where the sub-index $i$ denotes the $i$-th entry in a vector, and the super-index $j$ indicates the $j$-th mutational perturbation. This variance quantifies how much the average stable behavior changes per mutation, akin to the "surprise at the phenotype when mutations are combined" [52, 53]. A high variance indicates that, upon mutation, a genotype yields many different stable phenotypes.

We now address the case in which a phenotype is unstable. Analogous to the average stable behavior, we obtain the probabilities of instability for the $K$[3] mutational perturbations and average them into $\langle p_u(\tilde{\boldsymbol{W}}) \rangle$. We obtain the *instability difference* as

$$\Delta p_u(\boldsymbol{W}) = p_u(\boldsymbol{W}) - \langle p_u(\tilde{\boldsymbol{W}}) \rangle,$$

which measures how mutations affect the *stability* of phenotype expression itself, independently of which phenotype is expressed. Low values of $\Delta p_u(\boldsymbol{W})$ indicate that mutations do not significantly alter the likelihood of producing unstable dynamics, whereas higher values signal that stability tends to greatly change after mutations. Negative values indicate that mutations increase developmental stability beyond the original genotype, which is unexpected because organisms should preserve their function. We also want to know the variability of this probability upon mutations, for which we obtain the *instability sample variance*,

$$\rho^2(\boldsymbol{W}) = \frac{1}{K-1} \sum_{i=1}^{K} \left( p_u(\tilde{\boldsymbol{W}}^i) - \langle p_u(\tilde{\boldsymbol{W}}) \rangle \right)^2,$$

where each super-script $i$ indicates the $i$-th mutational perturbation. The instability sample variance $\rho^2(\boldsymbol{W})$ measures the variability in instability across different mutational scenarios. A higher variance indicates that, upon mutation, genotypes tend to show different probabilities of instability. These four quantities roughly describe the mutational landscape of matrices Mut($\boldsymbol{W}$).

---

[2] For the population averages and variances, we might define the probability $p_i$ associated with each $\boldsymbol{P}^i \in \{\pm 1\}^N$, and make a weighted average across them with a normalization constant $1/(1 - p_u)$. However, computing each $p_i$ is intractable, so we rely on samples

[3] We chose $K$ and $k$ by trial and error, aiming to balance statistical resolution against computational feasibility. Larger $K$ yields better results in exploring the mutational landscape, and larger $k$ yields a better resolution of each phenotype distribution. However, sampling each mutational landscape requires $O(kKN^2)$ operations due to matrix multiplication.

Notice that Mut($\boldsymbol{W}$) is defined at the organism level. We denote the population average as,

$$\langle \text{Mut}(\boldsymbol{W}) \rangle = \left( \langle \Delta \bar{\boldsymbol{v}}(\boldsymbol{W}) \rangle, \ \langle \sigma^2(\boldsymbol{W}) \rangle, \ \langle \Delta p_u(\boldsymbol{W}) \rangle, \ \langle \rho^2(\boldsymbol{W}) \rangle \right).$$

Furthermore, we compare each mutational robustness metric before and after evolution by defining the *change* in each metric as the difference between the averages of the initial and evolved populations. These differences are plotted in Section 3.4, where a negative value indicates an increase in mutational robustness. Each metric captures a distinct aspect of this robustness: a negative change in $\langle \Delta \bar{\boldsymbol{v}}(\boldsymbol{W}) \rangle$ indicates that the average stable phenotype remains closer to the original upon mutation; a negative change in $\langle \sigma^2(\boldsymbol{W}) \rangle$ indicates a smaller spread in the average stable phenotype across mutations; a negative change in $\langle \Delta p_u(\boldsymbol{W}) \rangle$ indicates that the probability of instability stays closer to that of the original; and a negative change in $\langle \rho^2(\boldsymbol{W}) \rangle$ indicates that the probability of instability varies less across mutations. Together, these metrics capture two dimensions of mutational robustness: how mutations shift the phenotype, and how mutations affect the stability of phenotype expression.

---

**Box 1: Mutational Robustness Landscape Mut($\boldsymbol{W}$)**

Stable phenotype:

$\Delta \bar{\boldsymbol{v}}(\boldsymbol{W})$: **Stable expression shift.** $\ell_1$ distance between the original average stable phenotype and the mean across mutated variants. Small values indicate that mutations preserve the mean stable phenotype. Range: $[0, 2N]$.

$\sigma^2(\boldsymbol{W})$: **Stable phenotype variance.** Sample variance of the average stable phenotype over $K$ perturbations. High values indicate that different mutations yield very different stable phenotypes.

Developmental stability:

$\Delta p_u(\boldsymbol{W})$: **Instability shift.** Change in the probability of an unstable phenotype after mutation. Low values indicate that mutations preserve developmental stability. Negative values indicate that mutations increase stability beyond the original and are highly unexpected.

$\rho^2(\boldsymbol{W})$: **Instability variance.** Sample variance of $p_u$ over $K$ perturbations. High values indicate that different mutations produce very different probabilities of unstable dynamics.

---

## 3 Results

### 3.1 Increase in Fitness and Decrease in Path Length

Across varying noise levels $\theta$, populations tend to optimize their fitness and reduce the number of update steps required to find a stable phenotype. In Figure 4 A, we find that GRNs exhibit consistent phenotype distributions close to the target. Furthermore, Figure 4 B shows a decrease in the average path length, suggesting they acquire similar structures that rule their expression dynamics. These results replicate and extend those of Wagner's [33] and echo previous research showing that the space of possible architectures in a GRN is constrained by function [38].

Figure 4 A shows that higher noise levels lead to a lower attained fitness after evolution. The decrease in fitness is not driven by a few outliers in the simulation trials, as evidenced by the small error bars. At each generation, most individuals still reach a stable phenotype (figure not shown), so we attribute the drop in fitness to imperfections in the network structure. These imperfections only matter when noise causes a few interaction weights in a $\boldsymbol{W}'$ realization to become much larger than others, such that they dominate the update dynamics. This can happen because the gamma

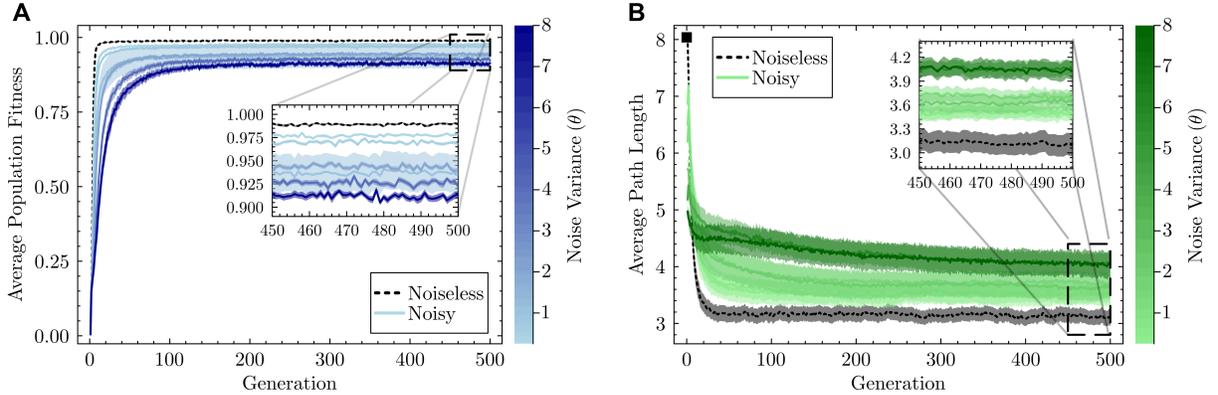

Figure 4: **(A) Evolution of the average population fitness and (B) average mean path length per generation**, across 30 independent replicates at different noise variances $\theta$. Shaded regions represent 95% confidence intervals around the mean in the zoomed-out panels and 68% in the zoomed-in panels. Lighter colors correspond to smaller noise variances. The noiseless case ($\theta = 0$) is shown with a dotted line. In panel (B), the mean path length at generation 0 for the noiseless scenario (8.04) is indicated by a black square.

distribution is right-skewed: most interactions shrink (are multiplied by values less than 1), but occasionally they grow (multiplied by values greater than 1). As noise increases, this skew becomes stronger, which explains both the greater asymmetry in the weights and the corresponding decrease in fitness across $\theta$.

We attribute this to an example of *developmental* epistasis as the "masking" of a genetic interaction by other genes [54]. In this case, the main driver is the nonlinearity in the dynamics caused by the sign activation function in Equation 1. With a high alignment (see Section 3.2), the chance of a misaligned weight to grow stochastically diminishes.[4] It is only through the aggregate behavior of genes in the network that one can observe significant effects on the phenotype expression. We conclude that, while noise might assign a single gene to disproportionately drive phenotype expression, alignment buffers against this effect by enhancing developmental epistasis: these flukes become much rarer, ensuring organisms approach optimal fitness.

In Figure 4 B, we observe a decrease in the average path length across noise levels, successfully replicating and extending Wagner's [33] results. Notably, increased noise levels increase the average path length over evolution; matrices might need to take extra steps to balance initial fluctuations. For instance, a different instance of $\boldsymbol{W'}$ can push the gene expression state in one direction before eventually converging to the stable phenotype. Yet, all evolved GRNs have an average path length lower than that of stable, non-evolved matrices (7.43). In the original publication, Wagner [33] discards the decreased path length as an explanation for mutational robustness since non-evolved matrices with shorter path lengths do not exhibit the same degree of robustness as evolved ones. Furthermore, Wagner shows that path length is a heritable trait by correlating parents' and offspring's path lengths. This heritability was surprising and non-intuitive since sexual recombination should disrupt any delicate structure established in a single genotype $W$. It is, however, less surprising if we consider that the evolved mechanism for consistently showing a phenotype is alignment, which depends only on the number of appropriately signed elements in the matrix (See Section 2.3). Along these lines, the recombination of two similarly aligned genotypes will produce a genotype that is similarly aligned.

## 3.2 Noise Variance Increases Alignment Score

We plot the evolution of alignment score across generations in Figure 5. The alignment score tends to increase across noise levels. To confirm this, we run a two-tailed Welch's t-test with the null hypothesis that there is no difference in the average alignment scores between populations evolved with ($n = 180$) and without noise ($n = 30$). We confidently reject the null hypothesis with a statistically significant difference between the two groups ($t = 33.942$, $P < 1.2 \times 10^{-67}$). Figure 5 B further illustrates the positive relationship between noise variance and the achieved alignment score. Although increasing, the growth is nonlinear and quickly approaches a plateau near its theoretical maximum $A(\boldsymbol{W}) = 1$.

---

[4]In a fully connected network, each gene has an effect on the order of $1/N$ to other genes. Large values of $N$ might diminish the likelihood that a mutation can severely affect stable phenotype expression, especially if most weights are properly aligned.

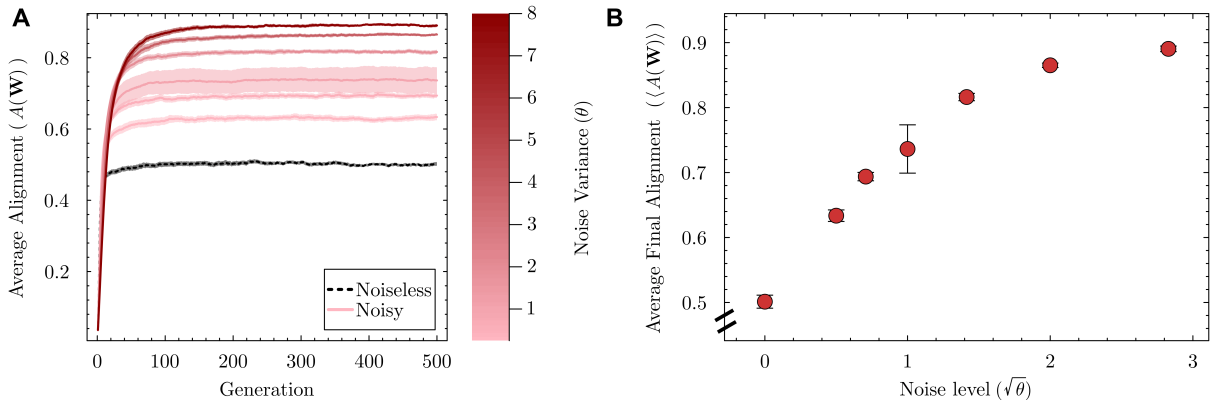

Figure 5: **Evolution of the average alignment scores**. (A) Average alignment score across 30 populations over generations. Each line represents the average alignment score for a population, and the color fillings represent 95% confidence intervals. Lighter colors represent smaller noise variances. The noiseless scenario is plotted with a dotted line. (B) Average alignment scores after evolution from 30 evolved populations across various noise levels $\sqrt{\theta}$. The 95% confidence intervals are plotted along the means, but they are barely visible.

We find it remarkable that the evolutionary algorithm favors matrices whose architecture approaches the optimal memory recall configuration described by Hopfield [16, 55, 56] when stochastic phenotype expression is introduced. The stronger the noise level, the closer the evolved matrices approach this configuration. Importantly, the evolutionary process does not explicitly minimize any energy function on the network's architecture—selection operates solely on phenotype stability and fitness. However, under stochastic expression dynamics, matrices with small basins of attraction around the target phenotype are more likely to fail to converge and tend to express suboptimal phenotypes, and are therefore selectively penalized. On the other hand, networks with sufficiently large basins of attraction around the target phenotype consistently produce stable phenotypes close to optimal. In Hopfield networks, such large basins correspond to deep energy minima. Because maximal alignment enforces the same sign structure as the Hopfield's learning rule [16], increasing alignment structurally approximates an energy-minimizing configuration. Thus, energy minimization emerges from selection towards stability under noisy developmental dynamics.

### 3.3 Evolution Enriches Coherent Feedforward and Positive Feedback Loops and Diminishes Incoherent Feedforward and Negative Feedback Loops

Since the alignment score increases in GRNs evolved without noise, we measure the concentration of coherent FFLs and positive FBLs in relation to non-evolved populations. In Figure 6 A, we see that there is no significant enrichment or depletion of any particular FFL due to the overlapping confidence interval bars. However, when aggregating FFLs into coherent and incoherent categories, we found that the evolved networks have a significantly higher proportion of coherent loops than their non-evolved counterparts (0.519 vs 0.504). We corroborated this result with a Welch's t-test, where the null hypothesis is that the proportion of coherent FFLs is equal between non-evolved ($n = 30$; stable) and evolved ($n = 30$; noiseless) GRNs, we find a statistically significant difference ($t = 10.66, P < 2.18 \times 10^{-13}$, after Bonferroni correction for 2 tests). Furthermore, Figure 6 B shows that the concentration of small-sized positive FBLs is slightly higher in evolved networks than in non-evolved populations, as indicated by the non-overlapping confidence intervals. Whereas stable matrices tend to have more positive FBLs, the evolutionary process led to higher concentrations.

We also compare populations of evolved matrices under very noisy ($\theta = 8$, $n = 30$) and noiseless ($n = 30$, the same population as before) conditions. In Figure 7 A, we find that every single coherent motif is enriched in comparison to its noiseless counterparts. Furthermore, the difference after aggregating coherent motifs is even larger (0.716 vs 0.519). Running a second Welch's t-test comparing the aggregated concentrations of coherent and incoherent feedforward motifs, we find a statistically significant difference between the populations evolved in noisy and noiseless conditions ($t = 31.46, P < 5.01 \times 10^{-25}$ after Bonferroni correction for 2 tests). Furthermore, Figure 7 B shows a much higher concentration of positive FBLs for all sizes, and its corresponding smaller concentration of negative ones.

To summarize, we find that populations evolved with and without noise exhibit higher concentrations of coherent FFLs and positive FBLs, and that the difference is larger at high noise levels. This echoes previous research on network motifs that protect function against noisy dynamics. For instance, [23] lists several functions, such as increased or decreased sensitivity to signals, time delays, and other diverse dynamical features. In particular, [57] finds that the coherent FFLs are the less-noisy motifs. Furthermore, Kadelka et al. [14] find a similar pattern in enrichment of coherent FFLs (and depletion of incoherent ones) across various expert-curated GRNs of diverse organisms when compared to null models. Whereas FBLs are usually neglected in models similar to those used in this work because they can introduce instability[41, 42], they may also play a crucial role in real-world GRNs as self-sustaining or self-regulating processes [4] and are associated with drug and acid resistance [58], but are extremely rare [58]. Our model, however, artificially imposes stability by restricting gene values to $\pm 1$, thereby neglecting very large expression values. Our findings suggest that enrichment of specific network motifs may be driven by cell-to-cell variability and the need for robust responses to it.

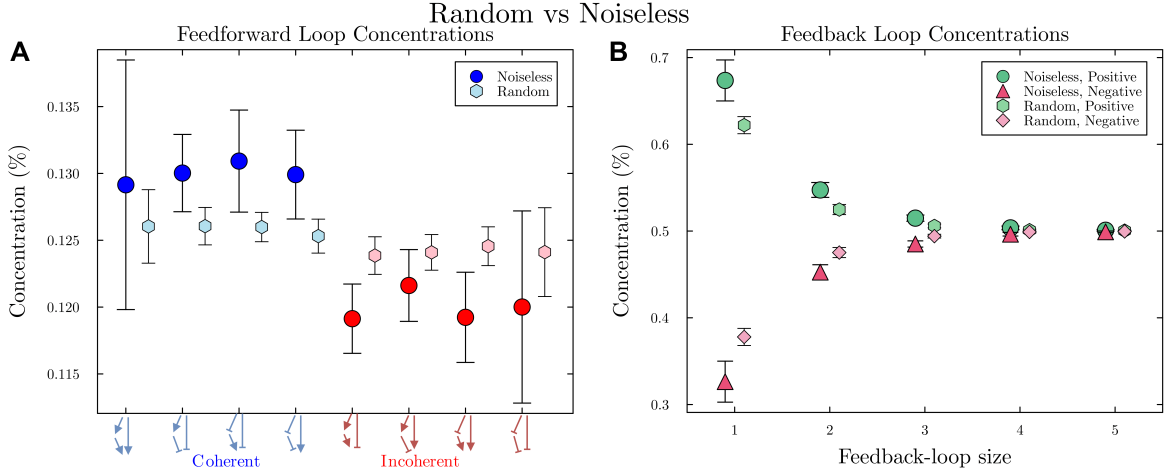

Figure 6: **Enrichment of network motifs as a result of noiseless evolution**. Comparison of 30 populations evolved without noise (darker hues) against non-evolved populations of stable matrices (lighter hues). (A) Average concentration per type of FFL. On the x-axis, we place the type of loop with a diagram. The error bars represent $95\%$ confidence intervals around the averages. (B) Average concentration of positive (green circles and hexagons) and negative (pink triangles and rhombuses) FBLs of different sizes with their $95\%$ confidence intervals around the averages.

### 3.4 Mutational Robustness

We find that, across noise levels, including the noiseless scenario, populations acquire greater mutational robustness across all four metrics (see Supplementary Material S1 for scatterplots of the data from which the changes in average were computed). Figure 8 A shows that evolved matrices tend to show an average stable behavior closer to their original average stable behavior after mutational perturbations, indicating that function is preserved against mutations. Figure 8 B shows that the distribution of stable phenotypes from mutational perturbations also tends to spread much less in evolved matrices. In both metrics, increased noise levels lead to a decrease in the shift and variability of the stable phenotypes, until it appears to plateau around $\sqrt{\theta} = 2.0$, which is consistent for matrices obtaining alignment scores closer to $A(\boldsymbol{W}) = 1$ (see Section 3.2). Figure 8 C demonstrates that evolved matrices show a stable phenotype more consistently, and that there is a minimum (most negative value) in the change of average instability for small values of $\sqrt{\theta}$.

Figure 8 D shows a non-trivial and different qualitative behavior with a similar overall effect in indicating increased mutational robustness. The horizontal line at $\langle \rho_W^2 \rangle = 0$ is only an artifact of the values in the choice of parameters (see Appendix A) and other conditions (see Supplementary Materials S2). Nevertheless, we find three regions delimited by $\sqrt{\theta}$. First, for small values of $\sqrt{\theta}$, the instability variance is not very different, suggesting that the mutational robustness is relatively similar, with minor fluctuations. As $\sqrt{\theta}$ increases, we observe a sharp decline in the variance, indicating that the underlying dynamics of the matrices shift significantly, yielding a more consistent probability of instability as their variability decreases. Finally, there is a threshold at which the variance plateaus as populations approach $A(\boldsymbol{W}) = 1$.

Intuitively, the stable phenotype distribution tends to narrow as populations evolve toward more stable phenotypic expression over evolutionary time, and this stability is subsequently preserved against mutational perturbations.

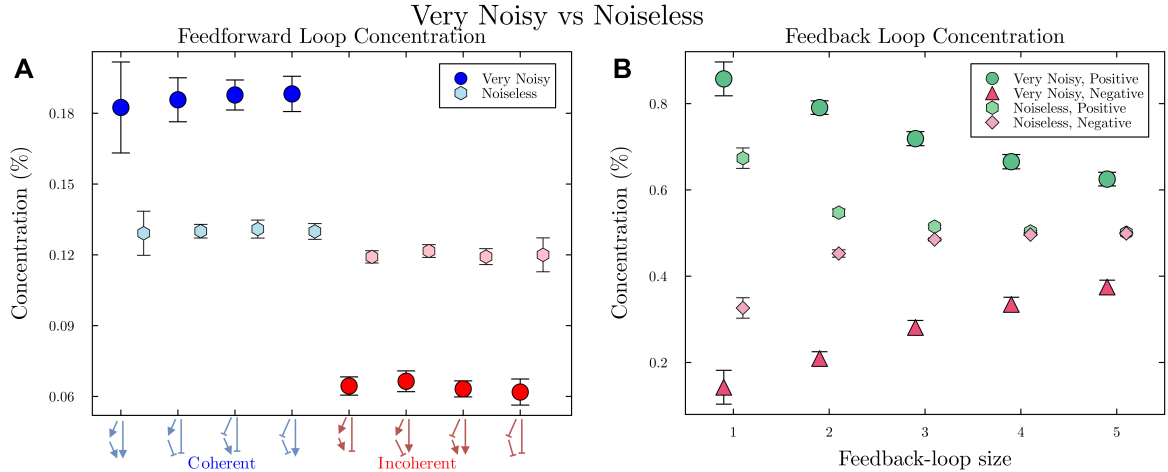

Figure 7: **Enrichment of network motifs as a result of noisy evolution, compared to noiseless evolution**. Comparison of 30 populations evolved with high noise variance ($\theta = 8$; darker hues) and no noise (lighter hues). (A) Average concentration per type of FFL. On the x-axis, we place the type of loop with a diagram. The error bars represent $95\%$ confidence intervals around the averages. (B) Average concentration of positive (green circles and hexagons) and negative (pink triangles and rhombuses) FBLs of different sizes with their $95\%$ confidence intervals around the averages.

One plausible interpretation is that the enrichment of coherent network motifs buffers against environmental noise (Section 3.3) and that this buffering capacity persists after mutations due to the over-representation of favorable network motifs. An alternative, yet equivalent, interpretation is that noise drives GRN architectures to align pairwise gene interactions so as to produce a more stable phenotype distribution: as more interactions become better aligned, random mutations are less likely to disrupt development since they would have to affect many weights simultaneously.

In the probability of expressing unstable phenotypes, higher noise levels drive matrices in which mutations do not change the probability of stable phenotype expression. Figure 8 C shows a decreased probability of instability, whereas 8 D shows that its variance is strongly nonlinear. We interpret these patterns through the geometry of basins of attraction when focusing on a single matrix. Because noise will not change the sign of interactions, the underlying structure of expression dynamics remains roughly the same. Coupled with stable expression landscapes, the probability of selecting a GRN with an unstable phenotype is extremely low in our parametrization ($e^{-10} = 4.5 \times 10^{-5}$). Then, we should not observe GRNs with different instability probabilities upon mutation. As for the variance, when alignment is high, the numerous appropriately signed gene-gene interactions create a larger basin of attraction for the expression dynamics, thereby reducing the variance in the probability of instability. As we decrease alignment, there are fewer gene-gene interactions buffering against mutations, and the basin of attraction becomes more sensitive to these perturbations. Alignment as an indicator of the basin of attraction around the target phenotype provides probabilistic intuition into how mutational perturbations affect stochastic phenotype expression.

## 4 Discussion

In this study, we examined how a probabilistic genotype-to-phenotype map shapes the evolution of GRN topology. Using a computational model, we found that noise in phenotype expression reduces average fitness by generating inaccurate or spurious stable expression patterns. At the same time, noise promotes the emergence of more optimal network configurations, in the Hopfield sense [16]. These architectures arise because stochastic phenotypes force the system to encode greater redundancy in its expression patterns through coherent network motifs or, equivalently, through increased alignment, resulting in a narrow spread of phenotype distributions and increased mutational robustness. Notably, coherent FFLs have been observed to be enriched in meta-analyses of curated biological networks [14], although our dynamical models differ.

Alignment connects developmental and mutational epistasis. During development, aligned matrices can buffer environmental noise because many gene interactions promote the coherent activation or inhibition of genes, while the nonlinear, bounded response prevents genes from being overexpressed [54]. Highly aligned matrices effectively mask the effects of misaligned gene-gene interactions during development. On the mutational timescale, we produced highly aligned matrices by imposing noisy phenotypic development, yielding consistent, mutationally robust distributions of phenotypes. Alignment reduces the impact of mutations by limiting the appearance of alternative phenotypes.

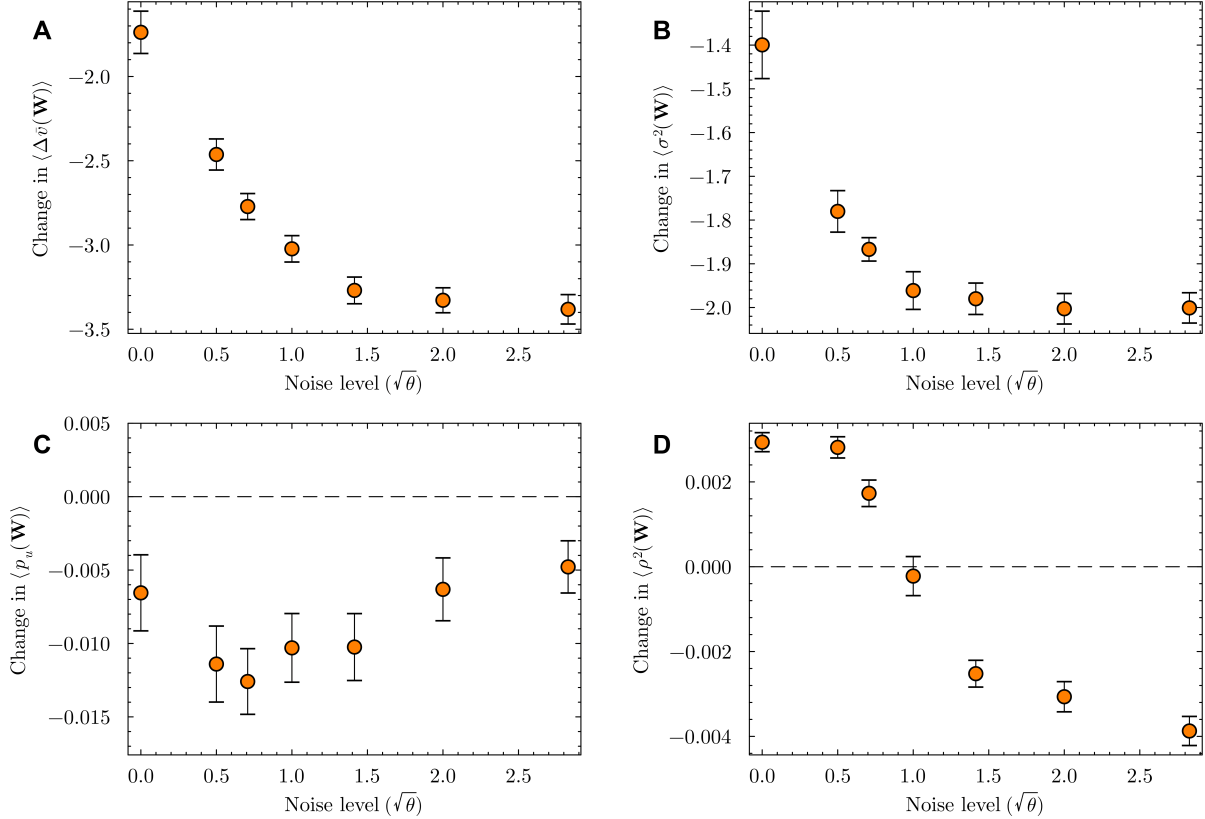

Figure 8: **Change in mutational robustness across evolved populations before and after evolution**. (A) Change in average *stable expression shift*, (B) average *stable expression variance*, (C) average *instability difference*, and (D) *instability variance* after the evolutionary processes for 30 populations across various noise levels (plotted as standard deviations in the x-axis). We fix the noise variance to $\theta_{\text{fixed}} = 1$ to compare all matrices against a single noise distribution. The $\theta = 0$ represents the noiseless scenario as it follows a Bernoulli distribution with no variance (always evaluates to 1). The error bars represent $95\%$ confidence intervals.

Enrichment of network motifs is ubiquitous across natural systems [50], and such motifs have been associated with robust expression dynamics and other functional properties [9, 23, 25, 37–39]. In contrast, Sole and Valverde [59] argue that the over-representation of network motifs might be an evolutionary spandrel rather than an adaptation arising from mechanisms such as gene duplication. Our model does not use any particular initialization structure; it self-organizes through mutations in signed gene-gene interactions that promote stable expression patterns. Moreover, the same evolutionary processes yield configurations with different alignment scores and more enriched topologies when evolving with and without noisy development. Another point of concern is whether network motifs are associated with plasticity or robustness in phenotype expression [22]. Our model does not reward plasticity. Instead, alignment emerges from selection for a narrow distribution of phenotypic expression under stabilizing conditions.

In this work, we do not study pleiotropy, but our model could be extended to do so. For example, Le Rouzic and Pouzet [41] use a continuous model of gene expression and measure how the direction of the phenotype changes after a mutation. In their approach, each element of the expression vector is treated as a separate trait, and their model can be extended to include stochastic expression. Alternatively, we could define a trait as a group of elements in the expression vector, similar to how an image depends on all its pixels. Then, pleiotropy can be understood as genes affecting multiple phenotypes, or as nodes that share edges with different sections of the network. Hopfield's rule [16] allows multiple target phenotypes to be encoded simultaneously, suggesting that a single gene regulatory network could support multiple functions. Future work could explore pleiotropy by following Le Rouzic and Pouzet's [41] methods, or by examining overlapping gene-gene interactions and segmented target phenotypes.

Recent evolutionary models have examined adaptability to changing environments to improve the theoretical understanding of phenomena such as antibiotic resistance and evolvability [29, 41]. There is also substantial theoretical work linking genetic diversity to adaptation under environmental shifts [28, 60]. Some studies suggest that noise-inducing alleles can promote access to alternative phenotypes [29], broader exploration of the adaptive landscape [31], and even induce stress-related phenotypes in development [61]. Our study instead focuses on a constant environment with fixed noise levels in phenotype expression, corresponding to stabilizing selection. Nevertheless, the framework could also be extended to incorporate dynamic target phenotypes [29, 41] or selection for noise-inducing alleles [29, 31]. We exclude these extensions here to maintain a clear focus on the core mechanisms of phenotypic evolution under stabilizing selection.

Several simplifying assumptions bind our conclusions. The model employs fully connected networks and binary gene expression, abstracting away from sparsity and structure observed in biological systems. Mutations are implemented as independent resampling of interaction strengths, excluding mechanisms such as gene duplication, insertion, deletion, and horizontal transfer. Noise is modeled as independent across interactions and lacks a covariance structure. Finally, network sizes are small, and scaling behavior remains unexplored. These abstractions define the scope of the present study while clarifying directions for future investigation.

In summary, our results support the view that stochastic expression of phenotypes is a primary driver of GRN evolution. Developmental noise promotes alignment, coherence, and redundancy in regulatory interactions, yielding architectures that are robust to both environmental fluctuations and genetic perturbations. Rather than acting solely as a destabilizing force, cell-to-cell variation emerges as a driver of GRN organization.

## Author Contributions

LIED and JH conceived the study. LIED wrote the manuscript with input from JH. LIED made the figures and implemented the computational analyses. LIED and JH designed the evolutionary algorithms together. JH supervised all aspects of the study and reviewed and edited the manuscript.

## Acknowledgments

LIED would like to thank peers and mentors, Minerva University, and the Santa Fe Institute for all their support and stimulating conversations. LIED and JH would specifically like to thank Chris Kempes, Sanjay Jain, and Daniel Muratore for useful discussions in the early stages of this project. JH would like to thank Marina Dubova, Andreas Wagner, Ricard Solé, and Rob Goldstone for stimulating discussions. JH would like to acknowledge the support of the National Science Foundation Grant Award Number EF–2133863. LIED was supported by funding from the Darla Moore Foundation and the McKinnon Family Foundation during a Undergraduate Complexity Researcher placement at the Santa Fe Institute.

## References

[1] Hamid Bolouri and Eric H Davidson. "Modeling transcriptional regulatory networks". In: *BioEssays* 24.12 (2002), pp. 1118–1129.

[2] Douglas H Erwin and Eric H Davidson. "The evolution of hierarchical gene regulatory networks". In: *Nature Reviews Genetics* 10.2 (2009), pp. 141–148.

[3] Eric H Davidson. "Emerging properties of animal gene regulatory networks". In: *Nature* 468.7326 (2010), pp. 911–920.

[4] Nathalia Almeida et al. "Employing core regulatory circuits to define cell identity". EN. In: *The EMBO Journal* (May 2021). DOI: 10.15252/embj.2020106785. URL: https://www.embopress.org/doi/10.15252/embj.2020106785.

[5] Daniel Kim et al. "Gene regulatory network reconstruction: harnessing the power of single-cell multi-omic data". en. In: *npj Systems Biology and Applications* 9.1 (Oct. 2023). Publisher: Nature Publishing Group, p. 51. ISSN: 2056-7189. DOI: 10.1038/s41540-023-00312-6. URL: https://www.nature.com/articles/s41540-023-00312-6.

[6] Piyush B. Madhamshettiwar et al. "Gene regulatory network inference: evaluation and application to ovarian cancer allows the prioritization of drug targets". In: *Genome Medicine* 4.5 (May 2012), p. 41. ISSN: 1756-994X. DOI: 10.1186/gm340. URL: https://doi.org/10.1186/gm340.

[7] Isabelle S Peter and Eric H Davidson. "A gene regulatory network controlling the embryonic specification of endoderm". In: *Nature* 474.7353 (2011), pp. 635–639.

[8] Juan M. Escorcia-Rodríguez et al. "Improving gene regulatory network inference and assessment: The importance of using network structure". en. In: *Front. Genet.* 14 (Feb. 2023), p. 1143382. ISSN: 1664-8021. DOI: 10.3389/fgene.2023.1143382. URL: https://www.frontiersin.org/articles/10.3389/fgene.2023.1143382/full.

[9] Julio A. Freyre-González et al. "System Principles Governing the Organization, Architecture, Dynamics, and Evolution of Gene Regulatory Networks". English. In: *Front. Bioeng. Biotechnol.* 10 (May 2022). Publisher: Frontiers. ISSN: 2296-4185. DOI: 10.3389/fbioe.2022.888732. URL: https://www.frontiersin.org/journals/bioengineering-and-biotechnology/articles/10.3389/fbioe.2022.888732/full.

[10] Mengyuan Zhao et al. "A comprehensive overview and critical evaluation of gene regulatory network inference technologies". In: *Briefings in Bioinformatics* 22.5 (Sept. 2021), bbab009. ISSN: 1477-4054. DOI: 10.1093/bib/bbab009. URL: https://doi.org/10.1093/bib/bbab009.

[11] Riet De Smet and Kathleen Marchal. "Advantages and limitations of current network inference methods". en. In: *Nat Rev Microbiol* 8.10 (Oct. 2010), pp. 717–729. ISSN: 1740-1526, 1740-1534. DOI: 10.1038/nrmicro2419. URL: https://www.nature.com/articles/nrmicro2419.

[12] Yujie Chen et al. "Gene regulatory landscape dissected by single-cell four-omics sequencing". In: *Nature* (2026), pp. 1–10.

[13] Adrian I. Campos and Julio A. Freyre-González. "Evolutionary constraints on the complexity of genetic regulatory networks allow predictions of the total number of genetic interactions". en. In: *Sci Rep* 9.1 (Mar. 2019). Publisher: Nature Publishing Group, p. 3618. ISSN: 2045-2322. DOI: 10.1038/s41598-019-39866-z. URL: https://www.nature.com/articles/s41598-019-39866-z.

[14] Claus Kadelka et al. "A meta-analysis of Boolean network models reveals design principles of gene regulatory networks". en. In: *Sci. Adv.* 10.2 (Jan. 2024), eadj0822. ISSN: 2375-2548. DOI: 10.1126/sciadv.adj0822. URL: https://www.science.org/doi/10.1126/sciadv.adj0822.

[15] Mark B. Gerstein et al. "Architecture of the human regulatory network derived from ENCODE data". en. In: *Nature* 489.7414 (Sept. 2012). Publisher: Nature Publishing Group, pp. 91–100. ISSN: 1476-4687. DOI: 10.1038/nature11245. URL: https://www.nature.com/articles/nature11245.

[16] J J Hopfield. "Neural networks and physical systems with emergent collective computational abilities." In: *Proceedings of the National Academy of Sciences* 79.8 (Apr. 1982). Publisher: Proceedings of the National Academy of Sciences, pp. 2554–2558. DOI: 10.1073/pnas.79.8.2554. URL: https://www.pnas.org/doi/10.1073/pnas.79.8.2554.

[17] Mukund Thattai and Alexander van Oudenaarden. "Intrinsic noise in gene regulatory networks". In: *Proceedings of the National Academy of Sciences of the United States of America* 98.15 (July 2001), pp. 8614–8619. ISSN: 0027-8424. DOI: 10.1073/pnas.151588598. URL: https://pmc.ncbi.nlm.nih.gov/articles/PMC37484/.

[18] Michael B. Elowitz et al. "Stochastic Gene Expression in a Single Cell". en. In: *Science* 297.5584 (Aug. 2002). Publisher: American Association for the Advancement of Science (AAAS), pp. 1183–1186. ISSN: 0036-8075, 1095-9203. DOI: 10.1126/science.1070919. URL: https://www.science.org/doi/10.1126/science.1070919.

[19] Vivek Kohar and Mingyang Lu. "Role of noise and parametric variation in the dynamics of gene regulatory circuits". en. In: *npj Systems Biology and Applications* 4.1 (Nov. 2018). Publisher: Nature Publishing Group, p. 40. ISSN: 2056-7189. DOI: 10.1038/s41540-018-0076-x. URL: https://www.nature.com/articles/s41540-018-0076-x.

[20] James Holehouse. *Do Distinct Subpopulations Signify Modes of Behavior in a Noisy Single Cell?* en. ISSN: 2692-8205 Pages: 2025.07.22.666238 Section: New Results. July 2025. DOI: 10.1101/2025.07.22.666238. URL: https://www.biorxiv.org/content/10.1101/2025.07.22.666238v1.

[21] Lucy Ham, Rowan D. Brackston, and Michael P. H. Stumpf. "Extrinsic Noise and Heavy-Tailed Laws in Gene Expression". en. In: *Phys. Rev. Lett.* 124.10 (Mar. 2020), p. 108101. ISSN: 0031-9007, 1079-7114. DOI: 10.1103/PhysRevLett.124.108101. URL: https://link.aps.org/doi/10.1103/PhysRevLett.124.108101.

[22] Apolline J R Petit, Anne Genissel, and Arnaud Le Rouzic. "Gene Expression Plasticity Is Associated with Regulatory Complexity but Not with Specific Network Motifs". en. In: *bioRxiv* (June 2025). DOI: https://doi.org/10.1101/2024.03.11.584403.

[23] Uri Alon. "Network motifs: theory and experimental approaches". en. In: *Nature Reviews Genetics* 8.6 (June 2007), pp. 450–461. ISSN: 1471-0056, 1471-0064. DOI: 10.1038/nrg2102. URL: https://www.nature.com/articles/nrg2102.

[24] Lesley T. MacNeil and Albertha J. M. Walhout. "Gene regulatory networks and the role of robustness and stochasticity in the control of gene expression". en. In: *Genome Research* 21.5 (May 2011). Company: Cold Spring Harbor Laboratory Press Distributor: Cold Spring Harbor Laboratory Press Institution: Cold Spring Harbor Laboratory Press Label: Cold Spring Harbor Laboratory Press Publisher: Cold Spring Harbor Lab, pp. 645–657. ISSN: 1088-9051, 1549-5469. DOI: 10.1101/gr.097378.109. URL: http://genome.cshlp.org/content/21/5/645.

[25] Claus Kadelka and David Murrugarra. "Canalization reduces the nonlinearity of regulation in biological networks". en. In: *npj Syst Biol Appl* 10.1 (June 2024). Publisher: Nature Publishing Group, p. 67. ISSN: 2056-7189. DOI: 10.1038/s41540-024-00392-y. URL: https://www.nature.com/articles/s41540-024-00392-y.

[26] Lori Layne, Elena Dimitrova, and Matthew Macauley. *Nested canalyzing depth and network stability*. arXiv:1111.2759 [q-bio]. Nov. 2011. DOI: 10.48550/arXiv.1111.2759. URL: http://arxiv.org/abs/1111.2759.

[27] R. Lande. "Adaptation to an extraordinary environment by evolution of phenotypic plasticity and genetic assimilation". In: *Journal of Evolutionary Biology* 22.7 (July 2009), pp. 1435–1446. ISSN: 1010-061X. DOI: 10.1111/j.1420-9101.2009.01754.x. URL: https://doi.org/10.1111/j.1420-9101.2009.01754.x.

[28] Reinhard Bürger and Michael Lynch. "EVOLUTION AND EXTINCTION IN A CHANGING ENVIRONMENT: A QUANTITATIVE-GENETIC ANALYSIS". In: *Evolution* 49.1 (Feb. 1995), pp. 151–163. ISSN: 0014-3820. DOI: 10.1111/j.1558-5646.1995.tb05967.x. URL: https://doi.org/10.1111/j.1558-5646.1995.tb05967.x.

[29] Bhaskar Kumawat et al. "Evolution takes multiple paths to evolvability when facing environmental change". In: *Proceedings of the National Academy of Sciences* 122.1 (Jan. 2025). Publisher: Proceedings of the National Academy of Sciences, e2413930121. DOI: 10.1073/pnas.2413930121. URL: https://www.pnas.org/doi/abs/10.1073/pnas.2413930121.

[30] Yifei Wang, Joanna Bryson, and Stephen Matthews. "Evolving Evolvability in the Context of Environmental Change: A Gene Regulatory Network (GRN) Approach". en. In: MIT Press, July 2014, pp. 47–53. DOI: 10.1162/978-0-262-32621-6-ch010. URL: https://dx.doi.org/10.1162/978-0-262-32621-6-ch010.

[31] Daniel M Weinreich et al. *The population genetics of biological noise*. en. Jan. 2025. DOI: 10.1101/2025.01.11.632402. URL: http://biorxiv.org/lookup/doi/10.1101/2025.01.11.632402.

[32] A Wagner. "Evolution of gene networks by gene duplications: a mathematical model and its implications on genome organization." In: *Proceedings of the National Academy of Sciences* 91.10 (May 1994). Publisher: Proceedings of the National Academy of Sciences, pp. 4387–4391. DOI: 10.1073/pnas.91.10.4387. URL: https://www.pnas.org/doi/10.1073/pnas.91.10.4387.

[33] Andreas Wagner. "DOES EVOLUTIONARY PLASTICITY EVOLVE?" en. In: *Evolution* 50.3 (June 1996), pp. 1008–1023. ISSN: 00143820. DOI: 10.1111/j.1558-5646.1996.tb02342.x. URL: https://academic.oup.com/evolut/article/50/3/1008/6870974.

[34] Lucy Ham, Rowan D Brackston, and Michael PH Stumpf. "Extrinsic noise and heavy-tailed laws in gene expression". In: *Physical review letters* 124.10 (2020), p. 108101.

[35] Ramon Grima and Pierre-Marie Esmenjaud. "Quantifying and correcting bias in transcriptional parameter inference from single-cell data". In: *Biophysical Journal* 123.1 (2024), pp. 4–30.

[36] James Holehouse. "Quantifying broken detailed balance in transcription". In: *npj Complexity* 3.1 (2026), p. 10.

[37] Yipei Guo and Ariel Amir. "Exploring the effect of network topology, mRNA and protein dynamics on gene regulatory network stability". en. In: *Nat Commun* 12.1 (Jan. 2021). Publisher: Nature Publishing Group, p. 130. ISSN: 2041-1723. DOI: 10.1038/s41467-020-20472-x. URL: https://www.nature.com/articles/s41467-020-20472-x.

[38] Z. Burda et al. "Motifs emerge from function in model gene regulatory networks". en. In: *Proc. Natl. Acad. Sci. U.S.A.* 108.42 (Oct. 2011), pp. 17263–17268. ISSN: 0027-8424, 1091-6490. DOI: 10.1073/pnas.1109435108. URL: https://pnas.org/doi/full/10.1073/pnas.1109435108.

[39] Albert-László Barabási and Zoltán N. Oltvai. "Network biology: understanding the cell's functional organization". en. In: *Nat Rev Genet* 5.2 (Feb. 2004), pp. 101–113. ISSN: 1471-0056, 1471-0064. DOI: 10.1038/nrg1272. URL: https://www.nature.com/articles/nrg1272.

[40] Nadav Kashtan and Uri Alon. "Spontaneous evolution of modularity and network motifs". In: *Proceedings of the National Academy of Sciences* 102.39 (Sept. 2005), pp. 13773–13778. DOI: 10.1073/pnas.0503610102. URL: https://www.pnas.org/doi/10.1073/pnas.0503610102.

[41] Arnaud Le Rouzic and Sylvain Pouzet. *Gene network topology drives the mutational landscape of gene expression*. en. Nov. 2024. DOI: 10.1101/2024.11.28.625874. URL: http://biorxiv.org/lookup/doi/10.1101/2024.11.28.625874.

[42] Mark L. Siegal and Aviv Bergman. "Waddington's canalization revisited: Developmental stability and evolution". In: *Proceedings of the National Academy of Sciences* 99.16 (Aug. 2002). Publisher: Proceedings of the National Academy of Sciences, pp. 10528–10532. DOI: 10.1073/pnas.102303999. URL: https://www.pnas.org/doi/full/10.1073/pnas.102303999.

[43] Peter S Swain, Michael B Elowitz, and Eric D Siggia. "Intrinsic and extrinsic contributions to stochasticity in gene expression". In: *Proceedings of the National Academy of Sciences* 99.20 (2002), pp. 12795–12800.

[44] William J Blake et al. "Phenotypic consequences of promoter-mediated transcriptional noise". In: *Molecular cell* 24.6 (2006), pp. 853–865.

[45] Yuichi Taniguchi et al. "Quantifying E. coli proteome and transcriptome with single-molecule sensitivity in single cells". In: *science* 329.5991 (2010), pp. 533–538.

[46] Nico Battich, Thomas Stoeger, and Lucas Pelkmans. "Control of transcript variability in single mammalian cells". In: *Cell* 163.7 (2015), pp. 1596–1610.

[47] Peter D Tonner et al. "A regulatory hierarchy controls the dynamic transcriptional response to extreme oxidative stress in archaea". In: *PLoS Genetics* 11.1 (2015), e1004912.

[48] James Holehouse, Zhixing Cao, and Ramon Grima. "Stochastic Modeling of Autoregulatory Genetic Feedback Loops: A Review and Comparative Study". English. In: *Biophysical Journal* 118.7 (Apr. 2020). Publisher: Elsevier, pp. 1517–1525. ISSN: 0006-3495, 1542-0086. DOI: 10.1016/j.bpj.2020.02.016. URL: https://www.cell.com/biophysj/abstract/S0006-3495(20)30165-X.

[49] Ioannis Lestas et al. "Noise in Gene Regulatory Networks". In: *IEEE Transactions on Automatic Control* 53.Special Issue (Jan. 2008), pp. 189–200. ISSN: 1558-2523. DOI: 10.1109/TAC.2007.911347. URL: https://ieeexplore.ieee.org/abstract/document/4439816.

[50] R Milo et al. "Network Motifs: Simple Building Blocks of Complex Networks". en. In: (2002).

[51] E. Wong et al. "Biological network motif detection: principles and practice". en. In: *Briefings in Bioinformatics* 13.2 (Mar. 2012), pp. 202–215. ISSN: 1467-5463, 1477-4054. DOI: 10.1093/bib/bbr033. URL: https://academic.oup.com/bib/article-lookup/doi/10.1093/bib/bbr033.

[52] Daniel M Weinreich et al. "Should evolutionary geneticists worry about higher-order epistasis?" In: *Current Opinion in Genetics & Development*. Genetics of system biology 23.6 (Dec. 2013), pp. 700–707. ISSN: 0959-437X.

[53] C. Brandon Ogbunugafor et al. "Adaptive Landscape by Environment Interactions Dictate Evolutionary Dynamics in Models of Drug Resistance". en. In: *PLOS Computational Biology* 12.1 (Jan. 2016), e1004710. ISSN: 1553-7358.

[54] Patrick C. Phillips. "Epistasis — the essential role of gene interactions in the structure and evolution of genetic systems". en. In: *Nature Reviews Genetics* 9.11 (Nov. 2008), pp. 855–867. ISSN: 1471-0056, 1471-0064. DOI: 10.1038/nrg2452. URL: https://www.nature.com/articles/nrg2452.

[55] Dmitry Krotov. "A new frontier for Hopfield networks". In: *Nature Reviews Physics* 5.7 (2023), pp. 366–367.

[56] Hubert Ramsauer et al. "Hopfield networks is all you need". In: *arXiv preprint arXiv:2008.02217* (2020).

[57] Bhaswar Ghosh, Rajesh Karmakar, and Indrani Bose. "Noise characteristics of feed forward loops". en. In: *Physical Biology* 2.1 (Mar. 2005), p. 36. ISSN: 1478-3975. DOI: 10.1088/1478-3967/2/1/005. URL: https://doi.org/10.1088/1478-3967/2/1/005.

[58] Luca Albergante, J Julian Blow, and Timothy J Newman. "Buffered Qualitative Stability explains the robustness and evolvability of transcriptional networks". In: *eLife* 3 (Sept. 2014). Ed. by Detlef Weigel, e02863. ISSN: 2050-084X. DOI: 10.7554/eLife.02863. URL: https://doi.org/10.7554/eLife.02863.

[59] R Sole and S Valverde. "Are network motifs the spandrels of cellular complexity?" en. In: *Trends in Ecology & Evolution* 21.8 (Aug. 2006), pp. 419–422. ISSN: 01695347. DOI: 10.1016/j.tree.2006.05.013. URL: https://linkinghub.elsevier.com/retrieve/pii/S0169534706001674.

[60] Sebastian Matuszewski, Joachim Hermisson, and Michael Kopp. "Catch Me if You Can: Adaptation from Standing Genetic Variation to a Moving Phenotypic Optimum". In: *Genetics* 200.4 (Aug. 2015), pp. 1255–1274. ISSN: 0016-6731. DOI: 10.1534/genetics.115.178574. URL: https://pmc.ncbi.nlm.nih.gov/articles/PMC4574244/.

[61] Christopher R. Evans et al. "Errors during Gene Expression: Single-Cell Heterogeneity, Stress Resistance, and Microbe-Host Interactions". In: *mBio* 9.4 (July 2018), e01018–18. ISSN: 2150-7511. DOI: [10.1128/mBio.01018-18](10.1128/mBio.01018-18). URL: <https://pmc.ncbi.nlm.nih.gov/articles/PMC6030554/>.

[62] Anil Ananthaswamy. "With a Little Help from Physics". en. In: *Why Machines Learn*. New York: DUTTON, 2024.

# A  Parameters and Code Availability

Table 1: Standard parameters used in the simulation.

| Symbol | Description | Value |
| --- | --- | --- |
| $G$ | Number of generations per run | 500 |
| $M$ | Max steps before declaring a state unstable | 100 |
| $s$ | Selection pressure | 10 |
| unstable_fitness | Fitness for unstable matrices | $\exp(-10)$ |
| mode | Type of initial population. Stable are random matrices that show a stable phenotype | "stable" |
| pop_size | Population size | 300 |
| $N$ | Number of genes | 10 |
| $c$ | Initial matrix density | 1.0 |
| $p_{\text{init}}$ | Proportion of +1 in initial gene expression state | 0.5 |
| $p_{\text{phen}}$ | Proportion of +1 in target optimal phenotype | 0.5 |
| $\mu_r$ | Weights distribution mean for sampling $W_{ij}$ | 0.0 |
| $\sigma_r$ | Weights distribution std. dev. for sampling $W_{ij}$ | 1.0 |
| $p_r$ | Regular mutation probability | 1.0 |
| $p_{\text{rec}}$ | Recombination probability | 0.5 |
| noise_prob | Probability of weight being affected by noise | 1.0 |
| noise_dist | Noise distribution | Bernoulli(1.0) |

# B  Pseudocode Evolutionary Algorithm

We describe the process incorporating Sections 2.1 and 2.2 using pseudocode.

---

**Algorithm 1** Evolutionary Algorithm for Stochastic Gene Regulatory Networks

---

1: **procedure** GRN EVOLUTION
2:    **set:** pop_size $= 300$, $N = 10$, $p_m = 0.01$, $s = 10$, $G_{\max} = 500$
3:    **sample:** $\boldsymbol{P}_0, \boldsymbol{P}_{\text{opt}} \in \{\pm 1\}^N$
4:    **while** $|\text{Pop}| < \text{pop\_size}$ **do**    ▷ Initial population with deterministic stability
5:      sample $W_{ij} \sim \mathcal{N}(0,1)$
6:      **if** phenotype from $\boldsymbol{W}' = \boldsymbol{W}$ is stable **then**
7:        add $(\boldsymbol{P}_0, W, \boldsymbol{P}_{\text{opt}})$ to Pop
8:      **end if**
9:    **end while**
10:   **for** $g = 1$ to $G_{\max}$ **do**    ▷ Run each generation
11:     initialize empty population Pop$'$
12:     **while** $|\text{Pop}'| < \text{pop\_size}$ **do**
13:       select parents $\boldsymbol{W}^{(1)}, \boldsymbol{W}^{(2)}$ uniformly from Pop    ▷ Sexual reproduction
14:       initialize offspring genotype $W$
15:       **for** $i = 1$ to $N$ **do**
16:         $\boldsymbol{W}_{i\cdot} \leftarrow \begin{cases} \boldsymbol{W}_{i\cdot}^{(1)} & \text{with probability } 1/2 \\ \boldsymbol{W}_{i\cdot}^{(2)} & \text{otherwise} \end{cases}$
17:       **end for**
18:       **for all** $i, j \leq N$ **do**    ▷ Mutation
19:         **if** $rand(0,1) < p_m$ **then**
20:           $W_{ij} \leftarrow \mathcal{N}(0,1)$
21:         **end if**
22:       **end for**
23:       **for all** $i, j \leq N$ **do**    ▷ Obtain combined matrix
24:         sample $X_{ij} \sim \Gamma(1/\theta, \theta)$
25:         $W'_{ij} \leftarrow W_{ij} X_{ij}$
26:       **end for**
27:       phenotype $\leftarrow$ unstable    ▷ Obtain the phenotype
28:       **for** $t = 0$ to $100$ **do**
29:         $\boldsymbol{P}_{t+1} = \text{sign}(\boldsymbol{W}' \boldsymbol{P}_t)$
30:         **if** $\boldsymbol{P}_{t+1} = \boldsymbol{P}_t$ **then**
31:           phenotype $\leftarrow$ stable; $t+1$
32:           $\boldsymbol{P} \leftarrow \boldsymbol{P}_{t+1}$
33:           **break**
34:         **end if**
35:         $\boldsymbol{P}_t \leftarrow \boldsymbol{P}_{t+1}$
36:       **end for**
37:       $F \leftarrow \begin{cases} \exp(-s\, d(\boldsymbol{P}, \boldsymbol{P}_{\text{opt}})) & \text{if stable} \\ \exp(-s) & \text{otherwise} \end{cases}$    ▷ Fitness-based acceptance
38:       **if** $rand(0,1) < F$ **then**
39:         add offspring to Pop$'$
40:       **end if**
41:     **end while**
42:     Pop $\leftarrow$ Pop$'$    ▷ Update population
43:   **end for**
44: **end procedure**

## C   Alignment and Coherence

Let $W \in \mathbb{R}^{N \times N}$ be a matrix with nonzero real numbers and let $\boldsymbol{P} \in \{\pm 1\}^N$ denote the target phenotype. Define the alignment score

$$A(\boldsymbol{W}) = \frac{1}{N} \sum_{i=1}^{N} \sum_{j=1}^{N} P_i P_j \, W_{ij}^{\text{norm}},$$

where $W_{ij}^{\text{norm}}$ denotes the row–normalized interaction strengths.

**Proposition:** Maximum alignment is achieved when all interactions have the same sign as the product of two elements in the target states.

In other words,

$$A(\boldsymbol{W}) = 1 \quad \iff \quad \text{sign}(W_{ij}) = \text{sign}(P_i P_j) \ \text{for all } i, j.$$

**Proof:** By construction, each row of $W^{\text{norm}}$ satisfies

$$\sum_{j=1}^{N} |W_{ij}^{\text{norm}}| = 1,$$

which implies

$$-1 \leq \sum_{j=1}^{N} P_i P_j W_{ij}^{\text{norm}} \leq 1 \quad \text{for all } i.$$

Hence $A(\boldsymbol{W}) \leq 1$.

The equality $A(\boldsymbol{W}) = 1$ holds if and only if every term in each inner sum is non–negative, i.e.,

$$P_i P_j W_{ij}^{\text{norm}} \geq 0 \quad \text{for all } i, j.$$

Since normalization does not change signs, this condition is equivalent to

$$\text{sign}(W_{ij}) = \text{sign}(P_i P_j) \quad \text{for all } i, j.$$

If this condition fails for any pair $(i, j)$, at least one row sum is strictly less than 1, implying $A(\boldsymbol{W}) < 1$. Conversely, if it holds for all pairs, each row sum equals 1, and therefore $A(\boldsymbol{W}) = 1$.

**Definition: Aligned Path**

Let $G$ be a directed signed graph with adjacency matrix $W$ and target phenotype $\boldsymbol{P}$. A path from node $a_1$ to node $a_n$ is a sequence $(a_1, a_2, \ldots, a_n)$ such that $W_{a_k a_{k+1}} \neq 0$ for all $k = 1, \ldots, n-1$. The path is *aligned* if

$$\text{sign}(P_{a_1} P_{a_n}) = \text{sign}\left(\prod_{k=1}^{n-1} W_{a_k a_{k+1}}\right).$$

**Corollary:** If a graph with adjacency matrix $W$ satisfies $A(\boldsymbol{W}) = 1$, then all feedforward and feedback loops in the graph are coherent.

From the proposition, $A(\boldsymbol{W}) = 1$ implies

$$\text{sign}(W_{ij}) = \text{sign}(P_i P_j) \ \forall \ i, j.$$

Consider any path $(a_1, \ldots, a_n)$. The product of edge signs along the path satisfies

$$\text{sign}\left(\prod_{k=1}^{n-1} W_{a_k a_{k+1}}\right) = \text{sign}\left(P_{a_1} P_{a_2}^2 \ldots P_{a_{n-1}}^2 P_{a_n}\right) = \text{sign}(P_{a_1} P_{a_n}),$$

so every path is aligned.

A feedforward loop consists of two paths sharing the same start and end nodes; since all such paths are aligned, their net signs agree, and the motif is coherent. A feedback loop is a path that begins and ends at the same node, for which

$$\text{sign}\left(\prod W_{a_k a_{k+1}}\right) = \text{sign}(P_{a_1} P_{a_1}) = +1,$$

implying that all feedback loops are reinforcing.

# Supplementary Information

## S1    Mutational Robustness Scatterplots

Figures S1, S2, S3, S4, S5, S6, and S7 present scatterplots of the four mutational robustness metrics $\langle \text{Mut}(\boldsymbol{W}) \rangle$ before and after evolution, across all 30 trials per noise condition (each point represents a single population). The diagonal line $y = x$ serves as a reference: points falling below it indicate a decrease in the respective metric after evolution, corresponding to a reduction in either the mean (panels A and C) or the spread (panels B and D) of the phenotype distribution upon mutation. Taken together, the results reveal a consistent, gradual decline in all mutational robustness metrics as environmental noise variability increases.

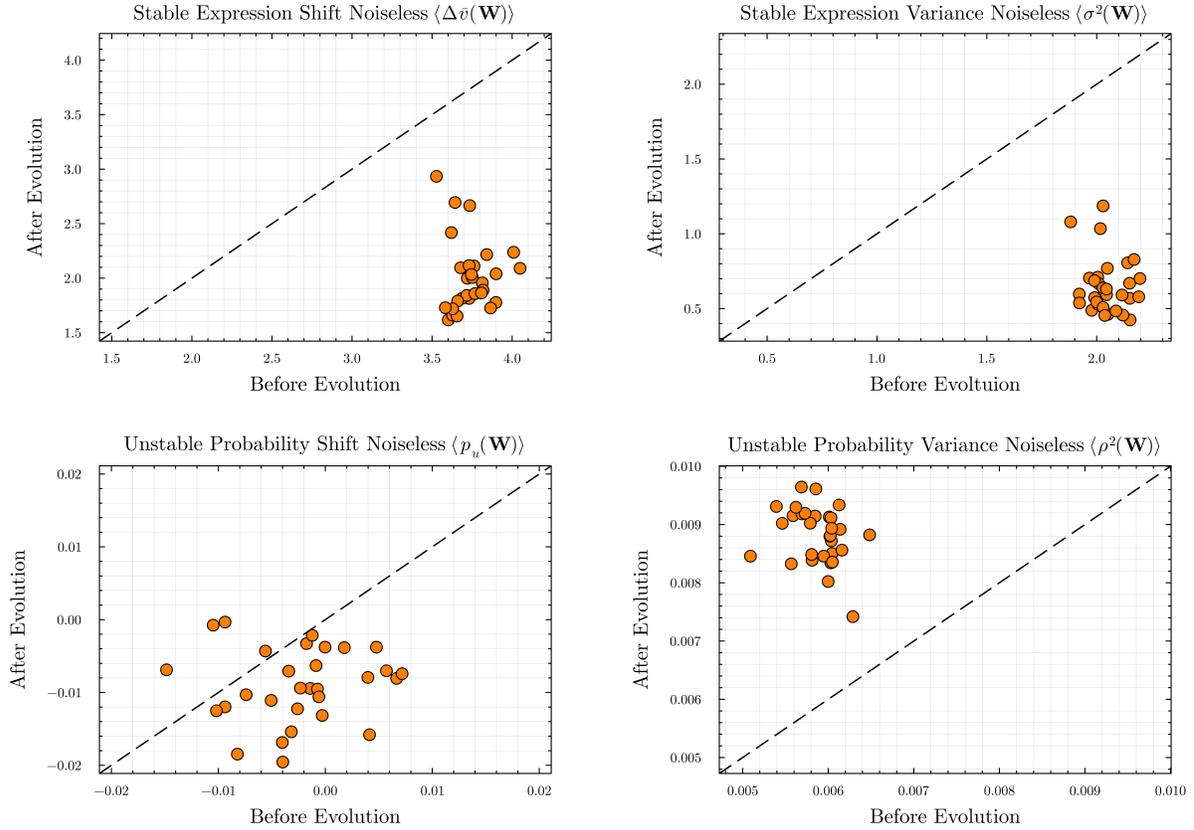

Figure S1: **Mutational robustness metrics** of 30 populations evolved with **noiseless** phenotype expression. Each dot is the average across a population, and the dashed line is the diagonal $y = x$: every dot below it indicates a decrease after evolution. (A) The average phenotype shift is the difference between the average stable phenotypes of mutated and non-mutated genotypes. (B) Stable-phenotype variability is the variance in the distance between the average stable phenotypes of mutated genotypes. (C) Average instability difference, which is the difference in the probability of showing stable phenotypes of mutated and non-mutated genotypes. (D) Instability variance is the spread of the sample instability probabilities upon mutation.

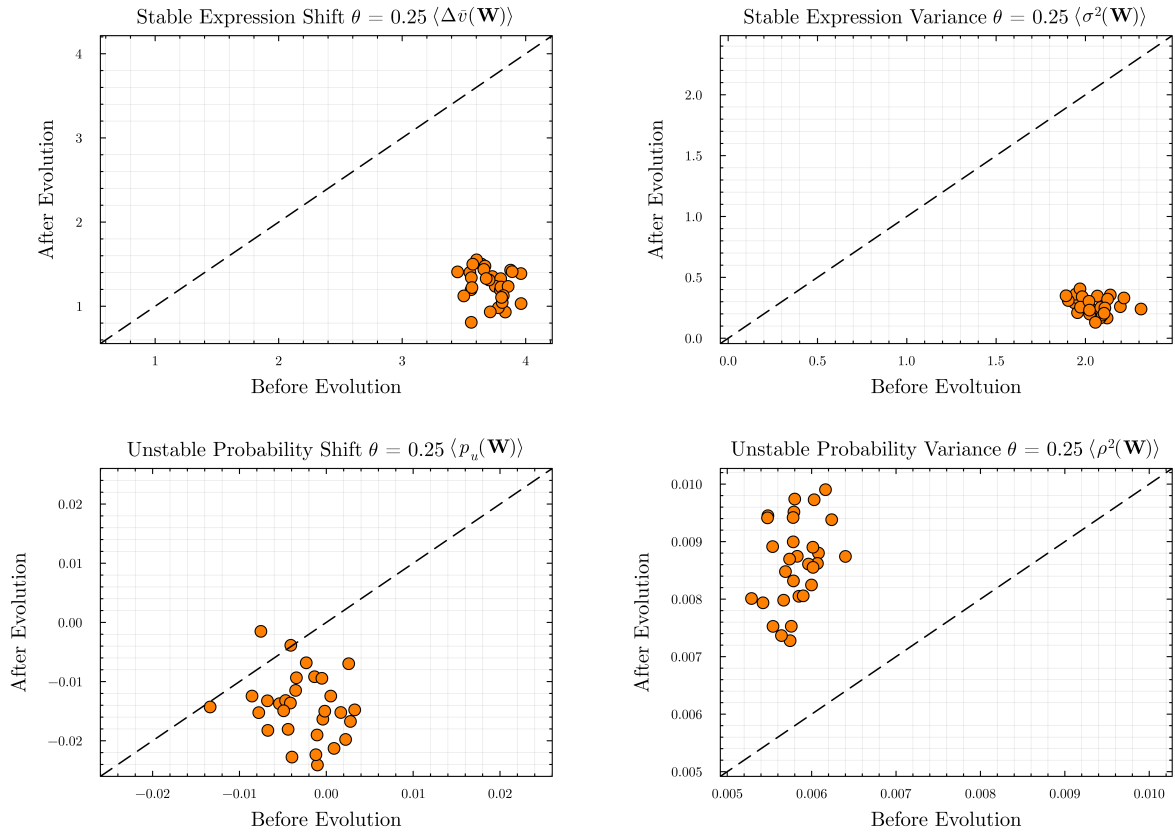

Figure S2: **Mutational robustness metrics** of 30 populations evolved with noisy phenotype expression at $\theta = 0.25$.

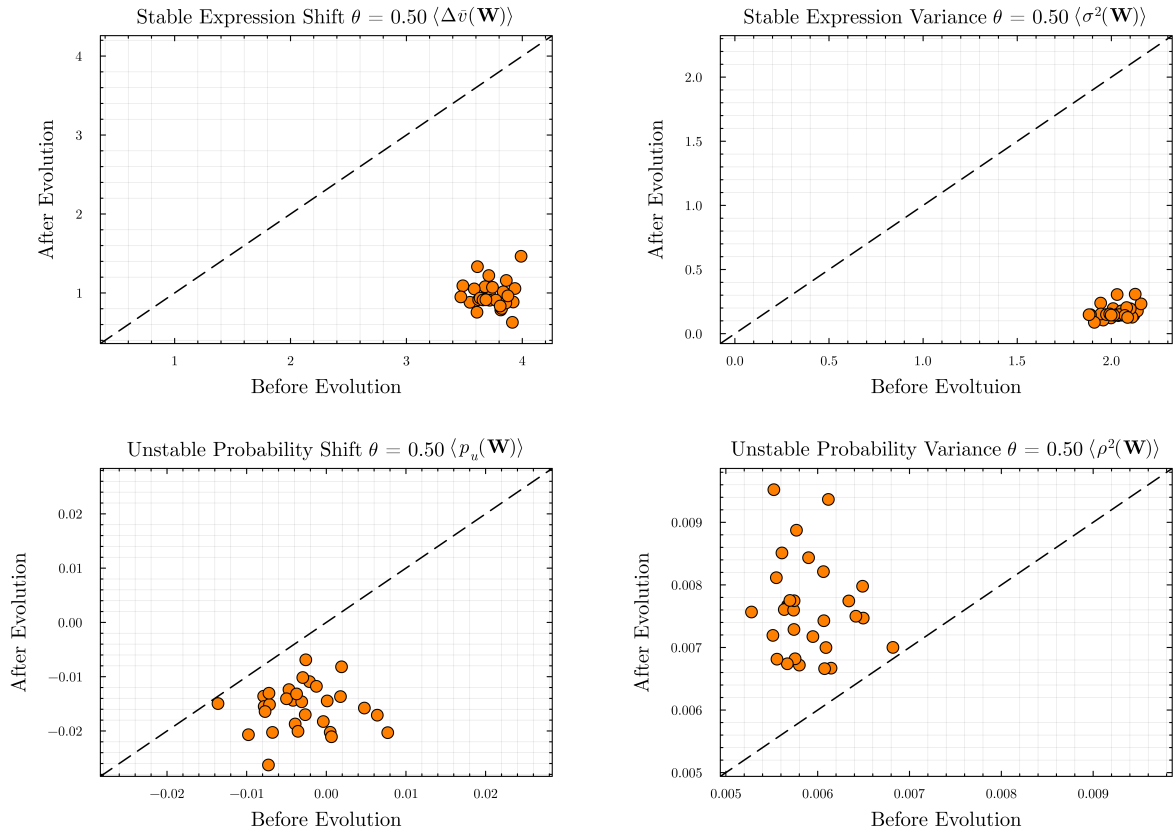

Figure S3: **Mutational robustness metrics** of 30 populations evolved with noisy phenotype expression at $\theta = 0.50$.

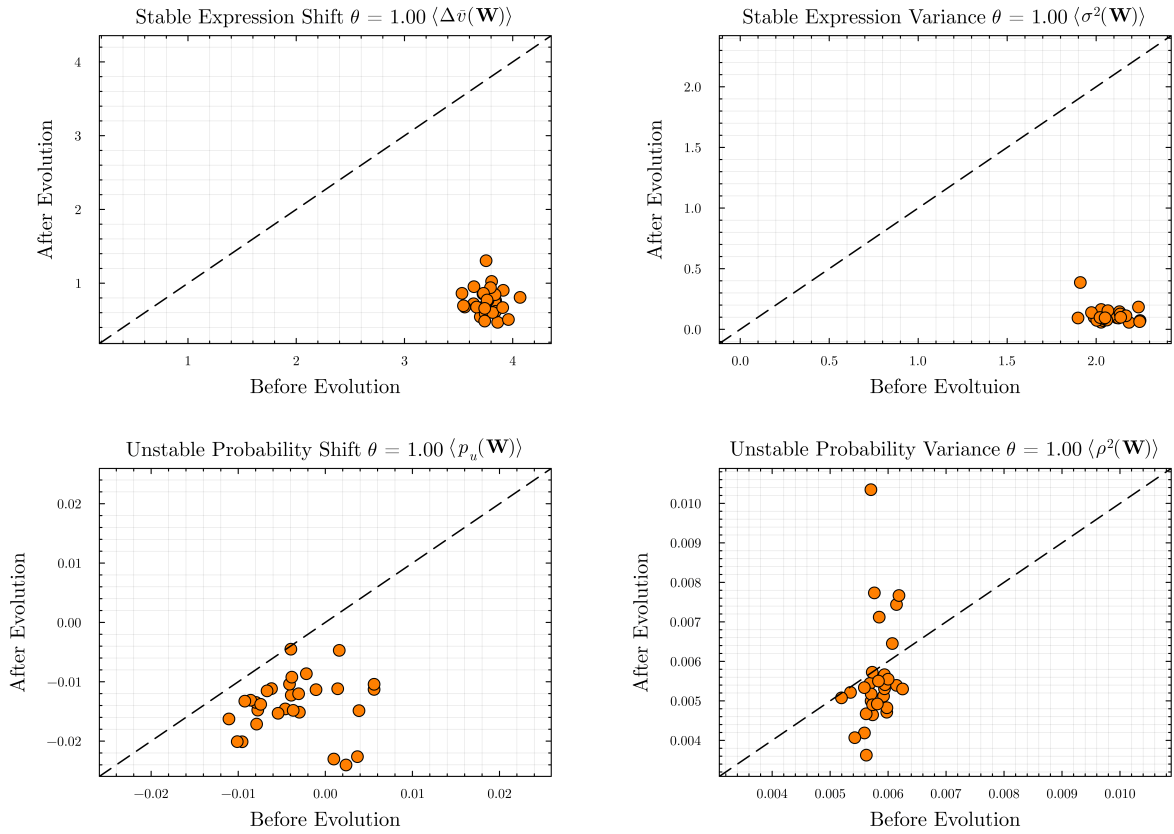

Figure S4: **Mutational robustness metrics** of 30 populations evolved with noisy phenotype expression at $\theta = 1.00$.

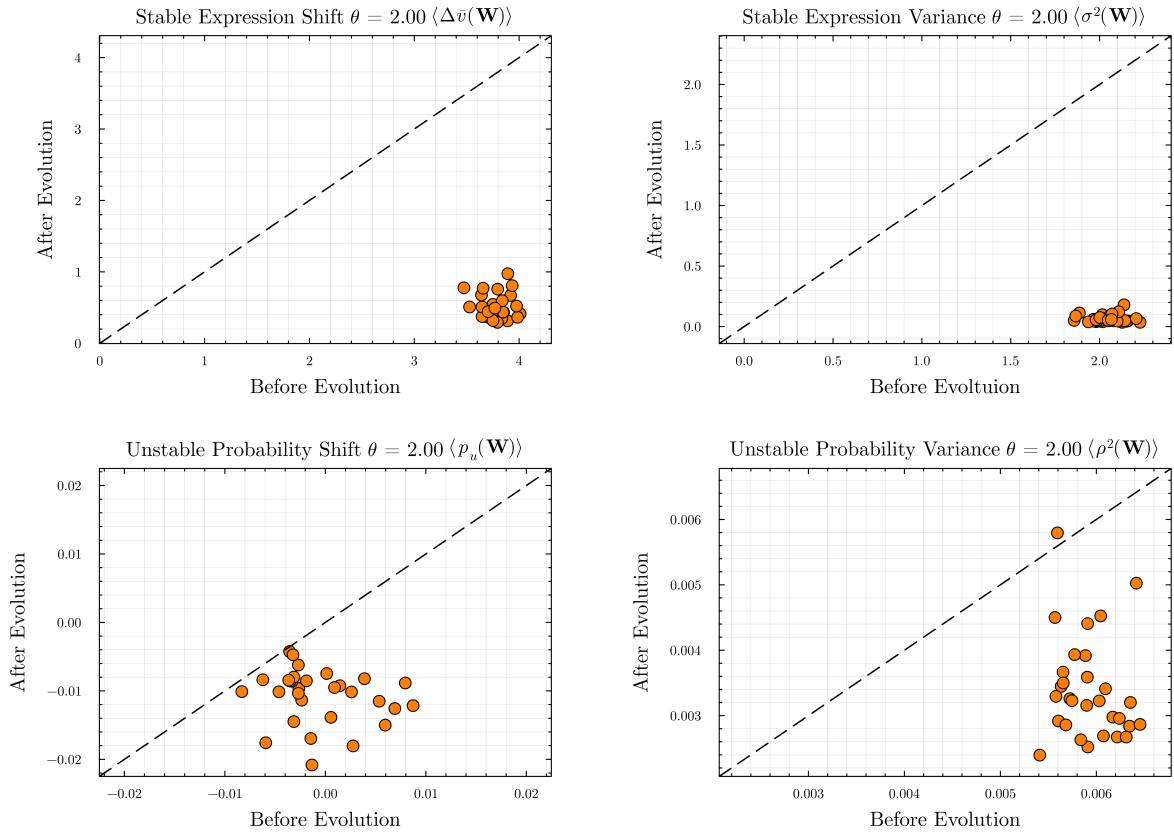

Figure S5: **Mutational robustness metrics** of 30 populations evolved with noisy phenotype expression at $\theta = 2.00$.

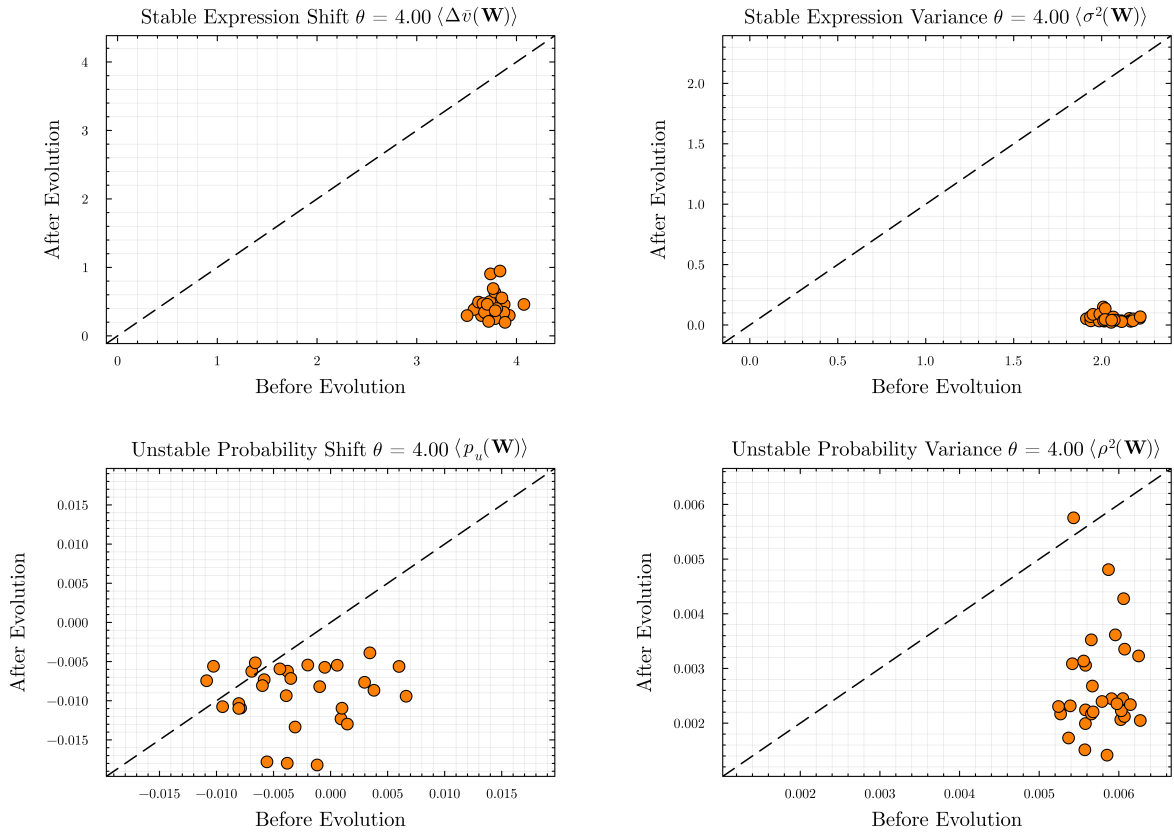

Figure S6: **Mutational robustness metrics** of 30 populations evolved with noisy phenotype expression at $\theta = 4.00$.

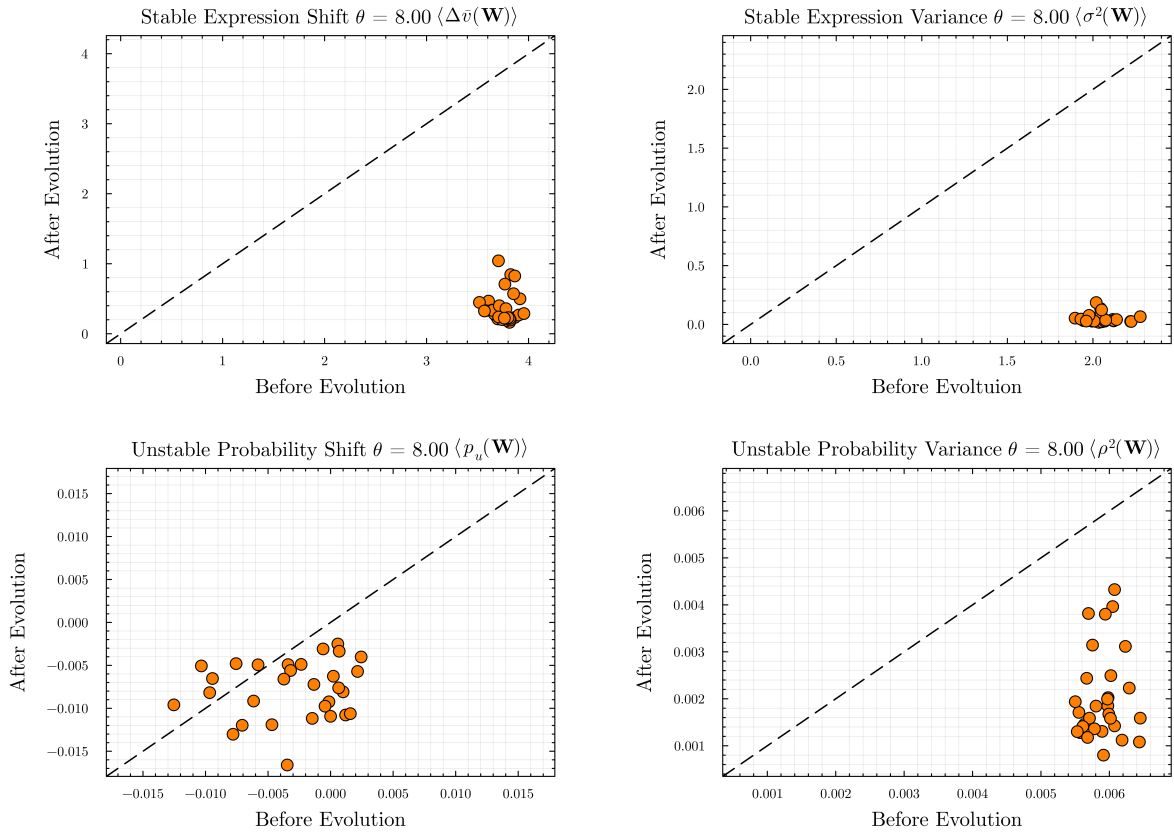

Figure S7: **Mutational robustness metrics** of 30 populations evolved with noisy phenotype expression at $\theta = 8.00$.

## S2 Robustness of the Results

### S2.1 Asynchronous Gene Dynamics

In the original work by Hopfield [16], gene states are updated asynchronously, meaning that at each timestep only one gene is updated based on its inputs, rather than updating all genes at once as in the main text. This update is typically applied either randomly or sequentially, so different genes can be updated at different times. A key property of these dynamics is that the system eventually reaches a stable state in which no further updates change any gene (see [62] for an accessible explanation).

In our implementation, we use a sequential version of asynchronous updates: we iterate over all genes, updating each in turn using its incoming interactions. We repeat this process in full sweeps over all genes until convergence. We define a state as stable when a complete sweep over all genes produces no changes in any gene state. In this case, the path length is the number of sweeps required for convergence. We note, however, that this implementation might fail to show stable phenotypes that take a long time to converge. Nevertheless, almost all matrices resulted in stable phenotype expression (figure not shown). We used the standard parameters in Appendix A with a stable initial population.

In Figure S8, we see that the average path length is smaller in these types of dynamics, but no qualitative difference in the decreasing trend. We find no difference in fitness evolution nor in the enrichment of network motifs. In Figure S12, we see the same behavior for almost all metrics, but not for $\langle \rho^2(\boldsymbol{W}) \rangle$. Instead, we see that it reaches a maximum for small values of $\sqrt{\theta}$ and then decreases sharply as the noise level increases. We do not have an explanation for this phenomenon. The variance in the probability of showing a stable phenotype seems to increase for genotypes evolved under relatively low noise levels $\sqrt{\theta}$.

In our statistical tests, we find that noiseless evolution results in a lower average alignment than noisy evolution ($t = 31.47, P = 6.04 \times 10^{-69}$). Additionally, coherent FFLs are enriched in evolved populations relative to non-evolved ones ($t = 4.71, P = 3.88 \times 10^{-5}$), and this enrichment is even stronger in populations evolved under high noise compared to those evolved without noise ($t = 36.58, P = 1.86 \times 10^{-27}$).

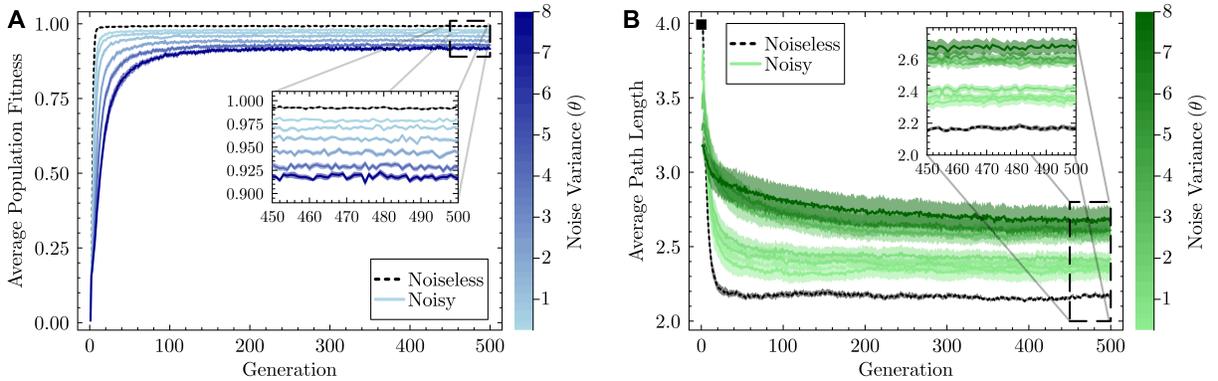

Figure S8: **(A) Evolution of the average population fitness and (B) average mean path length per generation**, across 30 independent replicates at different noise variances $\theta$. Shaded regions represent 95% confidence intervals around the mean in the zoomed-out panels and 68% in the zoomed-in panels. Lighter colors correspond to smaller noise variances. The noiseless case ($\theta = 0$) is shown with a dotted line.

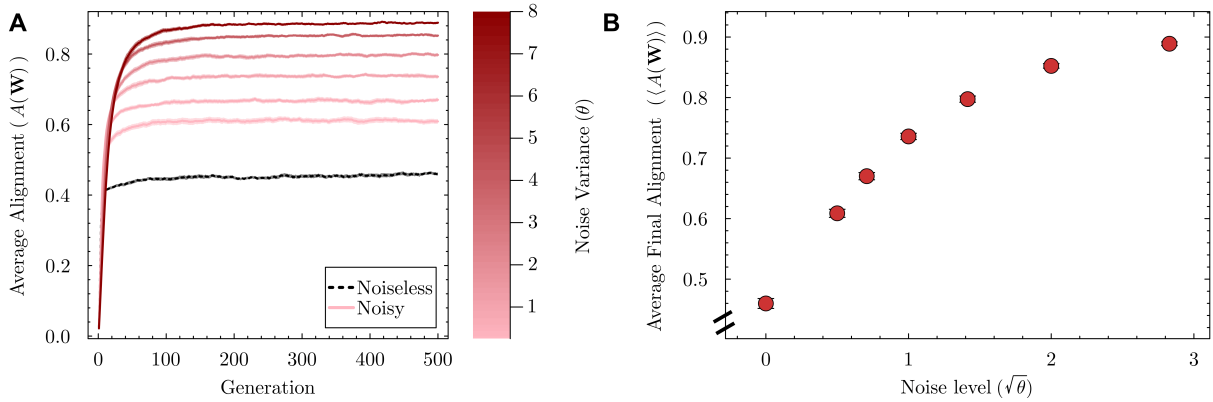

Figure S9: **Evolution of the average alignment scores**. (A) Average alignment score across 30 populations over generations. Each line represents the average alignment score for a population, and the color fillings represent 95% confidence intervals. Lighter colors represent smaller noise variances. The noiseless scenario is plotted with a dotted line. (B) Average alignment scores after evolution from 30 evolved populations across various noise levels $\sqrt{\theta}$. The 95% confidence intervals are plotted along the means.

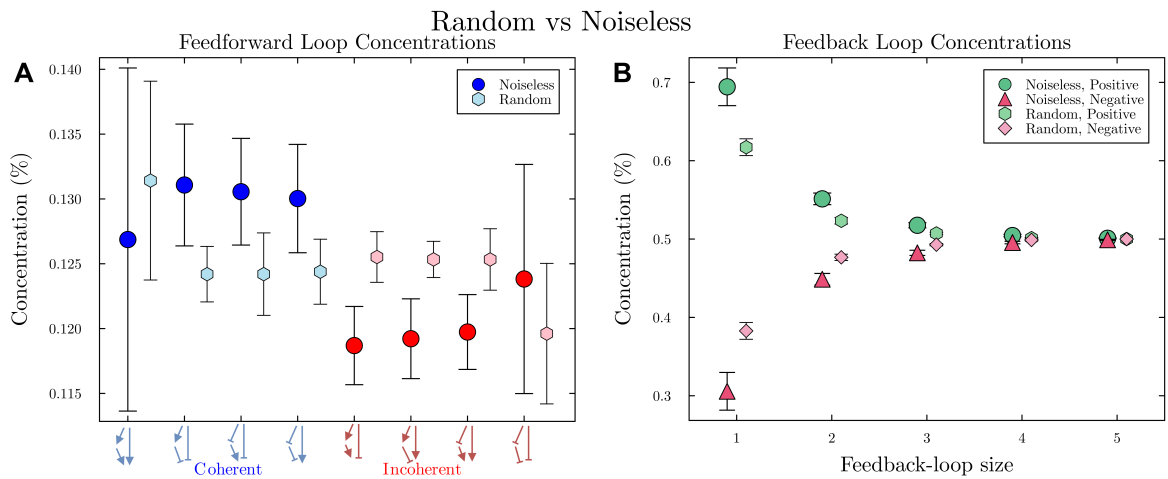

Figure S10: **Enrichment of network motifs as a result of noiseless evolution**. Comparison of 30 populations evolved without noise (darker hues) against non-evolved populations of stable matrices (lighter hues). (A) Average concentration per type of FFL. On the x-axis, we place the type of loop with a diagram. The error bars represent 95% confidence intervals around the averages. (B) Average concentration of positive (green circles and hexagons) and negative (pink triangles and rhombuses) FBLs of different sizes with their 95% confidence intervals around the averages.

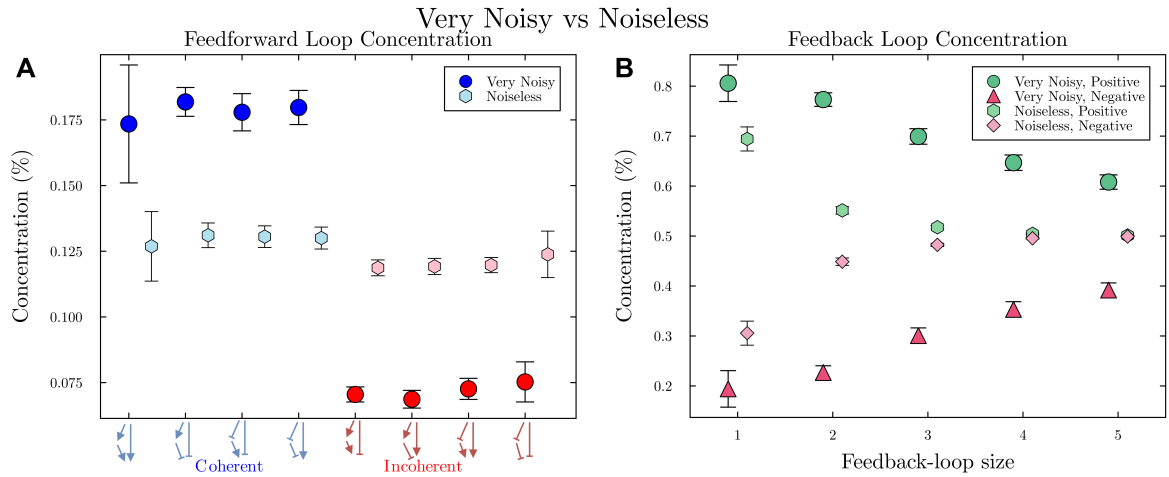

Figure S11: **Enrichment of network motifs as a result of noisy evolution, compared to noiseless evolution**. Comparison of 30 populations evolved with high noise variance ($\theta = 8$; darker hues) and no noise (lighter hues). (A) Average concentration per type of FFL. On the x-axis, we place the type of loop with a diagram. The error bars represent 95% confidence intervals around the averages. (B) Average concentration of positive (green circles and hexagons) and negative (pink triangles and rhombuses) FBLs of different sizes with their 95% confidence intervals around the averages.

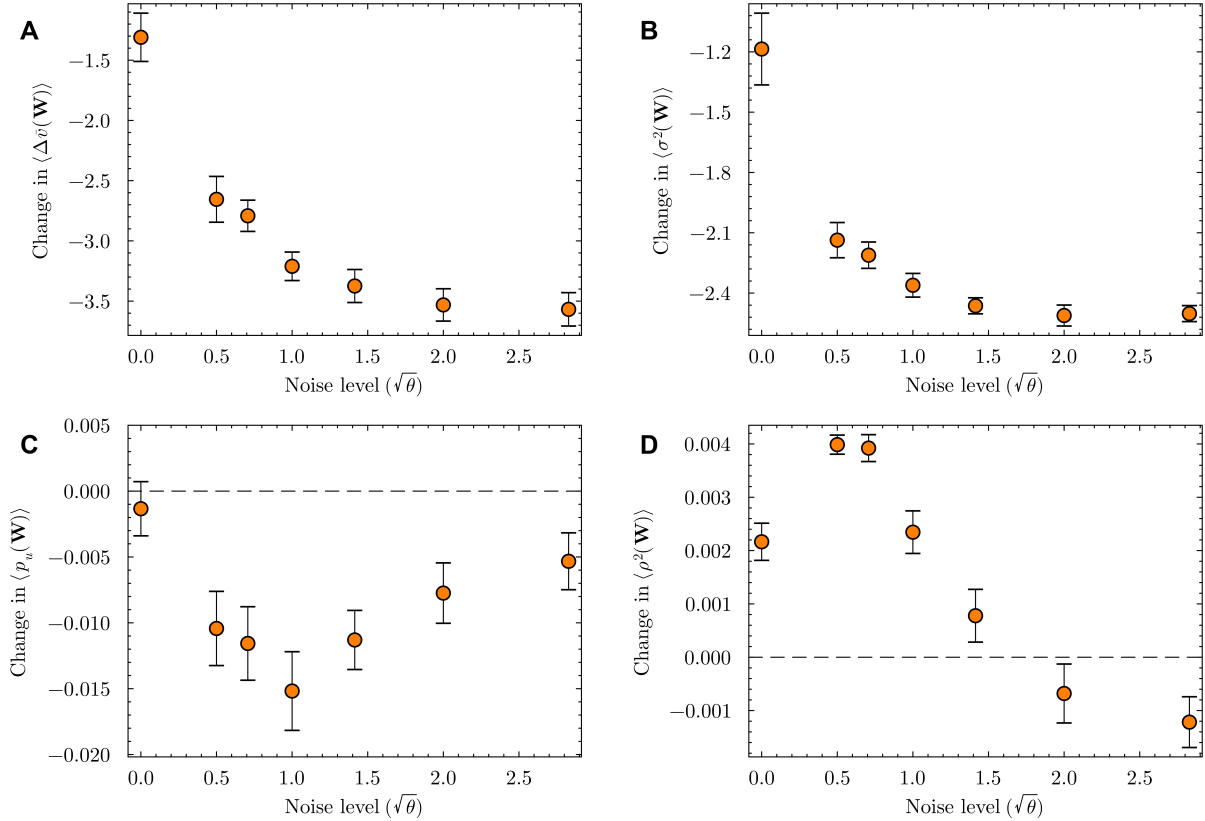

Figure S12: **Change in mutational robustness across evolved populations before and after evolution**. (A) Change in average *stable expression shift*, (B) average *stable expression variance*, (C) average *instability difference*, and (D) *instability variance* after the evolutionary processes for 30 populations across various noise levels (plotted as standard deviations in the x-axis). We fix the noise variance to $\theta_{\text{fixed}} = 1$ to compare all matrices against a single noise distribution. The $\theta = 0$ represents the noiseless scenario as it follows a Bernoulli distribution with no variance (always evaluates to 1). The error bars represent $95\%$ confidence intervals.

## S2.2  Initial Densities

We tested the robustness of our model against different initial densities $c$. In Wagner's [33] original publication, the model was consistent across values of $c$. Furthermore, the matrices appeared to acquire enhanced mutational robustness[1] with increased $c$. We argue that such a relationship is heavily influenced by the use of homogeneous initial populations (see Supplementary Material S2.3.1). Here, we test heterogeneous, non-optimal initial populations that show stable phenotypes under deterministic update dynamics across different initial densities.

## S2.2.1  High Density

In this configuration, we set $c = 0.9$, and find no qualitative differences with the results in the main text.

---

[1]Wagner used deterministic update dynamics to define mutational robustness.

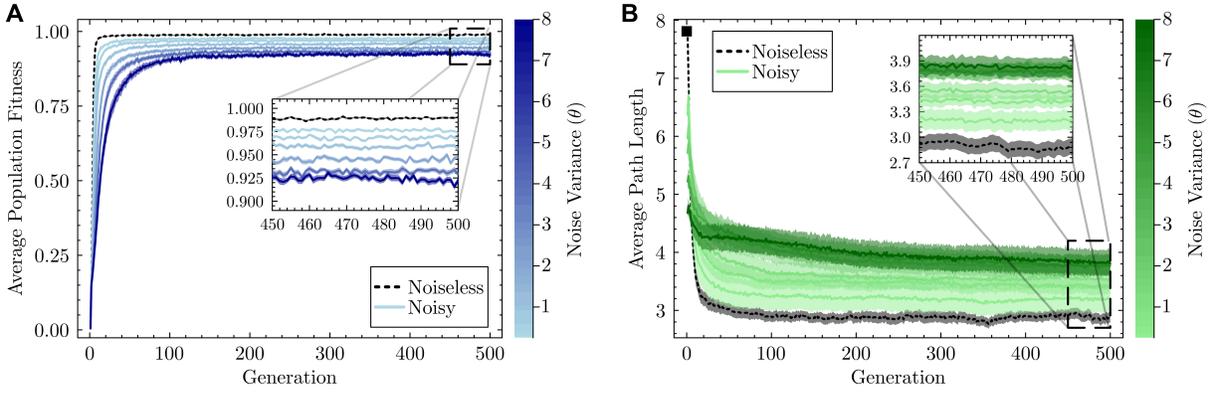

Figure S13: **(A) Evolution of the average population fitness and (B) average mean path length per generation**, across 30 independent replicates at different noise variances $\theta$. Shaded regions represent 95% confidence intervals around the mean in the zoomed-out panels and 68% in the zoomed-in panels. Lighter colors correspond to smaller noise variances. The noiseless case ($\theta = 0$) is shown with a dotted line.

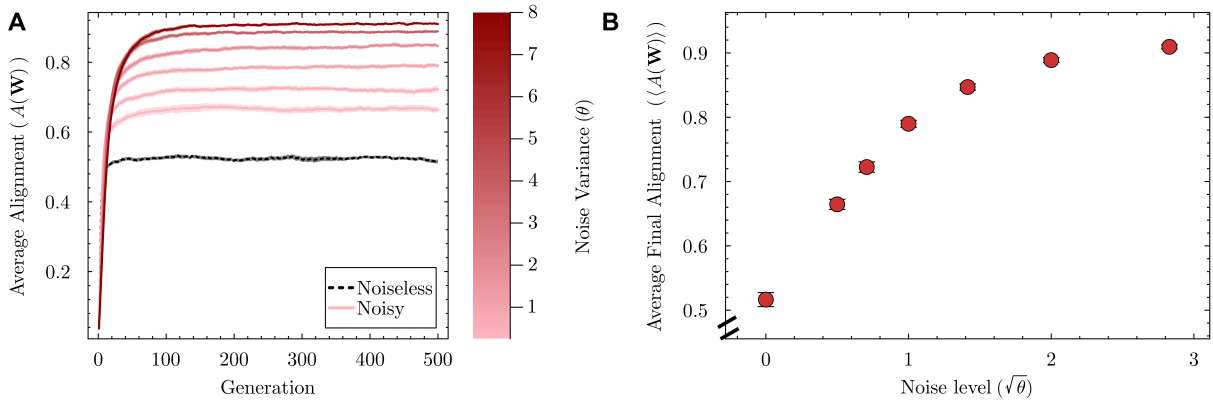

Figure S14: **Evolution of the average alignment scores**. (A) Average alignment score across 30 populations over generations. Each line represents the average alignment score for a population, and the color fillings represent 95% confidence intervals. Lighter colors represent smaller noise variances. The noiseless scenario is plotted with a dotted line. (B) Average alignment scores after evolution from 30 evolved populations across various noise levels $\sqrt{\theta}$. The 95% confidence intervals are plotted along the means.

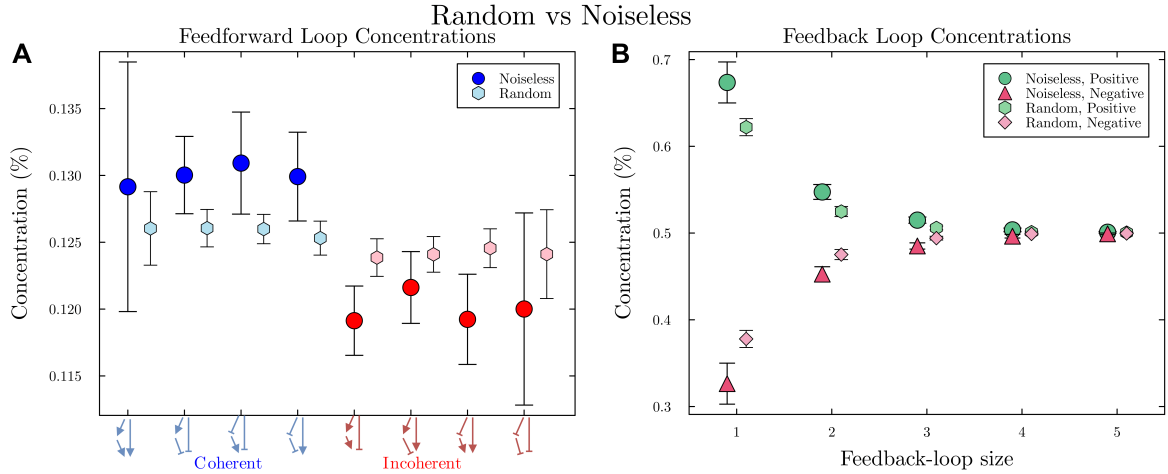

Figure S15: **Enrichment of network motifs as a result of noiseless evolution**. Comparison of 30 populations evolved without noise (darker hues) against non-evolved populations of stable matrices (lighter hues). (A) Average concentration per type of FFL. On the x-axis, we place the type of loop with a diagram. The error bars represent 95% confidence intervals around the averages. (B) Average concentration of positive (green circles and hexagons) and negative (pink triangles and rhombuses) FBLs of different sizes with their 95% confidence intervals around the averages.

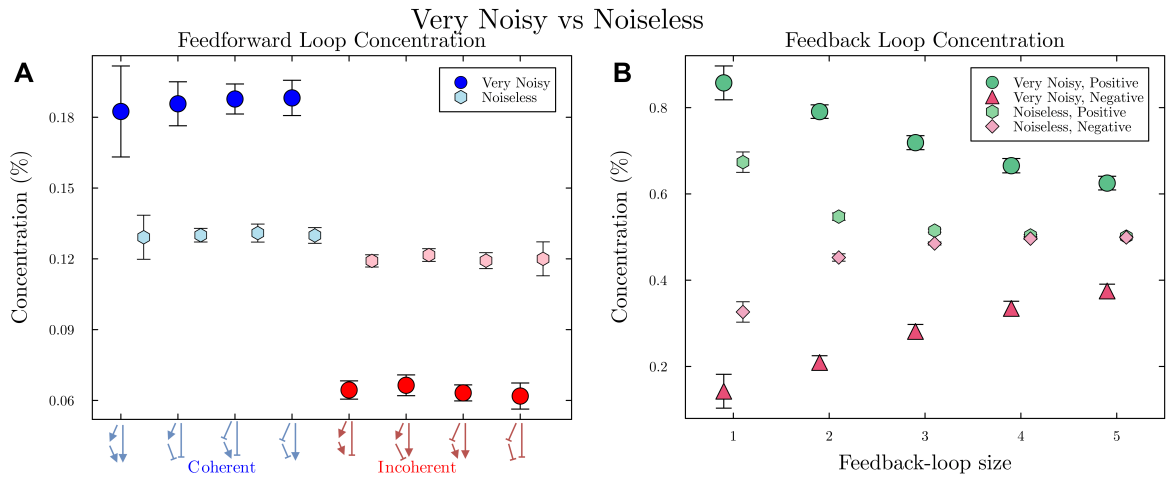

Figure S16: **Enrichment of network motifs as a result of noisy evolution, compared to noiseless evolution**. Comparison of 30 populations evolved with high noise variance ($\theta = 8$; darker hues) and no noise (lighter hues). (A) Average concentration per type of FFL. On the x-axis, we place the type of loop with a diagram. The error bars represent 95% confidence intervals around the averages. (B) Average concentration of positive (green circles and hexagons) and negative (pink triangles and rhombuses) FBLs of different sizes with their 95% confidence intervals around the averages.

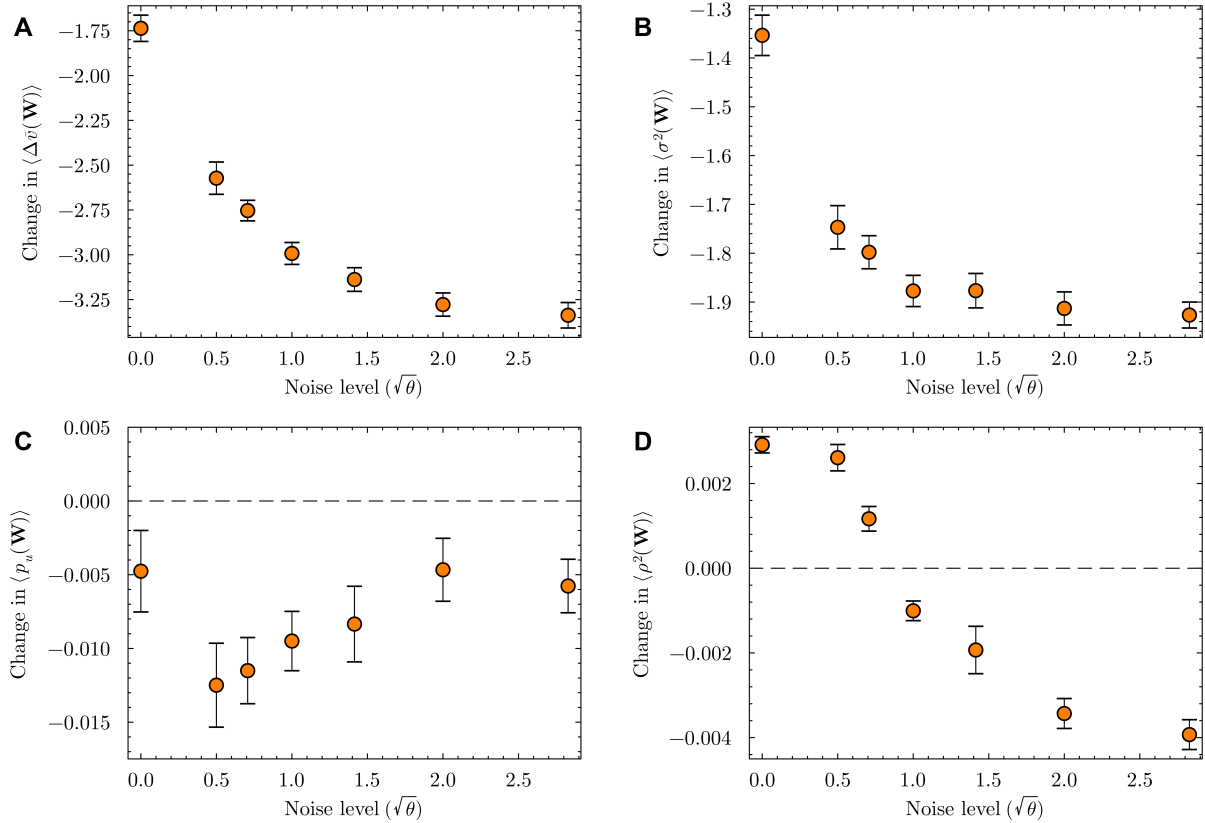

Figure S17: **Change in mutational robustness across evolved populations before and after evolution**. (A) Change in average *stable expression shift*, (B) average *stable expression variance*, (C) average *instability difference*, and (D) *instability variance* after the evolutionary processes for 30 populations across various noise levels (plotted as standard deviations in the x-axis). We fix the noise variance to $\theta_{\text{fixed}} = 1$ to compare all matrices against a single noise distribution. The $\theta = 0$ represents the noiseless scenario as it follows a Bernoulli distribution with no variance (always evaluates to 1). The error bars represent $95\%$ confidence intervals.

### S2.2.2 Medium Density

We find some differences for $c = 0.5$. In Figure S18, we find a much narrower average fitness distribution and a much wider average path length distribution across simulations per generation, which signals that these matrices show stable phenotypes closer to the target more consistently, but they do so by unpredictably taking one or two extra steps. In Figure S21, we see that the concentration of negative FBLs is decreased with high noise, but not as much as positive ones, which do not seem to be as enriched as with previous results. Negative feedback introduces instability in this model: if a gene is activated, it tends to suppress itself, and vice versa. Therefore, negative feedback might decrease the fitness of the GRNs, leading to their depletion over generations.

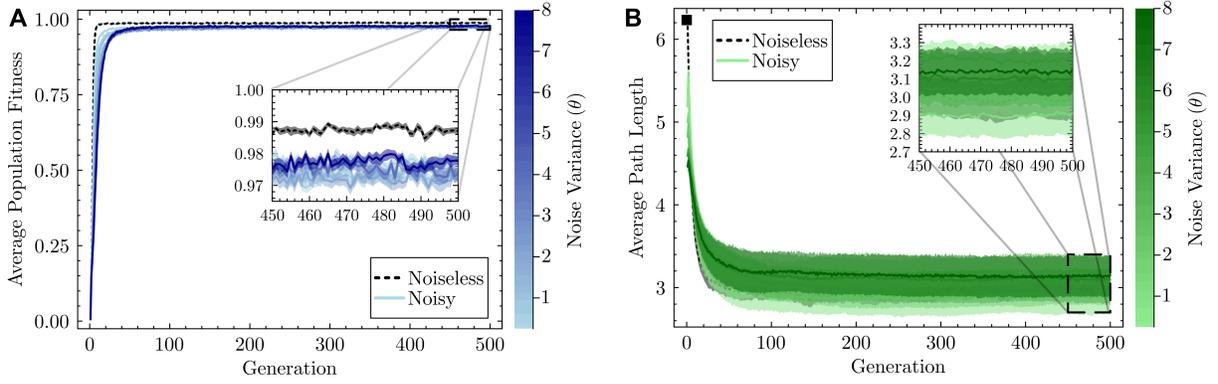

Figure S18: **(A) Evolution of the average population fitness and (B) average mean path length per generation**, across 30 independent replicates at different noise variances $\theta$. Shaded regions represent 95% confidence intervals around the mean in the zoomed-out panels and 68% in the zoomed-in panels. Lighter colors correspond to smaller noise variances. The noiseless case ($\theta = 0$) is shown with a dotted line.

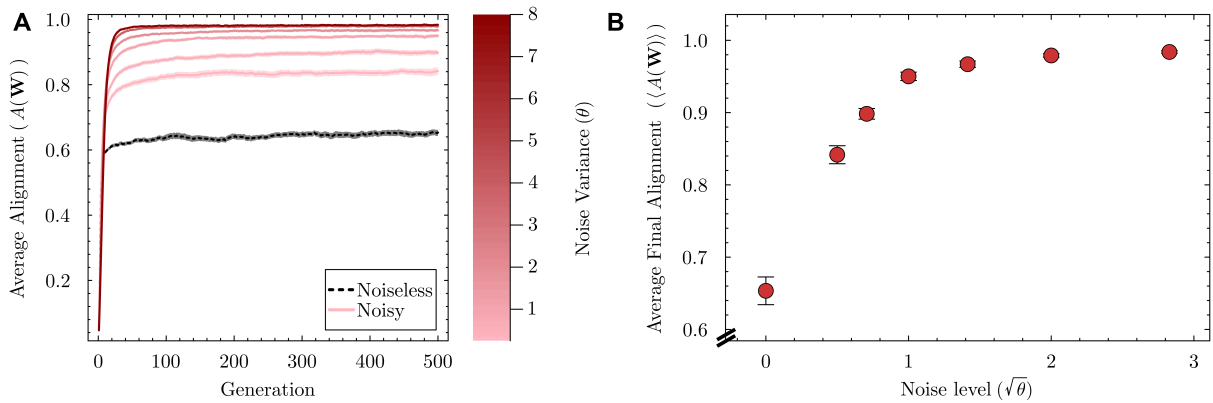

Figure S19: **Evolution of the average alignment scores**. (A) Average alignment score across 30 populations over generations. Each line represents the average alignment score for a population, and the color fillings represent 95% confidence intervals. Lighter colors represent smaller noise variances. The noiseless scenario is plotted with a dotted line. (B) Average alignment scores after evolution from 30 evolved populations across various noise levels $\sqrt{\theta}$. The 95% confidence intervals are plotted along the means.

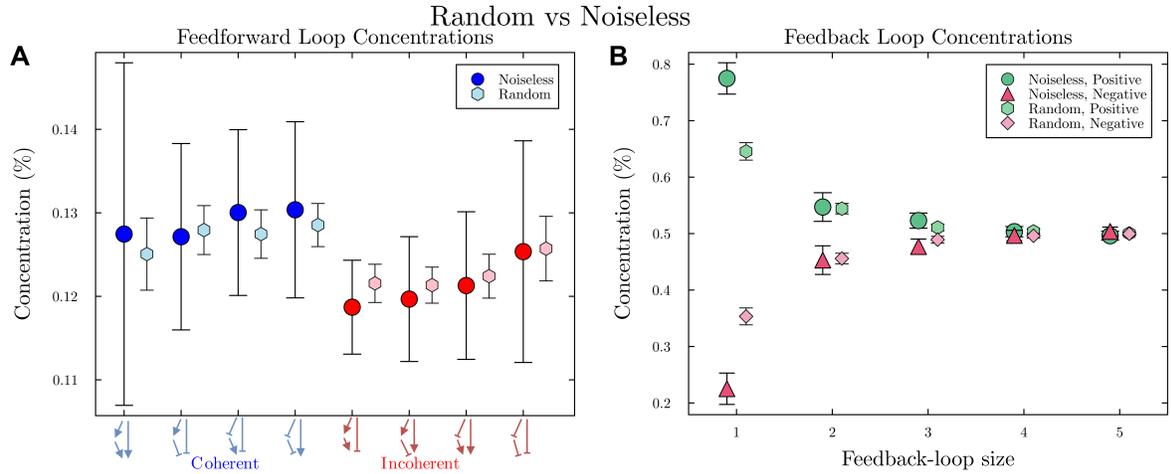

Figure S20: **Enrichment of network motifs as a result of noiseless evolution**. Comparison of 30 populations evolved without noise (darker hues) against non-evolved populations of stable matrices (lighter hues). (A) Average concentration per type of FFL. On the x-axis, we place the type of loop with a diagram. The error bars represent $95\%$ confidence intervals around the averages. (B) Average concentration of positive (green circles and hexagons) and negative (pink triangles and rhombuses) FBLs of different sizes with their $95\%$ confidence intervals around the averages.

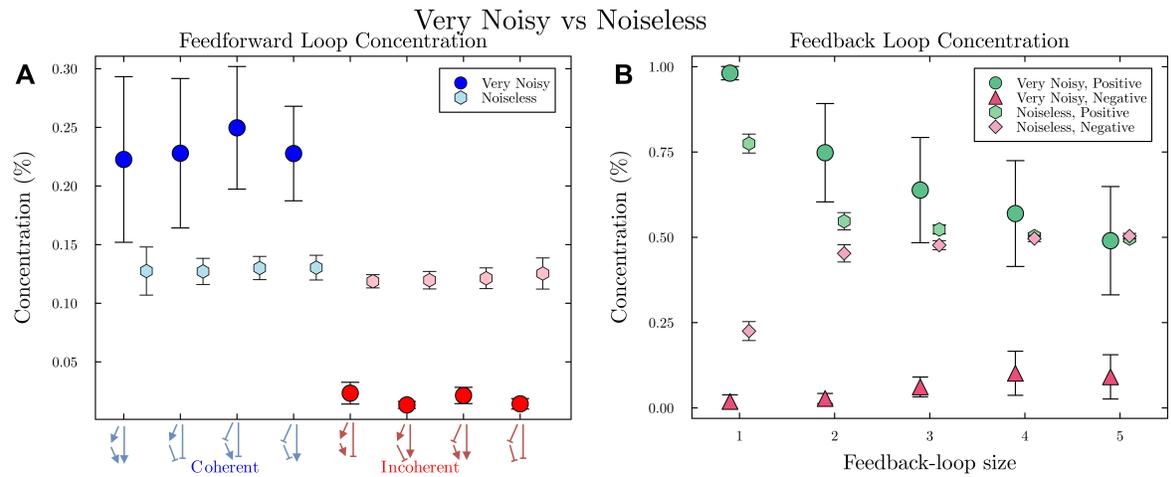

Figure S21: **Enrichment of network motifs as a result of noisy evolution, compared to noiseless evolution**. Comparison of 30 populations evolved with high noise variance ($\theta = 8$; darker hues) and no noise (lighter hues). (A) Average concentration per type of FFL. On the x-axis, we place the type of loop with a diagram. The error bars represent $95\%$ confidence intervals around the averages. (B) Average concentration of positive (green circles and hexagons) and negative (pink triangles and rhombuses) FBLs of different sizes with their $95\%$ confidence intervals around the averages.

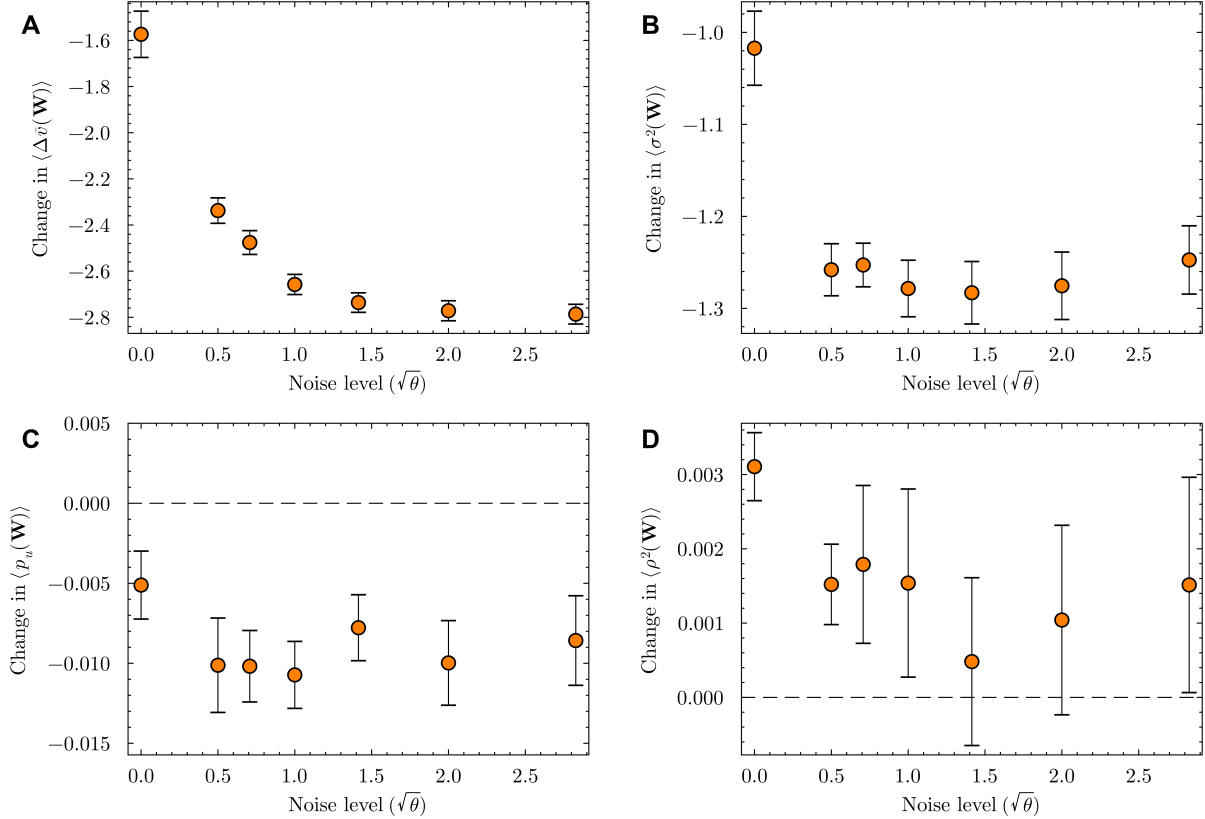

Figure S22: **Change in mutational robustness across evolved populations before and after evolution**. (A) Change in average *stable expression shift*, (B) average *stable expression variance*, (C) average *instability difference*, and (D) *instability variance* after the evolutionary processes for 30 populations across various noise levels (plotted as standard deviations in the x-axis). We fix the noise variance to $\theta_{\text{fixed}} = 1$ to compare all matrices against a single noise distribution. The $\theta = 0$ represents the noiseless scenario as it follows a Bernoulli distribution with no variance (always evaluates to 1). The error bars represent $95\%$ confidence intervals.

### S2.2.3 Low Density

In this region of the parameter space, we set $c = 0.1$, and we see even greater distortions. For instance, a row will be empty with a probability of $(1 - 0.1)^{10} = 0.349$. In our implementation, we assign $\text{sign}(0) = 0$, which means that a GRN with empty rows will never be optimal. Nevertheless, we find in Figure S23 A that the average fitness is very close to the maximum 1.0, possibly because non-empty rows rapidly fixate in the population. On Figure S23, the average path lengths have a wide spread and overlap across noise levels. Furthermore, we also find a much tighter adherence to a maximal alignment $\langle A(\boldsymbol{W}) \rangle = 1$ in a Figure S24, possibly because each row in a population has around one non-zero entry, which forces them into a smaller margin of alignment that sustains function. In Figure S27, we also find that matrices no longer decrease the spread of their phenotype distributions (B and D) as a function of the noise level $\sqrt{\theta}$, possibly because a single mutation disrupts the (likely) only edge in the row of a matrix. Nevertheless, the average stable (A) phenotype and the probability of instability (B) can increase mutational robustness.

We make a word of caution in observing Figures S25 and S26. In our implementation, we set the concentration to 0 if no network motif of a given type (classification and size) was found. For example, a network with no FBLs of size 1 would be recorded as having a concentration of $0\%$ in positive FBLs and $0\%$ in negative FBLs. This decision increases the observed noise in the error bars of both panels. We, however, point out the near-zero concentration of incoherent feedforward and negative FBLs in Figure S26.

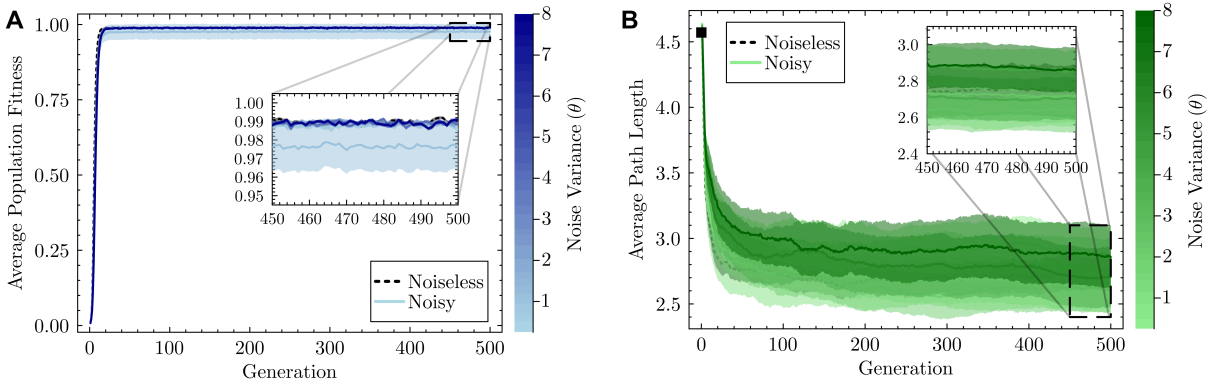

Figure S23: **(A) Evolution of the average population fitness and (B) average mean path length per generation**, across 30 independent replicates at different noise variances $\theta$. Shaded regions represent 95% confidence intervals around the mean in the zoomed-out panels and 68% in the zoomed-in panels. Lighter colors correspond to smaller noise variances. The noiseless case ($\theta = 0$) is shown with a dotted line.

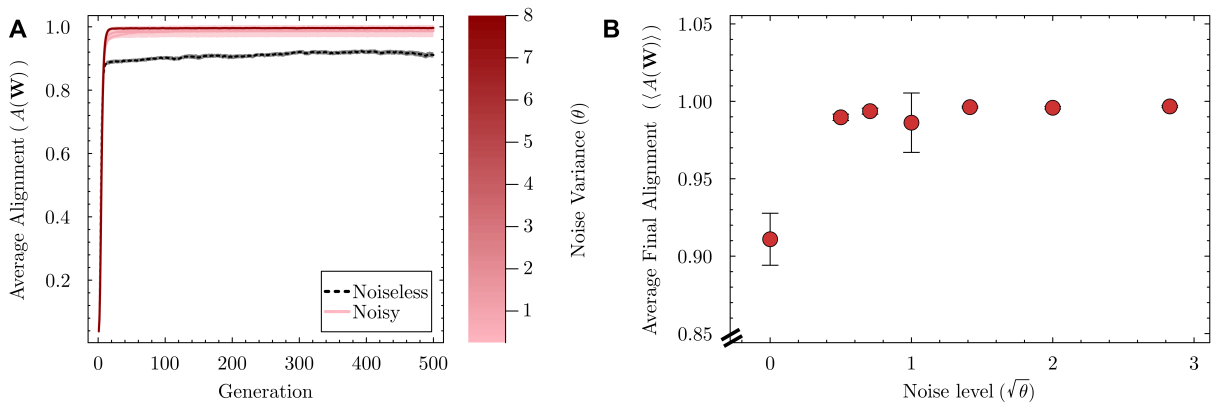

Figure S24: **Evolution of the average alignment scores**. (A) Average alignment score across 30 populations over generations. Each line represents the average alignment score for a population, and the color fillings represent 95% confidence intervals. Lighter colors represent smaller noise variances. The noiseless scenario is plotted with a dotted line. (B) Average alignment scores after evolution from 30 evolved populations across various noise levels $\sqrt{\theta}$. The 95% confidence intervals are plotted along the means.

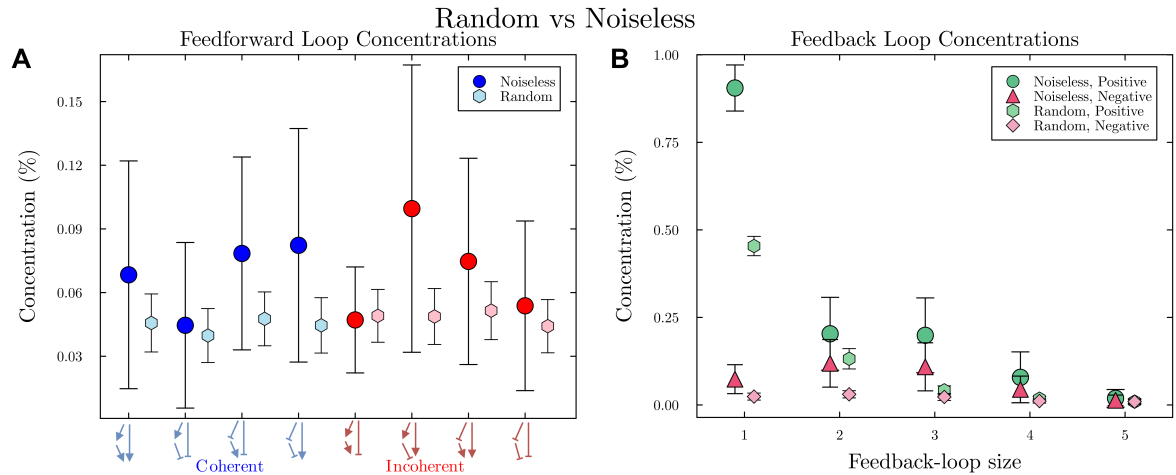

Figure S25: **Enrichment of network motifs as a result of noiseless evolution**. Comparison of 30 populations evolved without noise (darker hues) against non-evolved populations of stable matrices (lighter hues). (A) Average concentration per type of FFL. On the x-axis, we place the type of loop with a diagram. The error bars represent $95\%$ confidence intervals around the averages. (B) Average concentration of positive (green circles and hexagons) and negative (pink triangles and rhombuses) FBLs of different sizes with their $95\%$ confidence intervals around the averages.

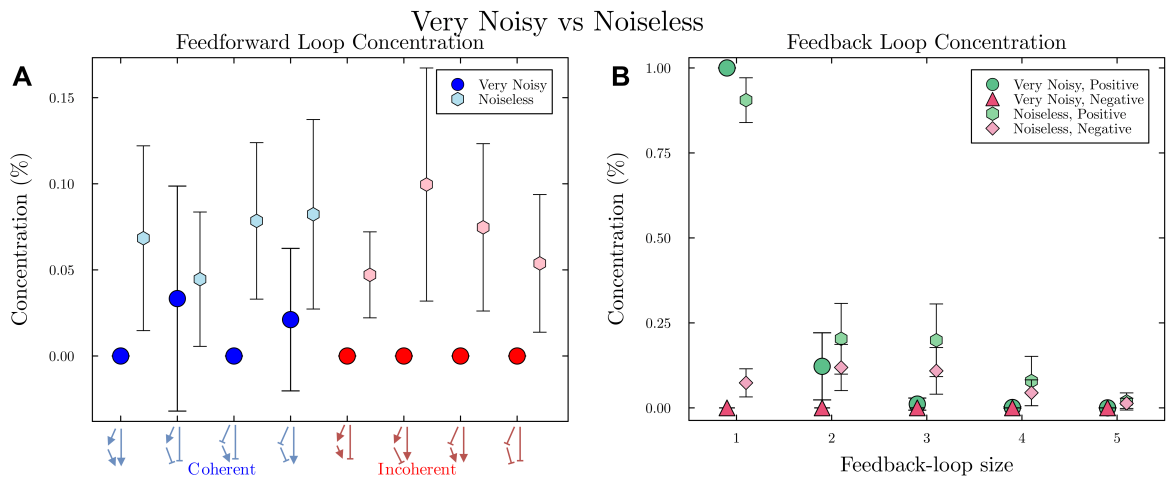

Figure S26: **Enrichment of network motifs as a result of noisy evolution, compared to noiseless evolution**. Comparison of 30 populations evolved with high noise variance ($\theta = 8$; darker hues) and no noise (lighter hues). (A) Average concentration per type of FFL. On the x-axis, we place the type of loop with a diagram. The error bars represent $95\%$ confidence intervals around the averages. (B) Average concentration of positive (green circles and hexagons) and negative (pink triangles and rhombuses) FBLs of different sizes with their $95\%$ confidence intervals around the averages.

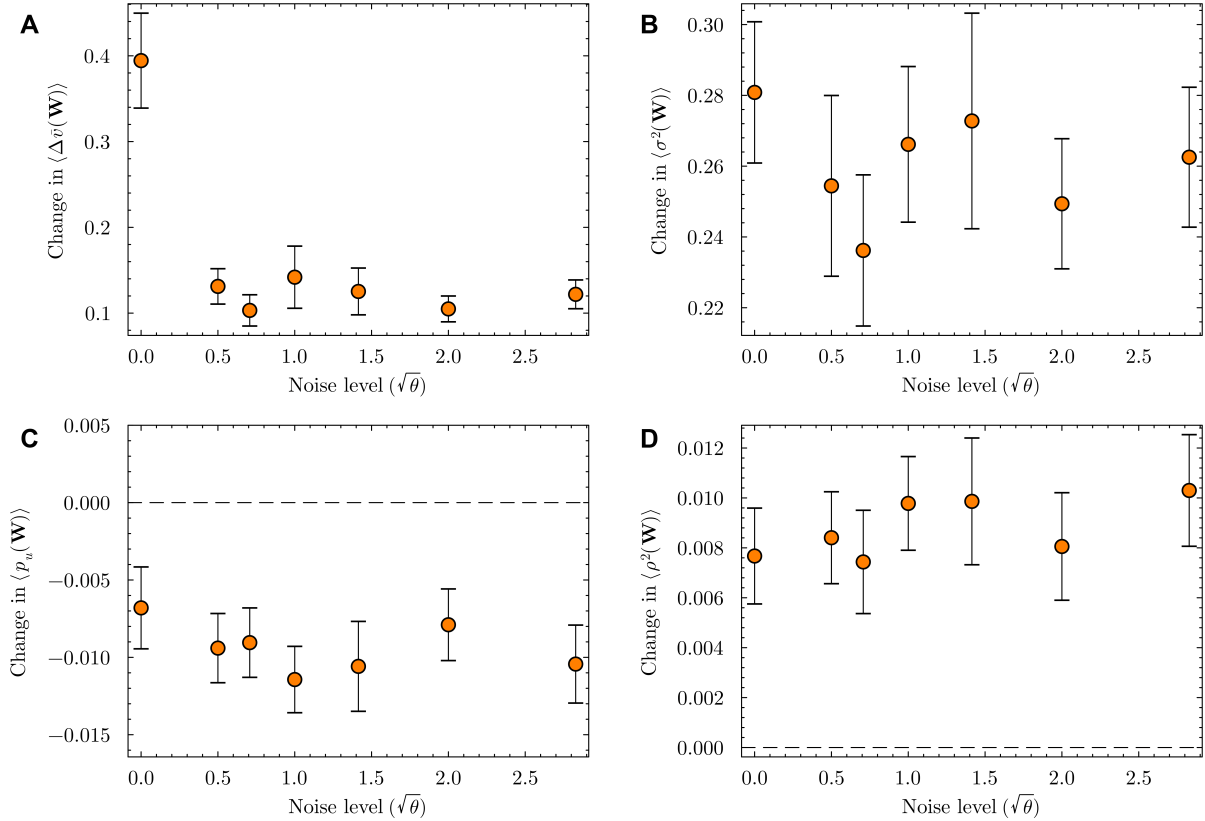

Figure S27: **Change in mutational robustness across evolved populations before and after evolution**. (A) Change in average *stable expression shift*, (B) average *stable expression variance*, (C) average *instability difference*, and (D) *instability variance* after the evolutionary processes for 30 populations across various noise levels (plotted as standard deviations in the x-axis). We fix the noise variance to $\theta_{\text{fixed}} = 1$ to compare all matrices against a single noise distribution. The $\theta = 0$ represents the noiseless scenario as it follows a Bernoulli distribution with no variance (always evaluates to 1). The error bars represent $95\%$ confidence intervals.

### S2.2.4 Very Low Density

In this configuration, we set $c = 0.05$, and we find even more extreme results than in Supplementary Material S2.2.3. In Figure S28, the fitness values are much closer to the noiseless scenario, and the spread in the average path length is increased. Furthermore, the maximal alignment in S29 is reached more often. Alignment is enhanced even in the noiseless scenario. The mutational robustness metrics in Figure S32 are also very similar to the case $c = 0.1$, possibly because they also acquire about one non-negative weight in each row. Following the word of caution from Supplementary Material S2.2.3, we deem Figures S30 and S31 uninformative, without any notable takeaways other than a lack of most network motifs.

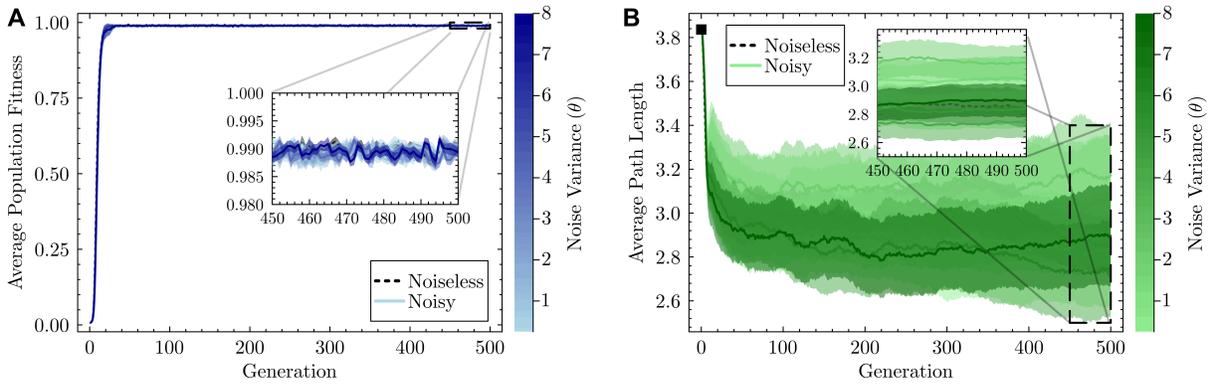

Figure S28: **(A) Evolution of the average population fitness and (B) average mean path length per generation**, across 30 independent replicates at different noise variances $\theta$. Shaded regions represent 95% confidence intervals around the mean in the zoomed-out panels and 68% in the zoomed-in panels. Lighter colors correspond to smaller noise variances. The noiseless case ($\theta = 0$) is shown with a dotted line.

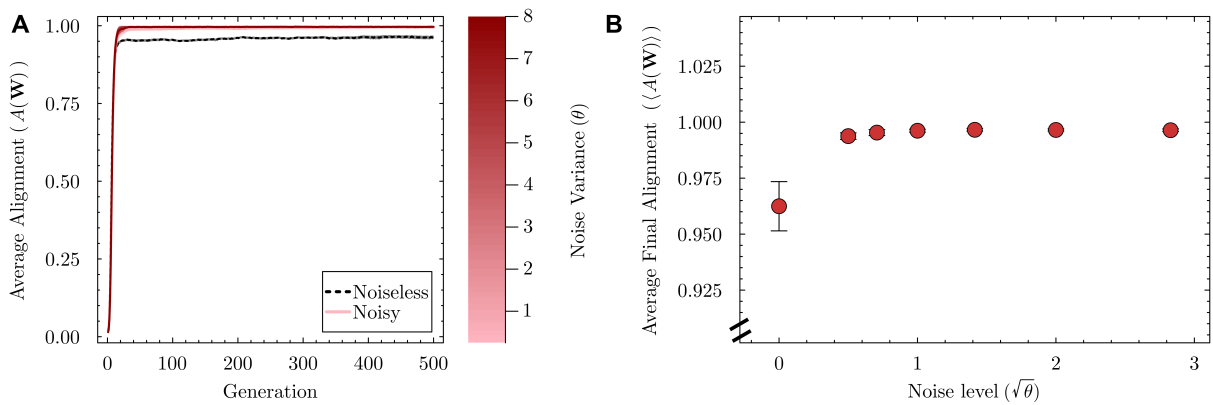

Figure S29: **Evolution of the average alignment scores**. (A) Average alignment score across 30 populations over generations. Each line represents the average alignment score for a population, and the color fillings represent 95% confidence intervals. Lighter colors represent smaller noise variances. The noiseless scenario is plotted with a dotted line. (B) Average alignment scores after evolution from 30 evolved populations across various noise levels $\sqrt{\theta}$. The 95% confidence intervals are plotted along the means.

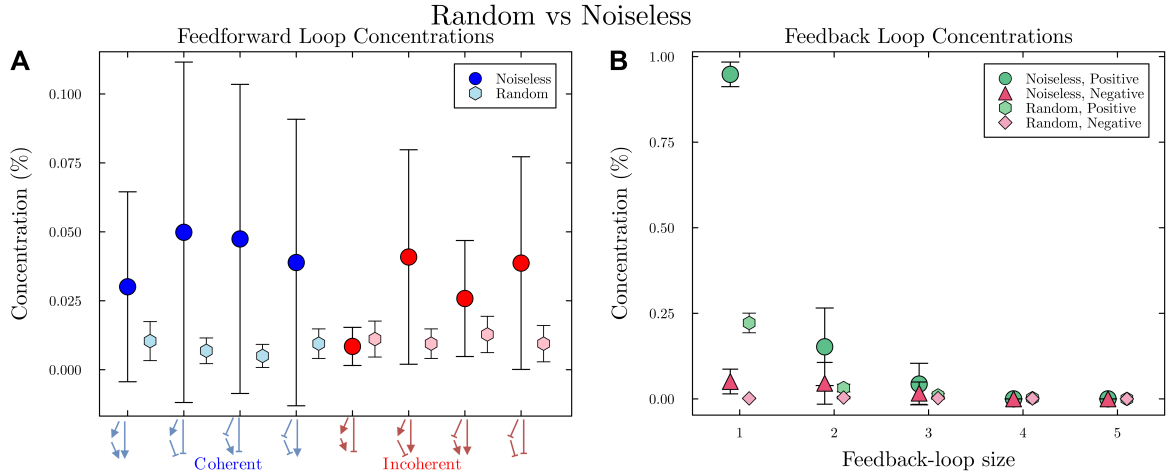

Figure S30: **Enrichment of network motifs as a result of noiseless evolution**. Comparison of 30 populations evolved without noise (darker hues) against non-evolved populations of stable matrices (lighter hues). (A) Average concentration per type of FFL. On the x-axis, we place the type of loop with a diagram. The error bars represent $95\%$ confidence intervals around the averages. (B) Average concentration of positive (green circles and hexagons) and negative (pink triangles and rhombuses) FBLs of different sizes with their $95\%$ confidence intervals around the averages.

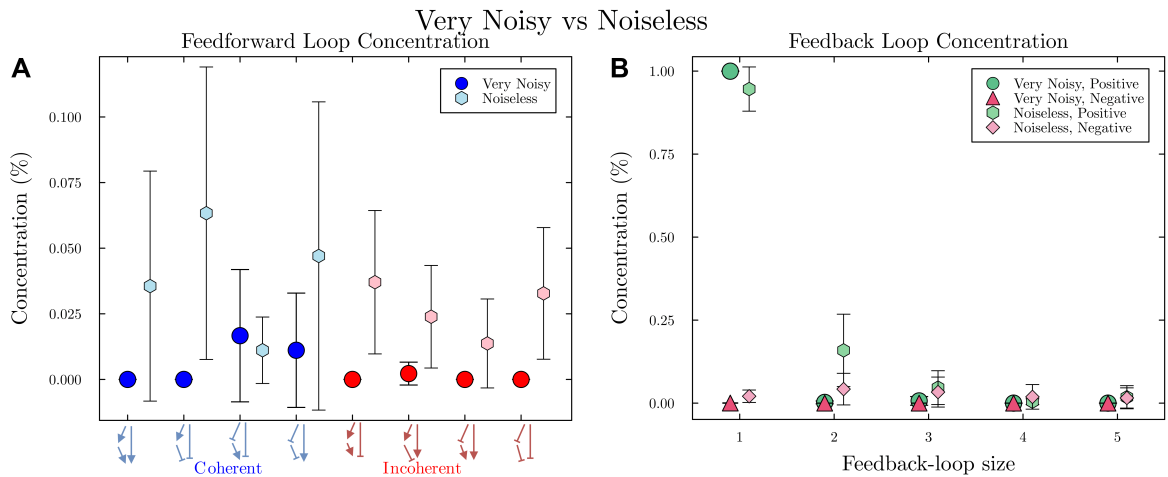

Figure S31: **Enrichment of network motifs as a result of noisy evolution, compared to noiseless evolution**. Comparison of 30 populations evolved with high noise variance ($\theta = 8$; darker hues) and no noise (lighter hues). (A) Average concentration per type of FFL. On the x-axis, we place the type of loop with a diagram. The error bars represent $95\%$ confidence intervals around the averages. (B) Average concentration of positive (green circles and hexagons) and negative (pink triangles and rhombuses) FBLs of different sizes with their $95\%$ confidence intervals around the averages.

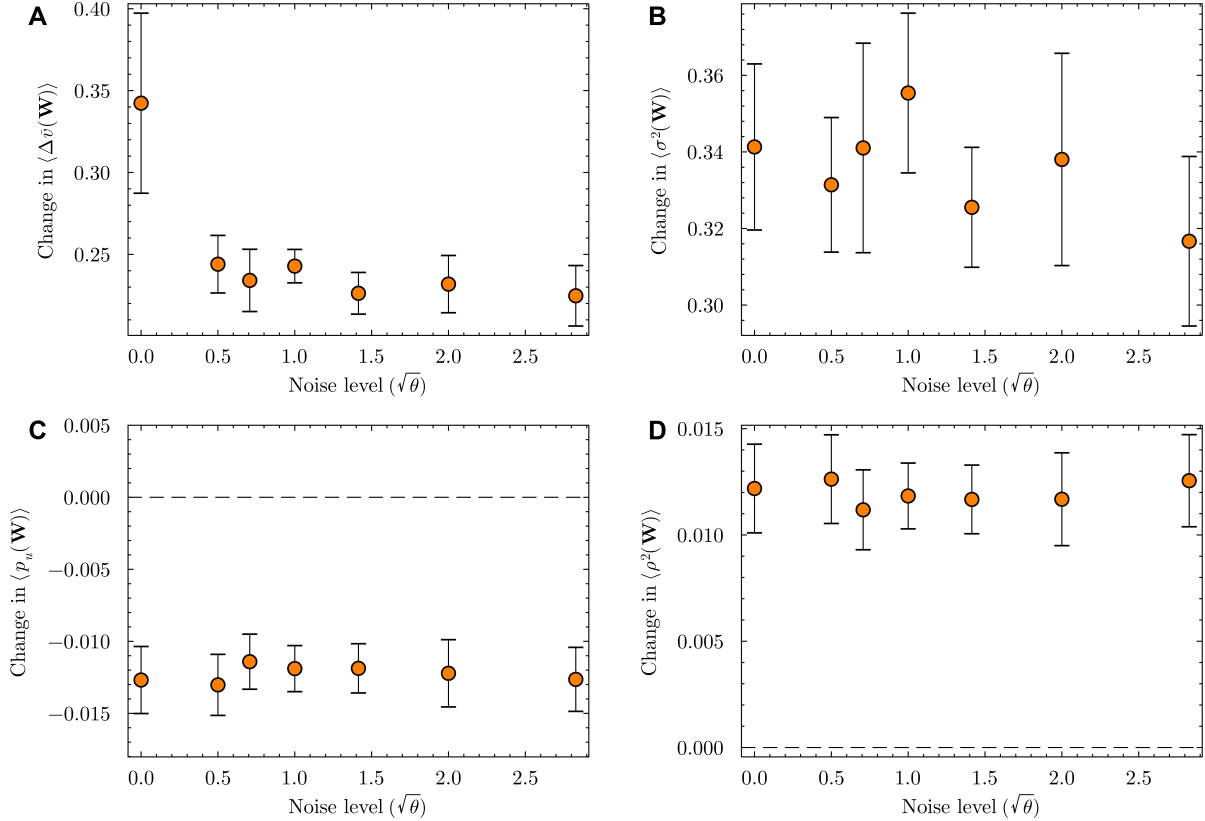

Figure S32: **Change in mutational robustness across evolved populations before and after evolution**. (A) Change in average *stable expression shift*, (B) average *stable expression variance*, (C) average *instability difference*, and (D) *instability variance* after the evolutionary processes for 30 populations across various noise levels (plotted as standard deviations in the x-axis). We fix the noise variance to $\theta_{\text{fixed}} = 1$ to compare all matrices against a single noise distribution. The $\theta = 0$ represents the noiseless scenario as it follows a Bernoulli distribution with no variance (always evaluates to 1). The error bars represent $95\%$ confidence intervals.

## S2.3 Initial Populations

We tested the robustness of our results across different initial populations. We initialize them in four different settings, in addition to the "stable" described in the main text.

### S2.3.1 Optimal Clones

This is the original configuration used by Wagner [33], which is referred to as a "founder" population. In it, the author assumes a homogeneous initial population subject to stabilizing selection.

We simplify the generative process, which we deem equivalent to Wagner's [33]. We first pick a randomly generated initial gene state $P_0$, where each entry can be $\pm 1$ with probability $1/2$, and a matrix $W$, where each element $W_{ij}$ has a probability $c$ (fixed to $c = 1$ in these simulations, but see Supplementary Material S2.2 for a check on robustness) to be non-zero. If an entry $W_{ij}$ is nonzero, we sample from the normal distribution $W_{ij} \sim \mathcal{N}(\mu_r, \sigma_r^2)$ independently of any other weights. Then, we deterministically develop $W$ into showing a stable phenotype $P$, which we declare as the target phenotype $P_{\text{opt}} = P$. If it shows an unstable phenotype, we discard this matrix and generate a new one. Then, we create `pop_size` copies of this matrix to fill our population. On the other hand, Wagner [33] randomly generated $P_0$ and $P_{\text{opt}}$, and tried finding one $W$ that showed a stable phenotype $P = P_{\text{opt}}$.

We argue that a population of clones lacks a mechanism to alter the underlying topology of GRNs throughout generations. Notice that a mutation will only re-sample the non-zero weights, possibly changing their sign but not the existence of an edge. Furthermore, recombining two matrices whose non-zero elements are in the same positions produces a matrix with non-zero elements in the same positions, making no changes to the network topology. Nevertheless, we find consistency in our results, with minor differences.

Generally, the results replicate across Figures S33, S34, S35, S36, and S37. Notice that, since initial populations are optimal under deterministic development, Figure S33 A shows an initial Fitness of 1.0 in the noiseless scenario, but a lower one in every other. We find differences in Figure S37, where the length of the error bars is increased in the change in $\langle \Delta \bar{v}(\boldsymbol{W}) \rangle$ and $\langle \sigma^2(\boldsymbol{W}) \rangle$, and the functional relationship with $\sqrt{\theta}$ is less clear. Finally, the change in $\langle \rho^2(\boldsymbol{W}) \rangle$ attains a maximum (most positive) with small values of $\sqrt{\theta}$.

In our statistical tests, we find that noiseless evolution results in a lower average alignment than noisy evolution ($t = 32.05, P = 8.02 \times 10^{-69}$). Additionally, coherent FFLs are enriched in evolved populations relative to non-evolved ones ($t = 5.66, P = 3.87 \times 10^{-6}$), and this enrichment is even stronger in populations evolved under high noise compared to those evolved without noise ($t = 29.57, P = 6.75 \times 10^{-24}$).

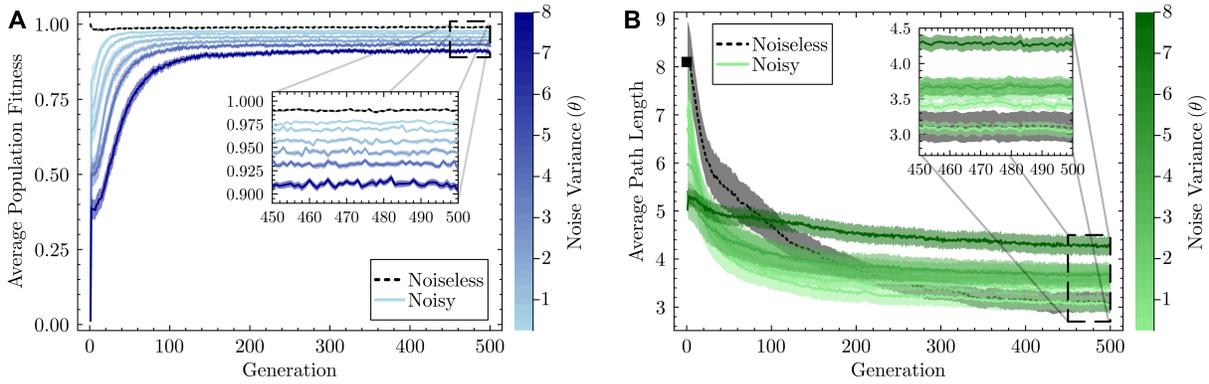

Figure S33: **(A) Evolution of the average population fitness and (B) average mean path length per generation**, across 30 independent replicates at different noise variances $\theta$. Shaded regions represent 95% confidence intervals around the mean in the zoomed-out panels and 68% in the zoomed-in panels. Lighter colors correspond to smaller noise variances. The noiseless case ($\theta = 0$) is shown with a dotted line.

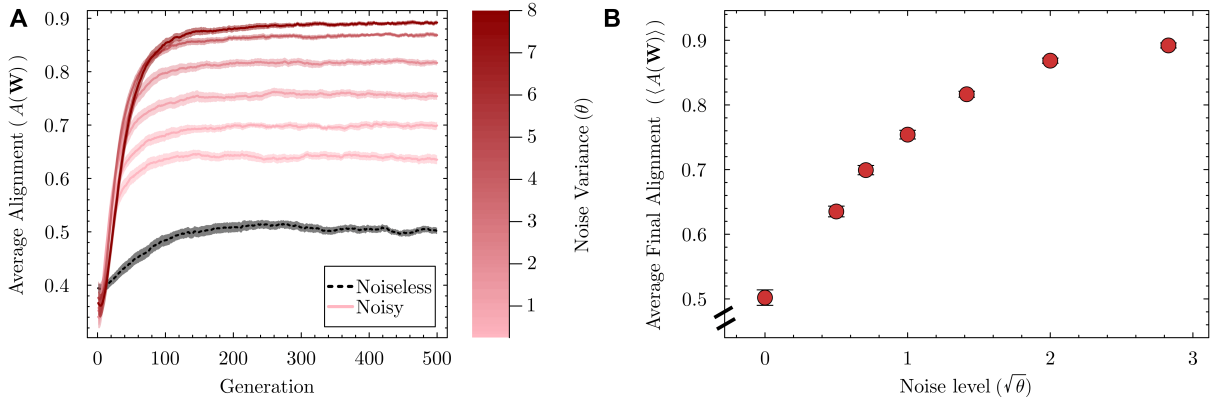

Figure S34: **Evolution of the average alignment scores**. (A) Average alignment score across 30 populations over generations. Each line represents the average alignment score for a population, and the color fillings represent 95% confidence intervals. Lighter colors represent smaller noise variances. The noiseless scenario is plotted with a dotted line. (B) Average alignment scores after evolution from 30 evolved populations across various noise levels $\sqrt{\theta}$. The 95% confidence intervals are plotted along the means.

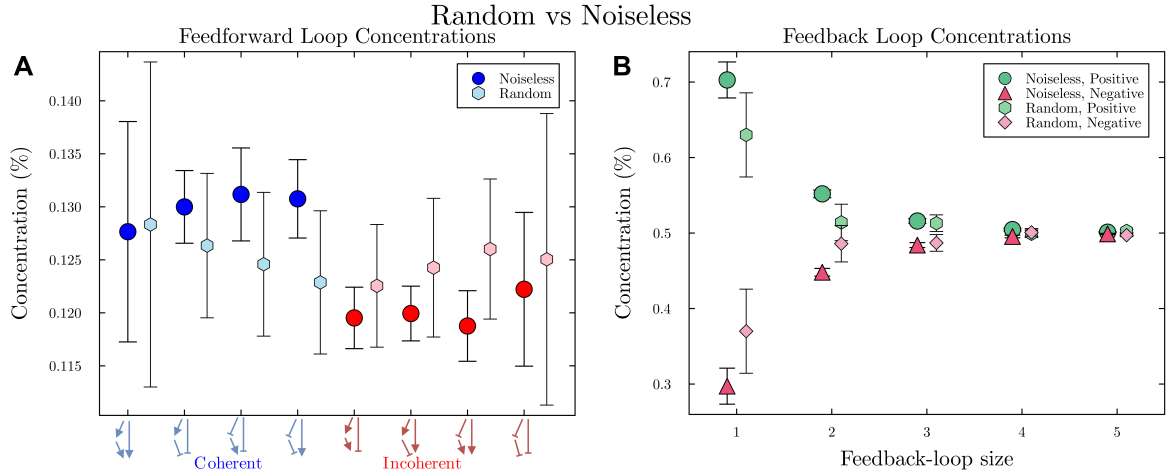

Figure S35: **Enrichment of network motifs as a result of noiseless evolution**. Comparison of 30 populations evolved without noise (darker hues) against non-evolved populations of stable matrices (lighter hues). (A) Average concentration per type of FFL. On the x-axis, we place the type of loop with a diagram. The error bars represent $95\%$ confidence intervals around the averages. (B) Average concentration of positive (green circles and hexagons) and negative (pink triangles and rhombuses) FBLs of different sizes with their $95\%$ confidence intervals around the averages.

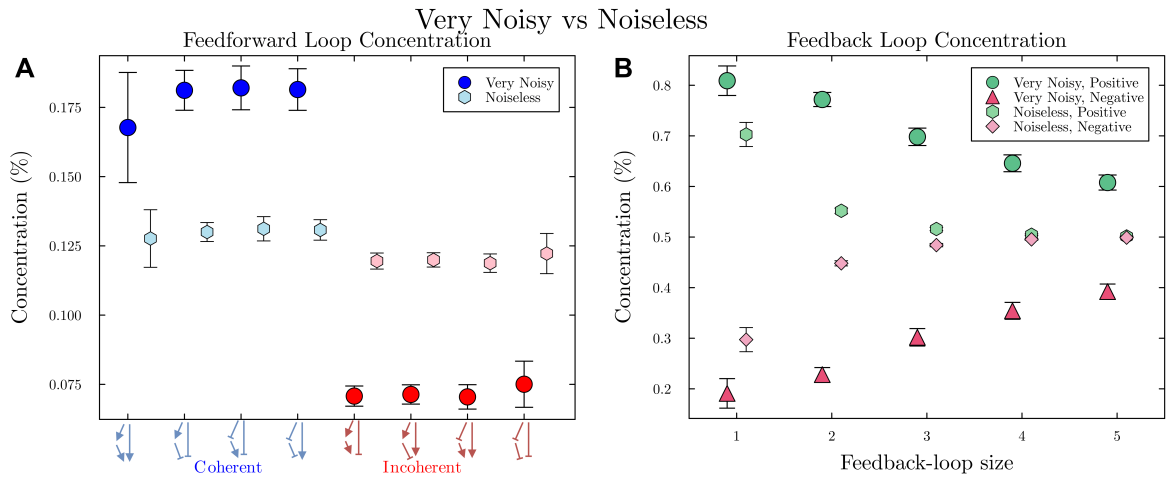

Figure S36: **Enrichment of network motifs as a result of noisy evolution, compared to noiseless evolution**. Comparison of 30 populations evolved with high noise variance ($\theta = 8$; darker hues) and no noise (lighter hues). (A) Average concentration per type of FFL. On the x-axis, we place the type of loop with a diagram. The error bars represent $95\%$ confidence intervals around the averages. (B) Average concentration of positive (green circles and hexagons) and negative (pink triangles and rhombuses) FBLs of different sizes with their $95\%$ confidence intervals around the averages.

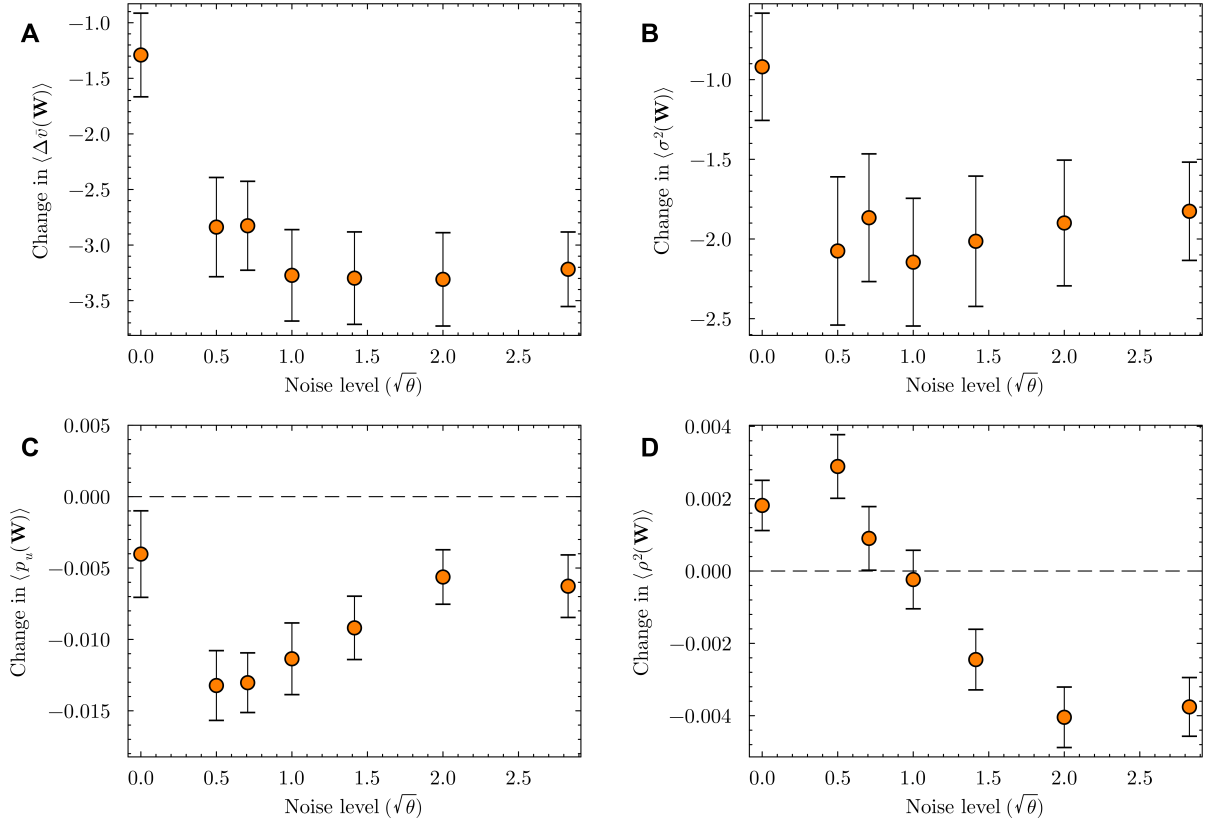

Figure S37: **Change in mutational robustness across evolved populations before and after evolution**. (A) Change in average *stable expression shift*, (B) average *stable expression variance*, (C) average *instability difference*, and (D) *instability variance* after the evolutionary processes for 30 populations across various noise levels (plotted as standard deviations in the x-axis). We fix the noise variance to $\theta_{\text{fixed}} = 1$ to compare all matrices against a single noise distribution. The $\theta = 0$ represents the noiseless scenario as it follows a Bernoulli distribution with no variance (always evaluates to 1). The error bars represent $95\%$ confidence intervals.

### S2.3.2 Nonoptimal Clones

In this configuration, we assume a homogeneous initial population subject to directional selection. To initialize this population, we generate a random starting gene expression $P_0$ and target phenotype $P_{\text{opt}}$, in which each entry has a probability $1/2$ to be $\pm 1$. We then pick a random matrix $W$, where each element $W_{ij}$ has a probability $c$ (fixed to $c = 1$ in these simulations, but see Supplementary Material S2.2 for a check on robustness) to be non-zero. If an entry $W_{ij}$ is nonzero, we sample from the normal distribution $W_{ij} \sim \mathcal{N}(\mu_r, \sigma_r^2)$ independently of any other weights. Then, we deterministically develop $W$ into showing a stable phenotype $P$. In this case, we do not assign the stable phenotype as the target phenotype. Instead, the deterministic phenotype expression serves to ensure stability. Finally, we generate `pop_size` copies of this stable genotype.

Interestingly, in Figure S38, we find that noiseless evolution proceeds in "evolutionary jumps" by having bursts in which the average population fitness quickly increases, whereas noisy evolution stays relatively smooth. In all other figures, the results stay very similar to Supplementary Material S2.3.1.

In our statistical tests, we find that noiseless evolution results in a lower average alignment than noisy evolution ($t = 29.48, P = 5.98 \times 10^{-64}$). Additionally, coherent FFLs are enriched in evolved populations relative to non-evolved ones ($t = 3.55, P = .0019$), and this enrichment is even stronger in populations evolved under high noise compared to those evolved without noise ($t = 32.28, P = 1.05 \times 10^{-25}$).

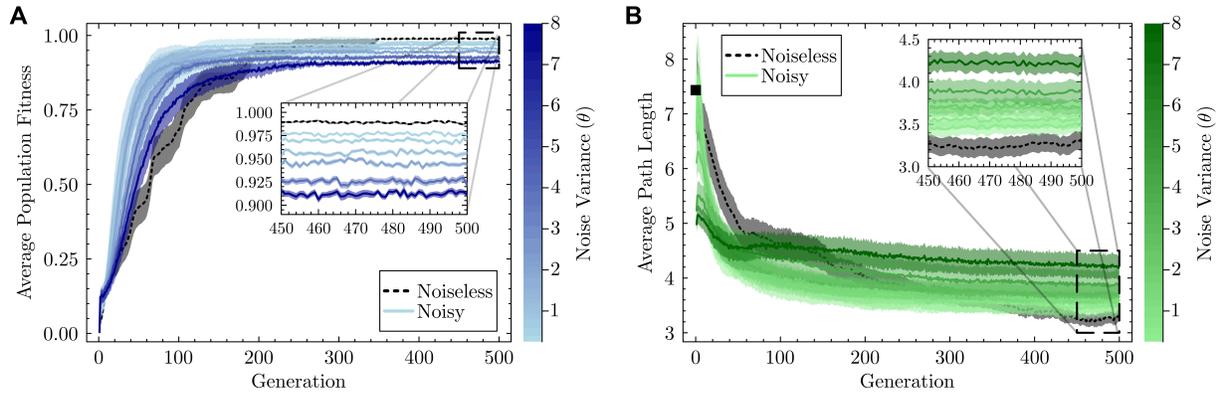

Figure S38: **(A) Evolution of the average population fitness and (B) average mean path length per generation**, across 30 independent replicates at different noise variances $\theta$. Shaded regions represent 95% confidence intervals around the mean in the zoomed-out panels and 68% in the zoomed-in panels. Lighter colors correspond to smaller noise variances. The noiseless case ($\theta = 0$) is shown with a dotted line.

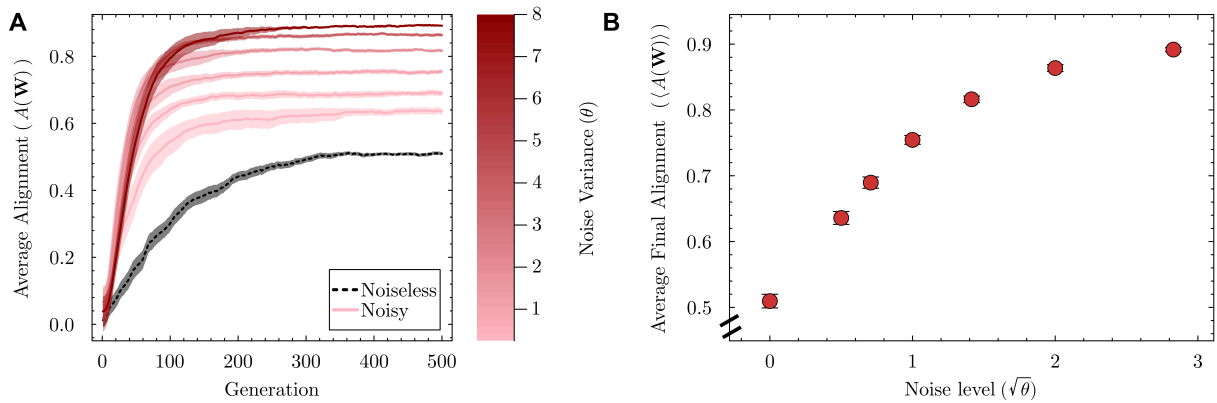

Figure S39: **Evolution of the average alignment scores**. (A) Average alignment score across 30 populations over generations. Each line represents the average alignment score for a population, and the color fillings represent 95% confidence intervals. Lighter colors represent smaller noise variances. The noiseless scenario is plotted with a dotted line. (B) Average alignment scores after evolution from 30 evolved populations across various noise levels $\sqrt{\theta}$. The 95% confidence intervals are plotted along the means.

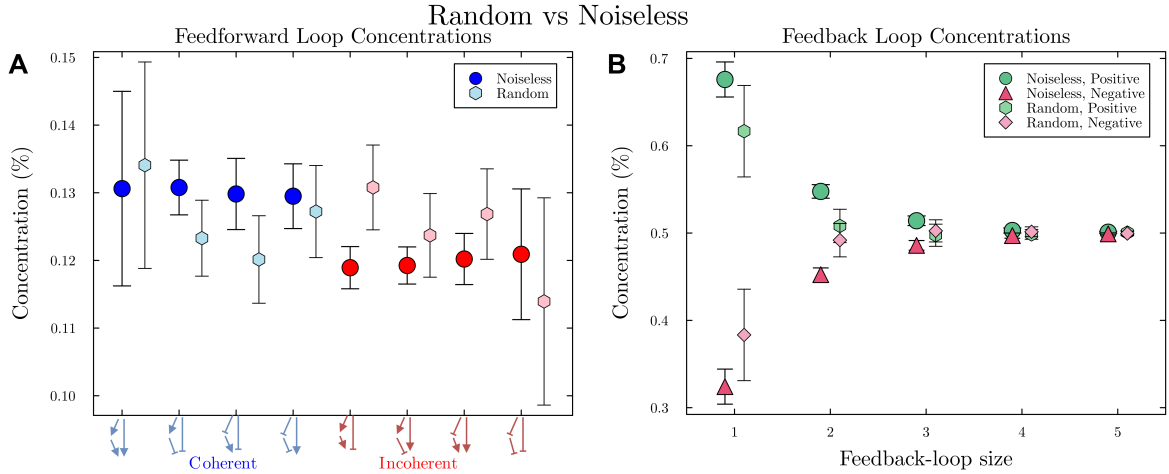

Figure S40: **Enrichment of network motifs as a result of noiseless evolution**. Comparison of 30 populations evolved without noise (darker hues) against non-evolved populations of stable matrices (lighter hues). (A) Average concentration per type of FFL. On the x-axis, we place the type of loop with a diagram. The error bars represent 95% confidence intervals around the averages. (B) Average concentration of positive (green circles and hexagons) and negative (pink triangles and rhombuses) FBLs of different sizes with their 95% confidence intervals around the averages.

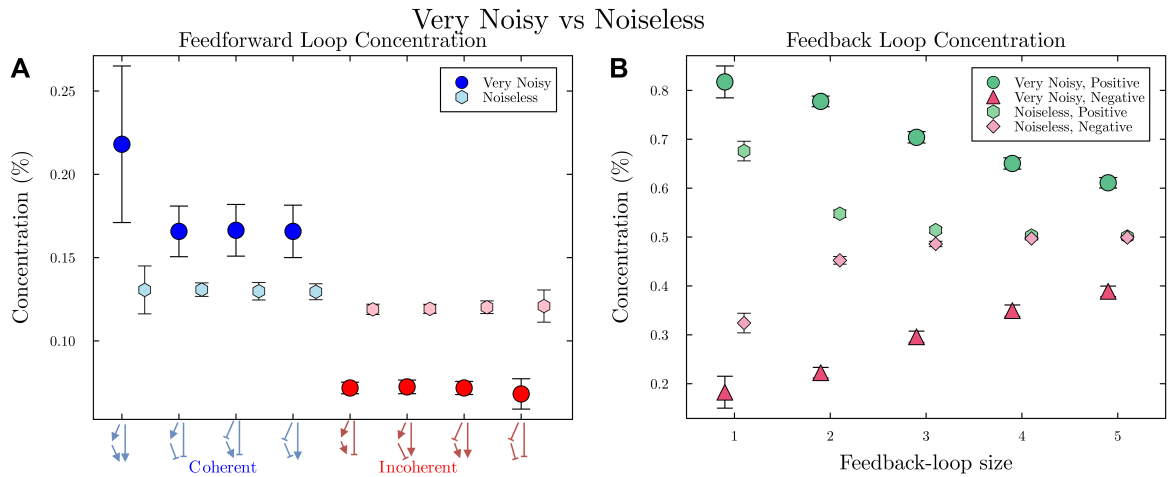

Figure S41: **Enrichment of network motifs as a result of noisy evolution, compared to noiseless evolution**. Comparison of 30 populations evolved with high noise variance ($\theta = 8$; darker hues) and no noise (lighter hues). (A) Average concentration per type of FFL. On the x-axis, we place the type of loop with a diagram. The error bars represent 95% confidence intervals around the averages. (B) Average concentration of positive (green circles and hexagons) and negative (pink triangles and rhombuses) FBLs of different sizes with their 95% confidence intervals around the averages.

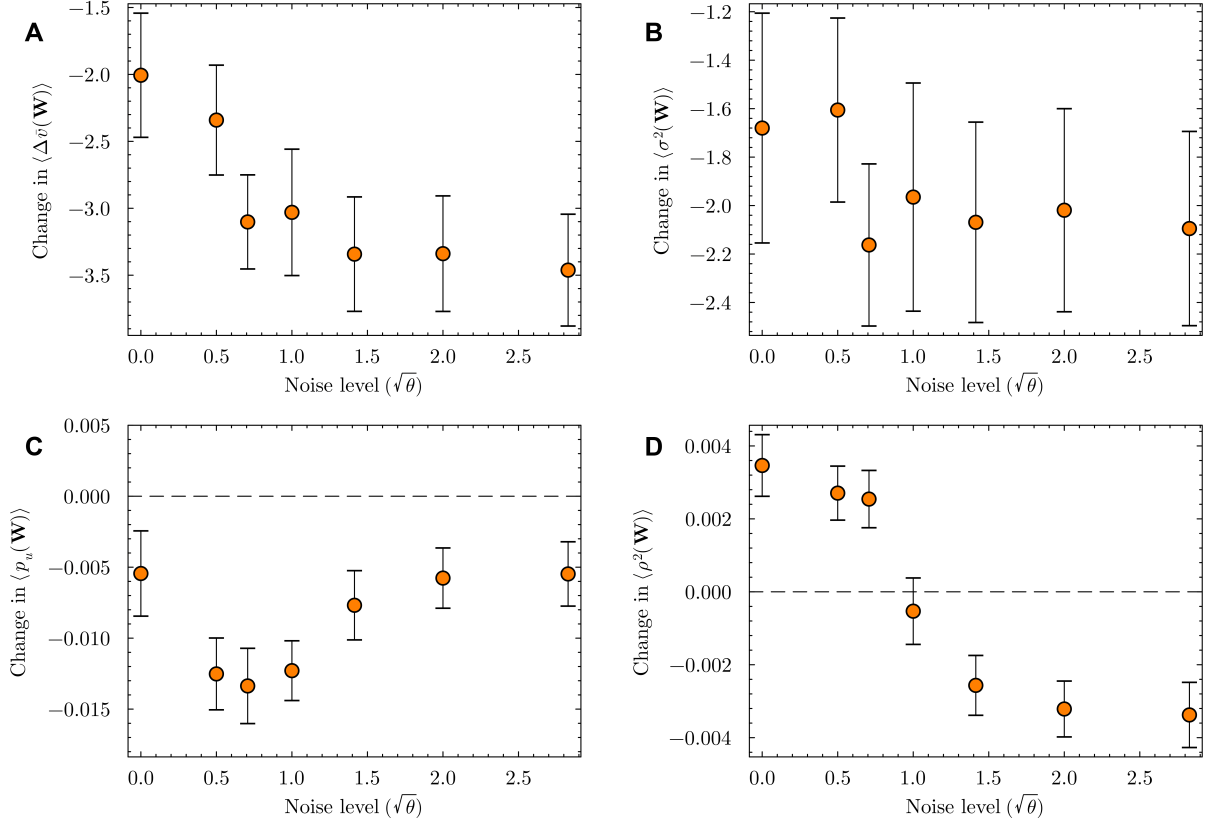

Figure S42: **Change in mutational robustness across evolved populations before and after evolution**. (A) Change in average *stable expression shift*, (B) average *stable expression variance*, (C) average *instability difference*, and (D) *instability variance* after the evolutionary processes for 30 populations across various noise levels (plotted as standard deviations in the x-axis). We fix the noise variance to $\theta_{\text{fixed}} = 1$ to compare all matrices against a single noise distribution. The $\theta = 0$ represents the noiseless scenario as it follows a Bernoulli distribution with no variance (always evaluates to 1). The error bars represent 95% confidence intervals.

### S2.3.3 Random Matrices

This setup assumes a heterogeneous initial population, including networks that may produce unstable phenotypes under deterministic dynamics, and subjects them to directional selection. To generate a population, we simply create `pop_size` random matrices $\boldsymbol{W}$, where each element $W_{ij}$ has a probability $c$ (fixed to $c = 1$ in these simulations, but see Supplementary Material S2.2 for a check on robustness) to be non-zero. If an entry $W_{ij}$ is nonzero, we sample from the normal distribution $W_{ij} \sim \mathcal{N}(\mu_r, \sigma_r^2)$ independently of any other weights. In contrast with the "stable" population in the main paper, we do not impose a stable phenotype expression.

We observe no qualitative changes in the results under this assumption. Specifically, noiseless evolution results in a lower average alignment than noisy evolution ($t = 33.40, P = 10^{-71}$). Additionally, coherent FFLs are enriched in evolved populations relative to non-evolved ones ($t = 12.82, P = 3.6 \times 10^{-16}$), and this enrichment is even stronger in populations evolved under high noise compared to those evolved without noise ($t = 30.81, P = 6.30 \times 10^{-25}$).

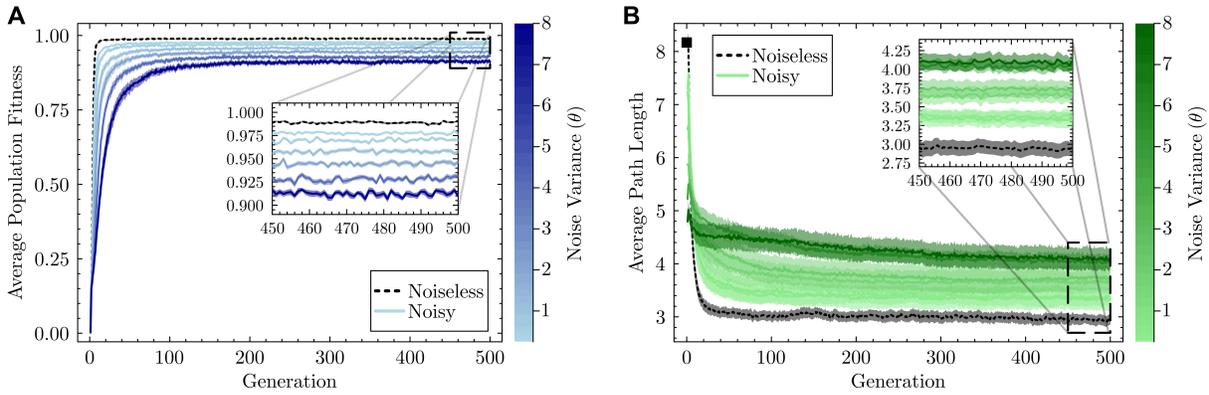

Figure S43: **(A) Evolution of the average population fitness and (B) average mean path length per generation**, across 30 independent replicates at different noise variances $\theta$. Shaded regions represent 95% confidence intervals around the mean in the zoomed-out panels and 68% in the zoomed-in panels. Lighter colors correspond to smaller noise variances. The noiseless case ($\theta = 0$) is shown with a dotted line.

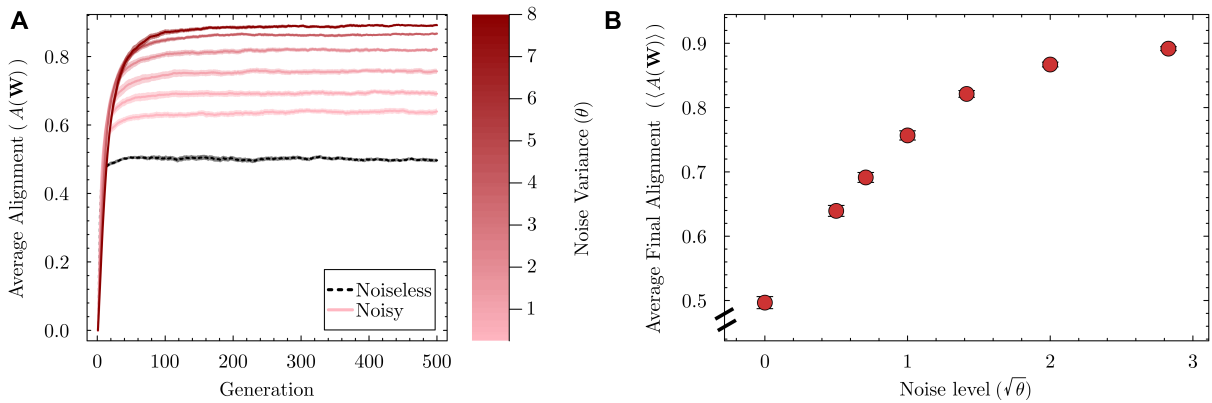

Figure S44: **Evolution of the average alignment scores**. (A) Average alignment score across 30 populations over generations. Each line represents the average alignment score for a population, and the color fillings represent 95% confidence intervals. Lighter colors represent smaller noise variances. The noiseless scenario is plotted with a dotted line. (B) Average alignment scores after evolution from 30 evolved populations across various noise levels $\sqrt{\theta}$. The 95% confidence intervals are plotted along the means.

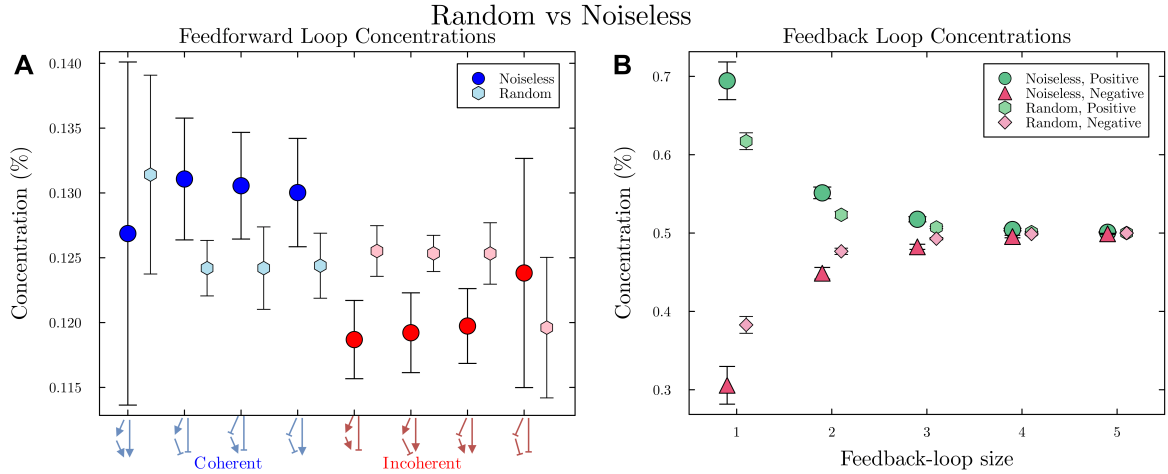

Figure S45: **Enrichment of network motifs as a result of noiseless evolution**. Comparison of 30 populations evolved without noise (darker hues) against non-evolved populations of stable matrices (lighter hues). (A) Average concentration per type of FFL. On the x-axis, we place the type of loop with a diagram. The error bars represent $95\%$ confidence intervals around the averages. (B) Average concentration of positive (green circles and hexagons) and negative (pink triangles and rhombuses) FBLs of different sizes with their $95\%$ confidence intervals around the averages.

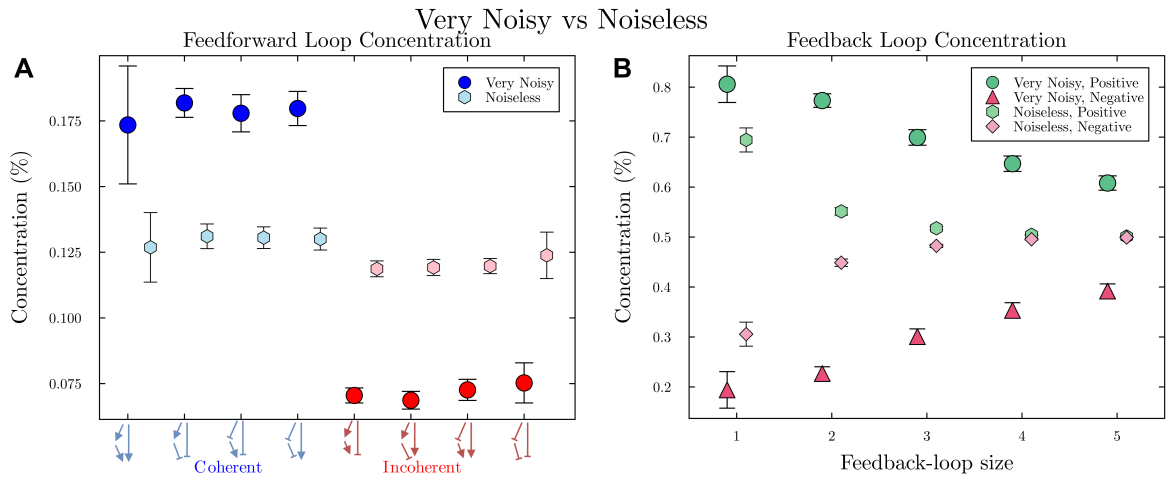

Figure S46: **Enrichment of network motifs as a result of noisy evolution, compared to noiseless evolution**. Comparison of 30 populations evolved with high noise variance ($\theta = 8$; darker hues) and no noise (lighter hues). (A) Average concentration per type of FFL. On the x-axis, we place the type of loop with a diagram. The error bars represent $95\%$ confidence intervals around the averages. (B) Average concentration of positive (green circles and hexagons) and negative (pink triangles and rhombuses) FBLs of different sizes with their $95\%$ confidence intervals around the averages.

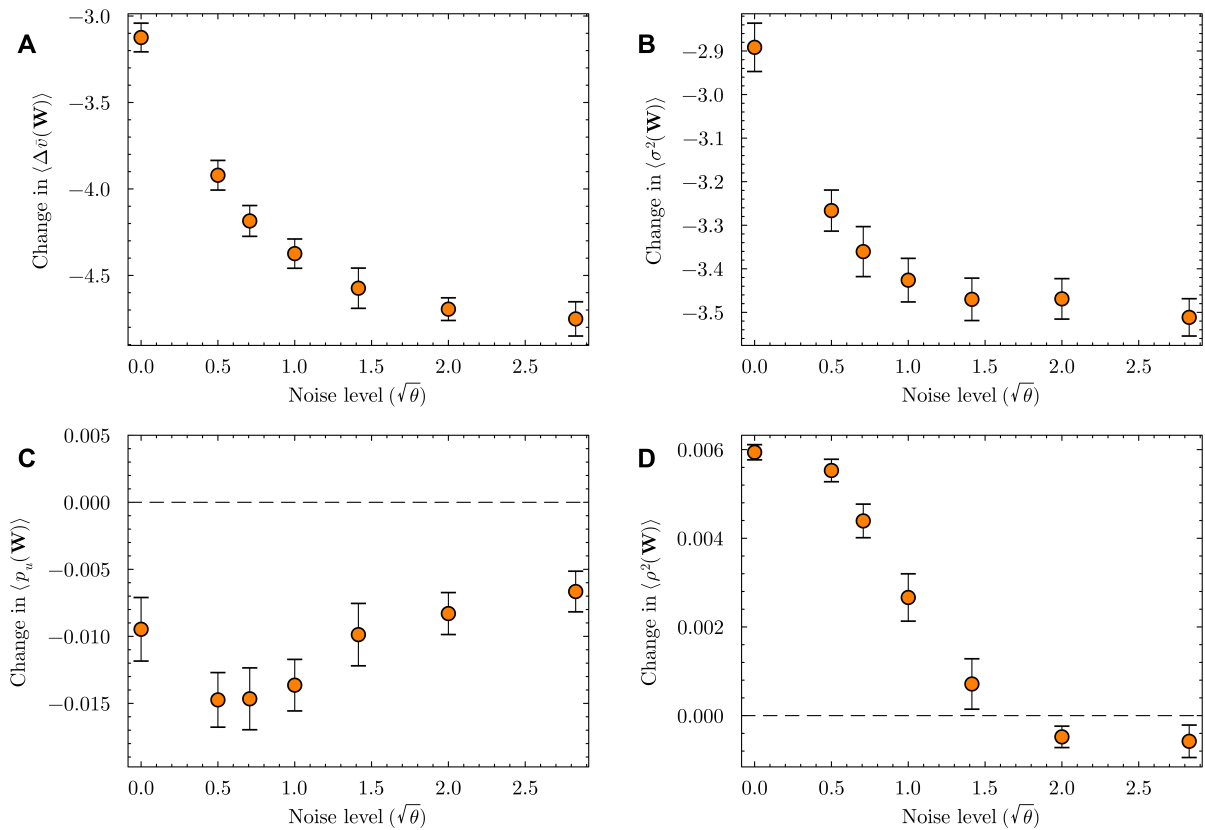

Figure S47: **Change in mutational robustness across evolved populations before and after evolution**. (A) Change in average *stable expression shift*, (B) average *stable expression variance*, (C) average *instability difference*, and (D) *instability variance* after the evolutionary processes for 30 populations across various noise levels (plotted as standard deviations in the x-axis). We fix the noise variance to $\theta_{\text{fixed}} = 1$ to compare all matrices against a single noise distribution. The $\theta = 0$ represents the noiseless scenario as it follows a Bernoulli distribution with no variance (always evaluates to 1). The error bars represent $95\%$ confidence intervals.

### S2.3.4 Ensemble Sample

This configuration assumes a heterogeneous, optimal initial population subject to stabilizing selection. To generate the population, we first define the initial gene expression $\boldsymbol{P}_0$ and the target phenotype $\boldsymbol{P}_{\text{opt}}$, where each element has a probability $1/2$ to be $\pm 1$. Then, we will generate pop_size *different* matrices that deterministically show a stable phenotype such that $\boldsymbol{P} = \boldsymbol{P}_{\text{opt}}$. For that, we generate a random matrix $\boldsymbol{W}$, where each element $W_{ij}$ has a probability $c$ (fixed to $c = 1$ in these simulations, but see Supplementary Material S2.2 for a check on robustness) to be non-zero. If an entry $W_{ij}$ is nonzero, we sample from the normal distribution $W_{ij} \sim \mathcal{N}(\mu_r, \sigma_r^2)$ independently of any other weights. We then develop $\boldsymbol{W}$ into showing a stable phenotype. We add it to the population only if the deterministic phenotype is equal to the target $\boldsymbol{P} = \boldsymbol{P}_{\text{opt}}$.

Although this method takes considerable time to run and generate an initial population, we do not find qualitative differences with the stable and random configurations. The only difference appears in Figure S48 A, where the noiseless scenario begins with a fitness of 1.0 in the first generation.

In our statistical tests, we find that noiseless evolution results in a lower average alignment than noisy evolution ($t = 33.09, P = 2.35 \times 10^{-74}$). Additionally, coherent FFLs are enriched in evolved populations relative to non-evolved ones ($t = 11.19, P = 1.35 \times 10^{-13}$), and this enrichment is even stronger in populations evolved under high noise compared to those evolved without noise ($t = 38.97, P = 3.50 \times 10^{-29}$).

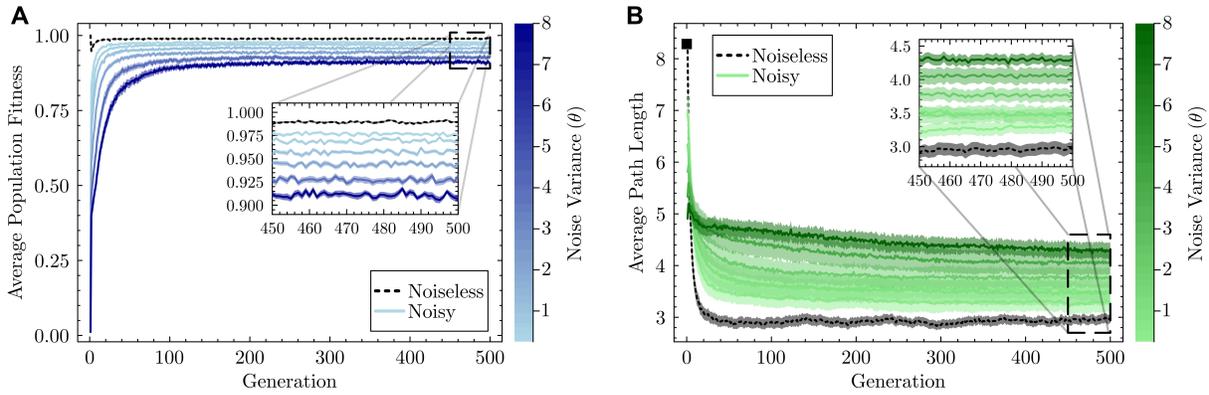

Figure S48: **(A) Evolution of the average population fitness and (B) average mean path length per generation**, across 30 independent replicates at different noise variances $\theta$. Shaded regions represent 95% confidence intervals around the mean in the zoomed-out panels and 68% in the zoomed-in panels. Lighter colors correspond to smaller noise variances. The noiseless case ($\theta = 0$) is shown with a dotted line.

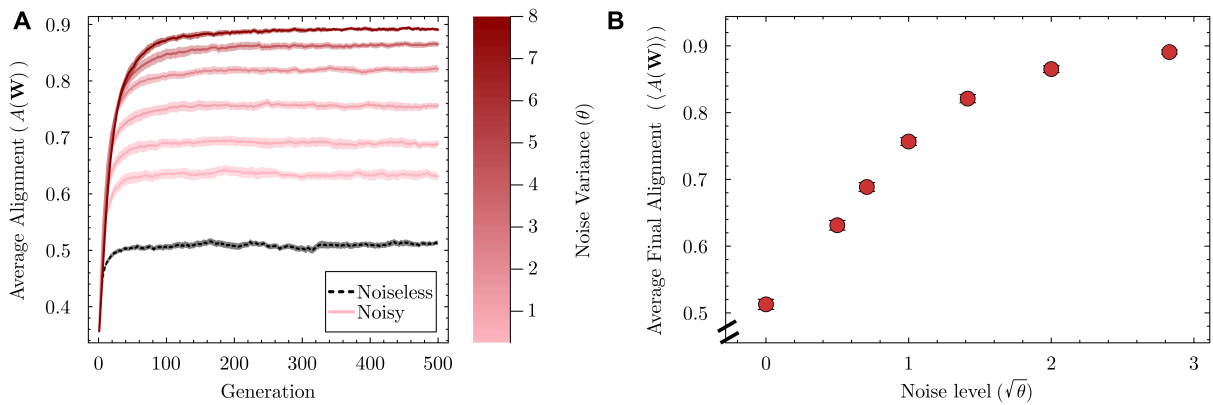

Figure S49: **Evolution of the average alignment scores**. (A) Average alignment score across 30 populations over generations. Each line represents the average alignment score for a population, and the color fillings represent 95% confidence intervals. Lighter colors represent smaller noise variances. The noiseless scenario is plotted with a dotted line. (B) Average alignment scores after evolution from 30 evolved populations across various noise levels $\sqrt{\theta}$. The 95% confidence intervals are plotted along the means.

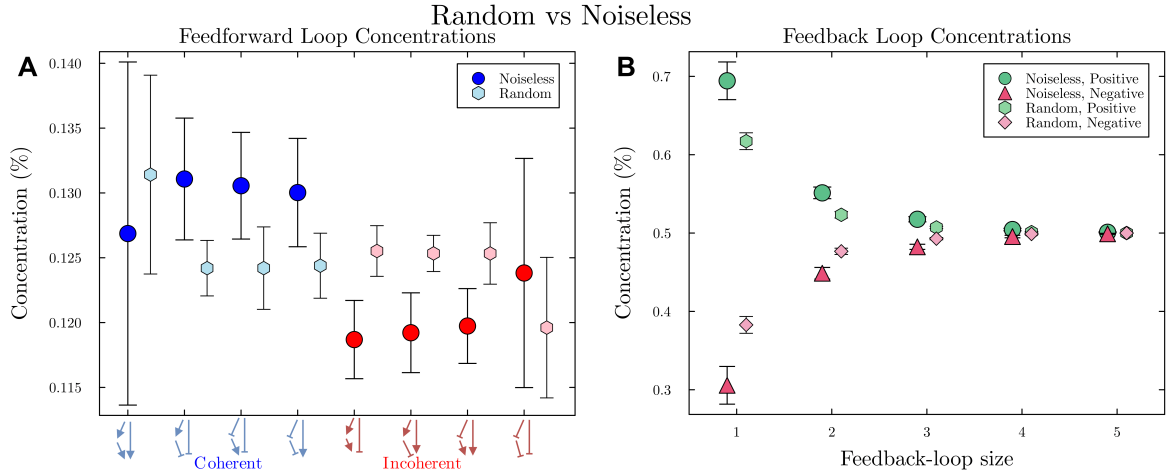

Figure S50: **Enrichment of network motifs as a result of noiseless evolution**. Comparison of 30 populations evolved without noise (darker hues) against non-evolved populations of stable matrices (lighter hues). (A) Average concentration per type of FFL. On the x-axis, we place the type of loop with a diagram. The error bars represent $95\%$ confidence intervals around the averages. (B) Average concentration of positive (green circles and hexagons) and negative (pink triangles and rhombuses) FBLs of different sizes with their $95\%$ confidence intervals around the averages.

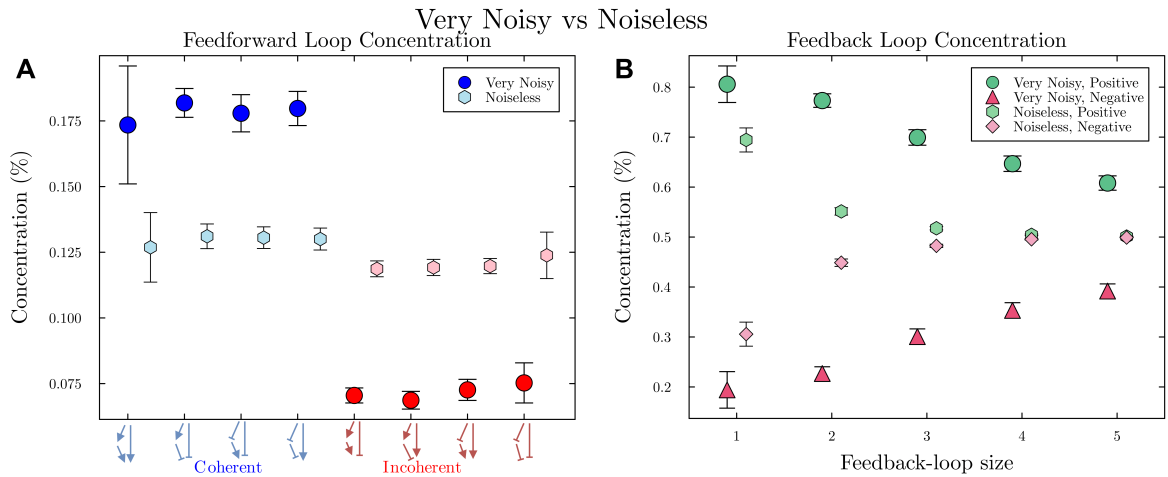

Figure S51: **Enrichment of network motifs as a result of noisy evolution, compared to noiseless evolution**. Comparison of 30 populations evolved with high noise variance ($\theta = 8$; darker hues) and no noise (lighter hues). (A) Average concentration per type of FFL. On the x-axis, we place the type of loop with a diagram. The error bars represent $95\%$ confidence intervals around the averages. (B) Average concentration of positive (green circles and hexagons) and negative (pink triangles and rhombuses) FBLs of different sizes with their $95\%$ confidence intervals around the averages.

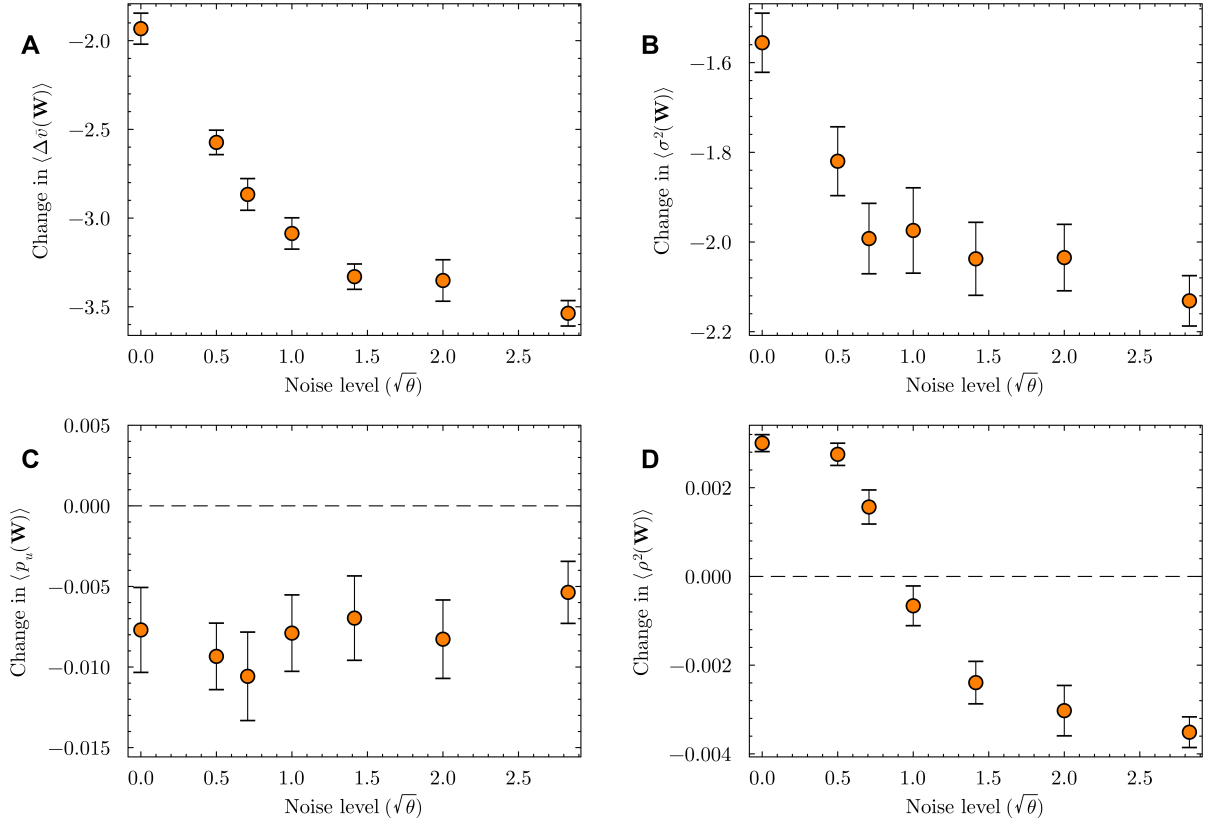

Figure S52: **Change in mutational robustness across evolved populations before and after evolution**. (A) Change in average *stable expression shift*, (B) average *stable expression variance*, (C) average *instability difference*, and (D) *instability variance* after the evolutionary processes for 30 populations across various noise levels (plotted as standard deviations in the x-axis). We fix the noise variance to $\theta_{\text{fixed}} = 1$ to compare all matrices against a single noise distribution. The $\theta = 0$ represents the noiseless scenario as it follows a Bernoulli distribution with no variance (always evaluates to 1). The error bars represent $95\%$ confidence intervals.